\documentclass[prb, notitlepage, superscriptaddress,  reprint, twocolumn]{revtex4-1}
\usepackage{comment}
\usepackage{enumerate}
\usepackage{amssymb}
\usepackage{amsmath}
\usepackage{braket}
\usepackage{graphicx}
\usepackage[usenames,dvipsnames]{color}
\usepackage{tensor}
\usepackage{empheq}
\usepackage{capt-of}
\usepackage[normalem]{ulem} 
\usepackage{soul} %
\usepackage{MnSymbol}

\usepackage[colorlinks,bookmarks=false,citecolor=NavyBlue,linkcolor=OliveGreen,urlcolor=blue]{hyperref}
\newcommand{\be}{\begin{equation}}
\newcommand{\ee}{\end{equation}}
\newcommand{\ba}{\begin{aligned}}
\newcommand{\ea}{\end{aligned}}
\newcommand{\bw}{\begin{widetext}}
\newcommand{\ew}{\end{widetext}}

\newcommand{\bea}{\begin{eqnarray}}
\newcommand{\eea}{\end{eqnarray}}

\def\doi{http://dx.doi.org/}

\begin{document}
\title{Noninteracting fermionic systems with localized losses: Exact results in the hydrodynamic limit}
\author{Vincenzo Alba}
\affiliation{Institute  for  Theoretical  Physics, Universiteit van Amsterdam,
Science Park 904, Postbus 94485, 1098 XH Amsterdam,  The  Netherlands}

\author{Federico Carollo}
\affiliation{Institut f\"ur Theoretische Physik, Universit\"at T\"ubingen, Auf der Morgenstelle 14, 72076 T\"ubingen, Germany}

\begin{abstract}
We investigate the interplay between unitary and nonunitary dynamics after a quantum quench 
in a noninteracting fermionic chain. In particular, we 
consider the effect of localized loss processes,  for which fermions are 
added and removed {\it incoherently} at the center of the chain. We focus on the 
hydrodynamic limit of large distances from the localized losses and of  long times, with their ratio being fixed. In this 
limit, the localized losses gives rise to an effective imaginary delta potential (nonunitary impurity), and the time-evolution 
of the local correlation functions admits a simple hydrodynamic description in terms 
of the fermionic occupations in the initial state and the reflection and transmission amplitudes 
of the impurity. We derive this hydrodynamic framework from the {\it ab initio} calculation of 
the microscopic dynamics. This allows us to analytically characterize the effect of losses 
for several theoretically relevant initial states, such as a uniform Fermi sea, 
homogeneous product states, or the inhomogeneous state obtained by joining two Fermi 
seas. In this latter setting, when both gain and loss processes are present, we observe the emergence of 
exotic nonequilibrium steady states with stepwise uniform density profiles. 
In all instances, for strong loss and gain rates the coherent dynamics of the system is 
arrested, which is a manifestation of the celebrated quantum Zeno effect. 

\end{abstract}

\maketitle

\section{Introduction}
\label{sec:intro}

The interaction between a many-body quantum system 
and its environment can give rise to exotic and 
counterintuitive out-of-equilibrium behavior. 
One of the most intriguing is the so-called quantum Zeno 
effect~\cite{degasperis-1974,misra-1977,facchi-2002}: 
As a consequence of the interaction with an environment, 
for instance performing some type of repeated measurement on the quantum system, the coherent Hamiltonian dynamics freezes. 
In nonequilibrium settings, this effect has been shown to be responsible for suppression of transport in quantum systems \cite{bernard2018,carollo2018,popkov2018}. 
On the other hand, dissipation can be also exploited to engineer desired quantum states~\cite{lin2013}, 
to perform quantum computation~\cite{verstraete-2009}, or even to 
prepare topological states of matter~\cite{diehl-2011}. 
The possibility of analyzing the 
interplay between dissipation and quantum 
criticality~\cite{vicari-2018,rossini-2019a,nigro-2019,rossini-2019,di-meglio-2020,rossini2021coherent} 
is also particularly intriguing. 
However, unfortunately, 
modeling the system-environment interaction  
within an analytic or numerical framework is in general 
a daunting task. 

In Markovian regimes,  the Lindblad equation provides a well-defined mathematical framework to 
treat open quantum systems~\cite{petruccione}. Still, exact results for the Lindblad equation are 
rare~\cite{prosen-2008,prosen-2011,prosen-2014,prosen-2015,
znidaric-2010,znidaric-2011,medvedyeva-2016,ilievski2017dissipation,buca-2020,bastianello-2020,essler-2020,ziolkowska-2020}, 
with the notable exception of noninteracting systems with {\it linear} dissipators~\cite{prosen-2008}. 
Interestingly, also a perturbative field-theoretical treatment of the Lindblad equation is  possible~\cite{sieberer-2016}. 
Furthermore, the recent discovery of Generalized Hydrodynamics~\cite{bertini-2016,olalla-2016} (GHD) 
triggered a lot of interest in understanding whether the hydrodynamic framework could be 
extended to open quantum systems~\cite{bouchoule-2020,bastianello-2020,Friedman_2020,deleeuw-2021,denardis2021}. 
Remarkably, for simple free-fermion setups it is possible 
to apply the so-called quasiparticle picture~\cite{calabrese-2005,fagotti-2008,alba-2017,alba-2018} to 
describe the quantum information spreading~\cite{alba-2021,maity-2020}
in the presence of global gain/loss dissipation. 

In this paper, we focus on the hydrodynamic description of the out-of-equilibrium dynamics of one-dimensional 
free-fermion systems in the presence of localized dissipation, namely a dissipative impurity. 
This setting is nowadays the focus of 
growing interest~\cite{dolgirev-2020,jin-2020,maimbourg-2020,
froml-2019,tonielli-2019,froml-2020,krapivsky-2019,krapivsky-2020,rosso-2020,vernier-2020}, since this type of dissipation can also be engineered 
in experiments with optical lattices~\cite{gericke-2008,brazhnyi-2009,zezyulin-2012,barontini-2013,patil-2015,labouvie-2016}. 
 Recent experiments also aim at investigating the effect of localized losses in quantum transport in fermionic systems~\cite{lebrat2019quantized,corman2019quantized}. 
In particular, we consider here the case of localized gain and loss of fermions. Our work 
takes inspiration from Ref.~\onlinecite{krapivsky-2019} (see also 
Ref.~\onlinecite{krapivsky-2020} for similar results in a bosonic chain) which deals with the case of a fully-occupied noninteracting fermionic chain subject to losses. (The effects of losses on a uniform Fermi sea have also been studied in Ref.~\onlinecite{froml-2019}.) Here, we consider several  homogeneous as well as  inhomogeneous out-of-equilibrium initial states.

The actual setup of interest is illustrated in Fig.~\ref{fig0:cartoon}. 
An infinite chain is subject to both gain and loss processes 
with rates $\gamma^+$ and $\gamma^-$, respectively. 
The dissipation acts at the center of the chain ($x=0$), removing or 
adding fermions {\it incoherently}. 
Here we consider the dynamics ensuing from a homogeneous initial state [see Fig.~\ref{fig0:cartoon} 
(a)], such as a uniform Fermi sea with generic filling, or initial product states, 
such as the fermionic N\'eel state, in which every other site of the chain 
is occupied. 
Furthermore, we also consider the dynamics from inhomogeneous initial states, as depicted in 
Fig.~\ref{fig0:cartoon} (b). We take as initial state the one obtained 
by joining two Fermi seas with different filling. This is a well-known 
setup to study quantum transport in one-dimensional 
systems. In the absence of dissipation it has been studied in Ref.~\onlinecite{viti-2016}. 
If one of the two chains is empty, this becomes the 
so-called geometric quench~\cite{mossel-2010}. If the left chain is fully-occupied 
the setup is that of the domain-wall quench~\cite{antal-1999}. 

In all these cases, we show that the evolution of the fermionic correlators 
$G_{x,y}:=\langle c_x^\dagger c_y\rangle$ is fully captured 
by a simple hydrodynamic picture, which we derive from the exact 
solution of the microscopic Lindblad equation. The hydrodynamic regime holds in the 
space-time scaling (or hydrodynamic) limit of large times and positions (see Fig.~\ref{fig0:cartoon}) 
$x,y,t\to\infty$ with their ratios $\xi_x:=x/(2t)$ and $\xi_y:=y/(2t)$ fixed. 
Crucially, in the hydrodynamic limit the local dissipation acts as an effective 
delta potential, with momentum-dependent reflection and transmission amplitudes 
that depend on the dissipation rate. 
This becomes manifest in the singular behavior at $x=0$ of the profile of 
local observables. For arbitrary $\xi_x$ and $\xi_y$ the hydrodynamic result contains detailed information 
about the model and the quench, and it can be derived easily only in a few cases. Interestingly, 
for $\xi_x\approx\xi_y$ the hydrodynamic result can be expressed entirely in terms of the 
initial fermionic occupations and the effective reflection and transmission amplitudes of the 
dissipative impurity. This is reminiscent of what happens in the absence of 
dissipation~\cite{viti-2016}.

Our findings demonstrate how a quantum Zeno effect~\cite{degasperis-1974,misra-1977} 
arises quite generically in the strong dissipation limit. 
In the presence of localized losses we show that the 
depletion of a uniform state, both at equilibrium as well as out-of-equilibrium 
after a quantum quench, is arrested for large dissipation rates. Similarly, 
quantum transport between two unequal Fermi seas is inhibited. What happens is that 
for strong dissipation, the central site is continuously subject to 
particle injection or ejection, and this determines a constant projection of its state  into the occupied or empty state. This projection effectively disconnects the central site from the rest of the chain. In turn, this effect hinders the depletion of the uniform state as well as the particle transport between the two halves of the chain \cite{carollo2018}. This interpretation can also be formalized by considering that for large rates $\gamma^{\pm}$, the Hamiltonian acts as a perturbative effect and the exchange of fermions between the central site and the rest can only take place at a rate $1/\gamma^{\pm}$ \cite{popkov2018}. This is a clear manifestation of a Zeno effect in dissipative nonequilibrium settings\cite{bernard2018,carollo2018,popkov2018}. Furthermore, in such a strong dissipation limit, the spatial profile 
of the fermionic density is expressed  in terms of the Wigner 
semicircle law, reflecting that the scattering with the impurity is 
``flat'' in energy. 

Finally, we discuss the dynamics starting from two unequal Fermi seas 
in the presence of balanced gain and loss dissipation, i.e., with $\gamma^+=\gamma^-$. 
It is well-known that in the absence of dissipation a 
Non-Equilibrium Steady State (NESS)~\cite{sabetta-2013,viti-2016} develops 
around $x=0$. The NESS exhibits the 
correlations of a boosted Fermi sea. 
For balanced loss/gain dissipation, an interesting ``broken" (piecewise homogeneous)
NESS appears. The corresponding density profile has a step-like 
structure with a discontinuity at $x=0$, 
reflecting once again that the local dissipation mimics an effective delta potential. 

The manuscript is organized as follows. In section~\ref{sec:model} we 
introduce the model, the Lindblad treatment of localized gain and 
losses, and the different quench protocols. In section~\ref{sec:warm-up} 
we focus on the effect of losses on homogeneous out-of-equilibrium 
states. In subsection~\ref{sec:ferro} we consider the case of 
localized losses in the fully-filled state, which was considered in 
Ref.~\onlinecite{krapivsky-2019}. In subsection~\ref{sec:neel} we 
discuss losses on the out-of-equilibrim state emerging after the 
quench from the N\'eel state. In section~\ref{sec:ness} we 
focus on the dynamics starting from inhomogeneous initial states. 
In subsection~\ref{sec:DW} we generalize the results of section~\ref{sec:warm-up} 
to the domain-wall quench. In subsection~\ref{sec:ness1} 
we discuss the quench from the two Fermi seas. We conclude 
and discuss future perspectives in section~\ref{sec:conc}. In Appendix~\ref{app:gain-losses} 
we present details on how to derive the solution of the problem with both gain and loss 
dissipation given the solution for dissipative loss only. In Appendix~\ref{sec:asy} 
we derive the reflection amplitude for the effective delta potential describing 
the dissipative impurity. In Appendix~\ref{sec:app-tech} we report the 
derivation of the results of section~\ref{sec:ness1}. Finally, 
in Appendix~\ref{sec:equal} we discuss the effect of losses on a uniform 
Fermi sea.

\section{Noninteracting fermions with gain and loss: The protocols}
\label{sec:model}

%
\begin{figure}[t]
\includegraphics[width=0.5\textwidth]{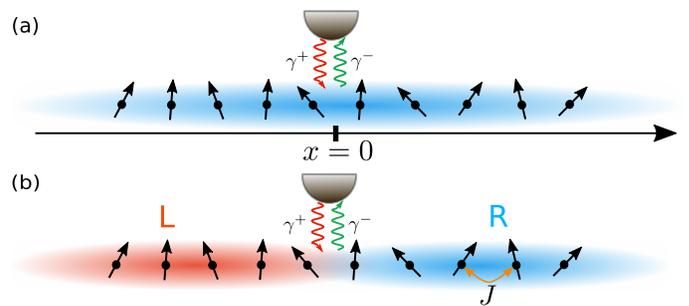}
\caption{An infinite free-fermion chain with localized  
 gain and loss processes acting at the center of the chain. Here 
 $\gamma^\pm$ are the dissipation rates and $J=1$ the hopping. 
 In (a) the chain is initially prepared in a homogeneous state 
 $|\Psi\rangle$. Here we consider the case with $|\Psi\rangle$ 
 being the fully occupied state $|F\rangle$, a Fermi sea with generic filling $k_F$ and the 
 fermionic N\'eel state. In (b) the initial state 
 is obtained by joining two semi-infinite homogeneous chains $L$ and 
 $R$, prepared in two different states.  
}
\label{fig0:cartoon}
\end{figure}
%
In this paper, we consider the infinite free-fermion chain defined by the tight-binding Hamiltonian 
\begin{equation}
\label{eq:ham}
	H=\sum_{x=-\infty}^\infty(c_x^\dagger c_{x+1}+c^\dagger_{x+1}c_x)\, ,
\end{equation}
where $c_x^\dagger,c_x$ are creation and annihilation operators at the different sites $x$ of the chain. They obey canonical anticommutation relations. The  Hamiltonian in Eq.~\eqref{eq:ham} becomes diagonal after taking a Fourier transform 
with respect to $x$. One can indeed define the fermionic operators $b_k$ as 
\begin{equation}
b_k:=\sum_{x=-\infty}^\infty e^{-ikx}c_x,\quad 
c_x=\int_{-\pi}^\pi \frac{dk}{2\pi} e^{i k x}b_k\, ,
\end{equation}
and in terms of these operators, Eq.~\eqref{eq:ham} is equivalent to   
\begin{equation}
\label{eq:ham-k}
H=\int_{-\pi}^\pi\frac{dk}{2\pi} \varepsilon_k b^\dagger_k b_k\, ,\quad 
\varepsilon_k:=2\cos(k)\, .
\end{equation}
The Hamiltonian $H$ conserves the particle number. At a fixed density $n_f=k_F/\pi$, the ground state can be obtained from the Fermi vacuum $\ket{0}$, by occupying the quasi-momenta $b_k$ with single-particle energies in $k\in[-k_F,k_F]$, where $k_F$ 
is the Fermi momentum.  For $n_f=1$ ($k_F=\pi$) one has the 
fully-filled state $|F\rangle$, which is a product 
state. For $0<k_F<\pi$ the ground state of~\eqref{eq:ham} is instead critical, 
i.e., with power-law decaying correlation functions. 
For later convenience, we define here the group velocity $v_k$ 
of the fermions as 
\begin{equation}
\label{eq:v-k}
v_k:=\frac{d\varepsilon_k}{dk}=-2\sin(k)\, . 
\end{equation}
In addition to the Hamiltonian contribution, we consider  a dynamics which is also affected by localized gain/loss processes at the center of the chain [see Fig.~\ref{fig0:cartoon}]. To account for these dissipative contributions, we exploit the formalism of quantum master equations \cite{petruccione}. The time-evolution of the system state $\rho_t$ is implemented by a Lindblad generator, through the following equation
\begin{equation}
\label{eq:lind}
\frac{d\rho_t}{dt}=-i[H,\rho_t]+\sum_{i=+,-}\left(L^{i}\rho_t L^{i\, \dagger}-\frac{1}{2}\{L^{i\,\dagger} L^{i},\rho_t\}\right)\, .
\end{equation}
Here, the so-called jump operators $L^{i}$ are given by  $L^+=\sqrt{\gamma^+}c^\dagger_0$ and 
$L^-=\sqrt{\gamma^-}c_0$ (see Fig.~\ref{fig0:cartoon} for a pictorial 
definition), and account for gain and loss, with rates $\gamma^+$ and $\gamma^-$, respectively. 

The relevant information about the system is contained in the fermionic two-point correlation functions
\begin{equation}
	G_{x,y}(t):=\mathrm{Tr}(c^\dagger_x c_y\rho(t))\, . 
\end{equation}
The dissipative dynamics of this {\it covariance matrix} is obtained as 
\begin{multline}
\label{eq:G}
G(t)=
e^{t\Lambda}G(0)e^{t\Lambda^\dagger}+\int_0^t dze^{(t-z)\Lambda}\Gamma^+ e^{(t-z)\Lambda^\dagger}, 
\end{multline}
with $G(0)$ being the matrix containing the initial correlations. 
The matrix $\Lambda$ is defined as 
\begin{equation}
	\label{eq:lambda}
	\Lambda=ih-\frac{1}{2}(\Gamma^++\Gamma^-), 
\end{equation}
where $h=\delta_{|x-y|,1}$ implements the Hamiltonian contribution while 
$\Gamma^\pm=\gamma^\pm\delta_{x,0}$ account for the localized dissipative effects.  
The correlation functions $G_{x,y}$ in~\eqref{eq:G} satisfy the 
linear system of equations  (we drop the explicit time dependence when this does not generate confusion)
\begin{multline}
\label{eq:one}
\frac{d G_{x,y}}{dt}=i(G_{x+1,y}+G_{x-1,y}-G_{x,y+1}-G_{x,y-1})\\
-\frac{\gamma^++\gamma^-}{2}(\delta_{x,0}G_{x,y}+\delta_{y,0}G_{x,y})+
	\gamma^+\delta_{x,0}\delta_{y,0}. 
\end{multline}
We mainly consider the loss process, setting 
$\gamma^+=0$ in~\eqref{eq:one} and~\eqref{eq:G}. This is not a severe 
limitation since the knowledge of $G_{x,y}$ for 
$\gamma^+=0$ is sufficient to reconstruct $G_{x,y}$ 
also in cases of a nonzero $\gamma^+$ (see Appendix~\ref{app:gain-losses}). 
We also notice that equations of the type of~\eqref{eq:one} can be efficiently numerically 
solved by standard iterative methods, such as the Runge-Kutta method~\cite{press2007numerical}. 
This is especially useful to treat the case of  non-quadratic Liouvillians, for instance, 
in the presence of dephasing or incoherent hopping~\cite{alba-2021}. In our case this is not 
necessary because the solution of~\eqref{eq:one} is given by~\eqref{eq:G}. Indeed, Eq.~\eqref{eq:G} can 
be evaluated numerically after noticing that the matrix $\Lambda$ is $L\times L$, and it can be diagonalized 
with a computational cost ${\mathcal O}(L^3)$. This allows to efficiently evaluate the integral in~\eqref{eq:G}.

In the following sections we discuss the effect of gain/loss dissipation in several theoretically and 
experimentally relevant 
situations. We consider both equilibrium as well as out-of-equilibrium systems, i.e. after 
a quantum quench~\cite{polkovnikov2011colloquium,eisert2015quantum,gogolin2016equilibration,dalessio2016quantum,calabrese-2016}. At equilibrium we are interested in understanding 
how the local dissipation affects the critical correlations of a 
homogeneous Fermi sea with arbitrary filling $0<k_F<\pi$. We also review the effect of losses 
in the non-critical state $|F\rangle$, which was discussed in 
Ref.~\onlinecite{krapivsky-2019}. 
Furthermore, we consider the case in which the initial state is a product state that 
is however not an eigenstate of the Hamiltonian~\eqref{eq:ham}. 
In the absence of dissipation this is one of the paradigm of quantum quenches. 
The generic out-of-equilibrium dynamics ensuing from an initial product 
state is in fact highly nontrivial, as for instance reflected by the 
ballistic growth of  bipartite entanglement. Interestingly, for integrable systems, this growth is due to the propagation 
of pairs of entangled quasiparticles~\cite{calabrese-2005,fagotti-2008,alba-2017,alba-2018}. 
Our setting thus allows us to investigate the interplay between localized 
dissipation and quench dynamics. For concreteness, we focus on the 
situation in which the initial state is the fermionic N\'eel state 
$|N\rangle:=\prod_{x\,\mathrm{even}} c^\dagger_x|0\rangle$, in which only every other site  is occupied. 

Finally, we consider quenches from inhomogeneous initial states 
obtained by joining two homogeneous Fermi seas with different 
Fermi levels $k_F^l$ and $k_F^r$ [see Fig.~\ref{fig0:cartoon}(b)]. 
The choice $k_F^l=\pi/2$ and $k_F^r=0$ corresponds 
to the so-called geometric quench~\cite{mossel-2010}, whereas $k_F^l=\pi$ and $k_F^r=0$ 
to the domain-wall quench~\cite{antal-1999,gobert-2005,antal-2008,allegra-2016,gamayun-2020}. The case with 
$k_F^l\ne k_F^r$ is particularly interesting since, in the absence of 
dissipation and at long times, a non-equilibrium steady state (NESS) emerges 
around the interface between the two parts of the chain. Such a NESS exhibits 
the critical correlations of a boosted Fermi sea~\cite{sabetta-2013,viti-2016}. 
Interestingly, the space-time profile of physical observables and of the 
von Neumann entropy in these setups admit an elegant field theory description 
in terms a  Conformal Field Theory in a curved space~\cite{allegra-2016,dubail-2017,brun-2017,dubail-2017a,brun-2018,ruggiero-2019,bastianello-2020,collura-2020}. 
For generic integrable systems without dissipation, 
similar inhomogeneous protocols can be studied by using the 
recently-developed Generalized Hydrodynamics~\cite{bertini-2016,olalla-2016} (GHD).

\section{HOMOGENEOUS out-of-equilibrium states}
\label{sec:warm-up}

In this section we discuss the effect of losses on homogeneous 
out-of-equilibrium states. To introduce the notation, we start by reviewing the quench from the fully-filled state \cite{krapivsky-2019} in section~\ref{sec:ferro}. Note that, in the absence of 
dissipation there is no dynamics since such state is an eigenstate 
of Eq.~\eqref{eq:ham}. We will obtain analogous results in the more general context 
of section~\ref{sec:ness}. In order to study the interplay between 
unitary and dissipative dynamics, in section~\ref{sec:neel} we 
consider the quench from the fermionic N\'eel state. 

\subsection{Fully-filled state}
\label{sec:ferro}

Let us consider the out-of-equilibrium dynamics starting from 
the fully-filled state $|F\rangle$ defined as 
\begin{equation}
	|F\rangle:=\prod_{x=-\infty}^\infty c_x^\dagger|0\rangle\, .
\end{equation}
The above state  is a product state, with 
diagonal correlator $G_{x,y}$, given by 
\begin{equation}
\label{eq:in-g}
G_{x,y}(0)=\delta_{x,y}\, . 
\end{equation}
To solve~\eqref{eq:one} with $\gamma^+=0$ we employ 
a product ansatz~\cite{krapivsky-2019}  for 
$G_{x,y}$. Specifically, we take $G_{x,y}$ of the form 
\begin{equation}
\label{eq:ansatz}
G_{x,y}=\sum_{k=-\infty}^\infty S_{k,x} \bar S_{k,y}\, , 
\end{equation}
where the bar denotes complex conjugation. 
A similar product ansatz will be used in section~\ref{sec:ness}. 
The factorization as in~\eqref{eq:ansatz} arises naturally when 
treating transport problems in free-fermion models~\cite{viti-2016}. 
Eq.~\eqref{eq:ansatz} is consistent with~\eqref{eq:one} provided that 
$S_{k,x}$ satisfies 
\begin{equation}
\label{eq:S-eq}
\frac{dS_{k,x}}{dt}=i[S_{k,x+1}+S_{k,x-1}]-\frac{\gamma^-}{2}\delta_{x,0}S_{k,x}\, . 
\end{equation}
From Eq.~\eqref{eq:in-g} we obtain as initial condition for $S_{k,x}$ 
\begin{equation}
\label{eq:S-in}
	S_{k,x}(0)=\delta_{x,k}\, . 
\end{equation}
Eq.~\eqref{eq:S-eq} is conveniently solved by a combination of 
Laplace transform with respect to time and Fourier transform with 
respect to the space coordinate $x$. Let us define the 
Laplace transform $\widehat S_{k,x}(s)$ as 
\begin{equation}
\label{eq:S-lap}
\widehat S_{k,x}(s)=\int_0^\infty dt e^{-st}S_{k,x}(t). 
\end{equation}
This allows us to rewrite~\eqref{eq:S-eq} as 
\begin{equation}
\label{eq:laplace}
s \widehat S_{k,x}-S_{k,x}(0)=
i[\widehat S_{k,x+1}+\widehat S_{k,x-1}]-\frac{\gamma^-}{2}\delta_{x,0}
\widehat S_{k,x}. 
\end{equation}
We can now perform the Fourier transform with respect to 
$x$, by defining 
\begin{equation}
\label{eq:S-ft}
	\widehat S_{k,q}=\sum_{x=-\infty}^\infty \widehat S_{k,x} e^{-iqx}, 
\end{equation}
with $q\in[-\pi,\pi]$ being the momentum. From now on, we will use $\widehat{S}_{k,q/p}$ to indicate the Laplace and Fourier transform of $S_{k,x}$; instead $\widehat{S}_{k,x/y}$  will stand for the Laplace transform of $S_{k,x/y}$ only.  After substituting in~\eqref{eq:laplace} and 
using the initial condition~\eqref{eq:S-in}, we obtain
\begin{equation}
\label{eq:lap-ft}
s\widehat S_{k,q}-
e^{-i q k}=i\widehat S_{k,q}(e^{iq}+e^{-iq})-\frac{\gamma^-}{2}\widehat S_{k,x=0}. 
\end{equation}
The solution of~\eqref{eq:lap-ft} is straightforward, yielding 
\begin{equation}
\label{eq:s0}
\widehat S_{k,q}=\big[e^{-iqk}-\frac{\gamma^-}{2}\widehat 
S_{k,x=0}\big]\frac{1}{s-2i\cos(q)}. 
\end{equation}
We note that $\widehat S_{k,x=0}$ is conveniently written as  
\begin{equation}
	\widehat S_{k,x=0}=\frac{1}{2\pi}\int_{-\pi}^{\pi} dq \widehat S_{k,q}. 
\end{equation}
We can now take the inverse Fourier transform  in~\eqref{eq:s0}, and using that 
\begin{equation}
	\label{eq:ff-nota}
	\frac{1}{2\pi}\int_{-\pi}^{\pi}dq\frac{e^{iqx}}{s-2i\cos(q)}= \frac{1}{\sqrt{s^2+4}}
	\Big(\frac{2i}{s+\sqrt{s^2+4}}\Big)^{|x|}, 
\end{equation}
we obtain 
\begin{multline}
\label{eq:f-final}
	\widehat S_{k,x}=
	\frac{1}{\sqrt{s^2+4}}\Big(\frac{2i}{s+\sqrt{s^2+4}}\Big)^{|k-x|}
	\\-\frac{\gamma^-/2}{(\gamma^-/2+\sqrt{s^2+4})\sqrt{s^2+4}}
	\Big(\frac{2i}{s+\sqrt{s^2+4}}\Big)^{|k|+|x|}. 
\end{multline}
Note the absolute value $|x|$ in the second term in~\eqref{eq:f-final}. 
The last step is to take the inverse Laplace transform of~\eqref{eq:f-final}. 
This is straightforward for the first term in Eq.~\eqref{eq:f-final}, which 
accounts for the unitary part of the evolution, and  gives 
a term $J_{|x-y|}(2t)$, with $J_x(t)$ the Bessel function of 
the first type. The second term in Eq.~\eqref{eq:f-final} encodes the 
effects of the losses. One can write~\cite{krapivsky-2019}
\begin{equation}
	\label{eq:f-final-1}
	S_{k,x}(t)=i^{|x-k|}J_{|x-k|}(2t)-\frac{\gamma^-}{2} i^{|x|+|k|}K_{|x|+|k|}(t). 
\end{equation}
Here $K_{|x|+|k|}$ is the inverse Laplace transform of the second 
term in Eq.~\eqref{eq:f-final}. To determine $K_{|k|+|x|}$ analytically, 
one can use the inverse Laplace transform 
\begin{multline}
\label{eq:dn}
D_{|x|}:={\mathcal L}^{-1}\Big(\frac{1}{\frac{\gamma^-}{2}+\sqrt{s^2+4}}
	\Big(\frac{2i}{s+\sqrt{s^2+4}}\Big)^{|x|}\Big)=
	i^{|x|}J_{|x|}(2t)\\-i^{|x|}\frac{\gamma^-}{2}\int_{0}^t dz e^{-\gamma^-z/2}
	\Big(\frac{t-z}{t+z}\Big)^{|x|/2}J_{|x|}(2\sqrt{t^2-z^2}), 
\end{multline}
together with the fact that 
\begin{equation}
{\mathcal L}^{-1}\Big(\frac{1}{{\sqrt{s^2+4}}}\Big)=J_0(2t). 
\end{equation}
This allows us to obtain the inverse Laplace of the second 
term in~\eqref{eq:f-final} as  the convolution 
\begin{multline}
	{\mathcal L}^{-1}\Big(\frac{1}{(\frac{\gamma^-}{2}+\sqrt{s^2+4})\sqrt{s^2+4}}
	\Big(\frac{2i}{s+\sqrt{s^2+4}}\Big)^{|k|+|x|}\Big) =\\ 
	\int_0^t d\tau J_0(2(t-\tau))D_{|x|+|k|}(\tau), 
\end{multline}
with $D_{|x|+|k|}$ defined in Eq.~\eqref{eq:dn}. 
We anticipate that we will also employ Eq.~\eqref{eq:dn} in 
section~\ref{sec:ness}. 

Here we are interested in the space-time 
scaling limit $x,k,t\to\infty$, with their ratio fixed. 
We define the two scaling variables $u,\xi_x$ as 
\begin{equation}
	u:=\frac{k}{2t},\quad \xi_x:=\frac{x}{2t}. 
\end{equation}
Since the initial state is homogeneous and the dissipation acts 
at $x=0$ we expect local observables, such as the fermionic 
density, to be even functions of $x$. Thus, we can restrict 
ourselves to $\xi_x>0$. 
The asymptotic behaviour of $J_{|x-k|}$ and $K_{|x|+|k|}$ is 
derived analytically~\cite{krapivsky-2019} and is given by 
\begin{multline}
 \label{eq:asyJ} 
   J_{|x-k|}(2t)\simeq \\ 
   \frac{ \cos \big[2t \sqrt{1 - (u -\xi_x)^2} - 2 t |u-\xi_x|  
   \arccos |u-\xi_x| -\frac{\pi}{4} \big]}
   {\sqrt{\pi t}\, \left[1 - (u -\xi_x)^2\right]^{1/4}}, 
\end{multline}
which holds for  $-1\le u-\xi_x\le1$. For $u,\xi_x$ outside of this interval the 
asymptotic behavior of $J_{|x-k|}$ is subleading in the scaling limit. 
Similarly, one can show that~\cite{krapivsky-2019} 
\begin{multline}
 \label{eq:asyK}     
   K_{|x| + |k|}(t)\simeq \\
   \frac{ \cos \big[2t \sqrt{1 - (\xi_x +|u|)^2} - 2  t (\xi_x +|u|) 
	  \arccos (\xi_x +|u|)
      -\frac{\pi}{4} \big]}
      {\sqrt{\pi t}\, \left[\gamma^-/2 + 2(\xi_x + |u|)\right]\left[1 - 
(\xi_x + |u|)^2
      \right]^{1/4}},
\end{multline}
which holds in the interval  $-1\le |u|+\xi_x\le1$. 
As it will be clear in section~\ref{sec:ness}, the 
scaling behavior of $G_{x,y}$ will be given by a simple formula, 
in the limit $x,y,t\to\infty$ with $x/t$, $y/t$ fixed and 
$|x-y|/t\to0$. 
On the other hand, we should stress that by using~\eqref{eq:f-final-1} together 
with the asymptotic expansions~\eqref{eq:asyJ} and~\eqref{eq:asyK} 
it is possible to obtain the behavior of $G_{x,y}$ for large 
$x,y,t$ with {\it arbitrary} fixed ratios $\xi_x=x/(2t),\xi_y=y/(2t)$. However, as it is 
clear from~\eqref{eq:asyJ} and~\eqref{eq:asyK} the result contains 
detailed information about the quench parameters and dissipation. 

Let us consider the dynamics of the density profile $n_{x,t}$. We have 
\begin{equation}
\label{eq:den1}
n_{x,t}=\sum_{k=-\infty}^\infty |S_{k,x}|^2.
\end{equation}
%
\begin{figure}[t]
\includegraphics[width=0.45\textwidth]{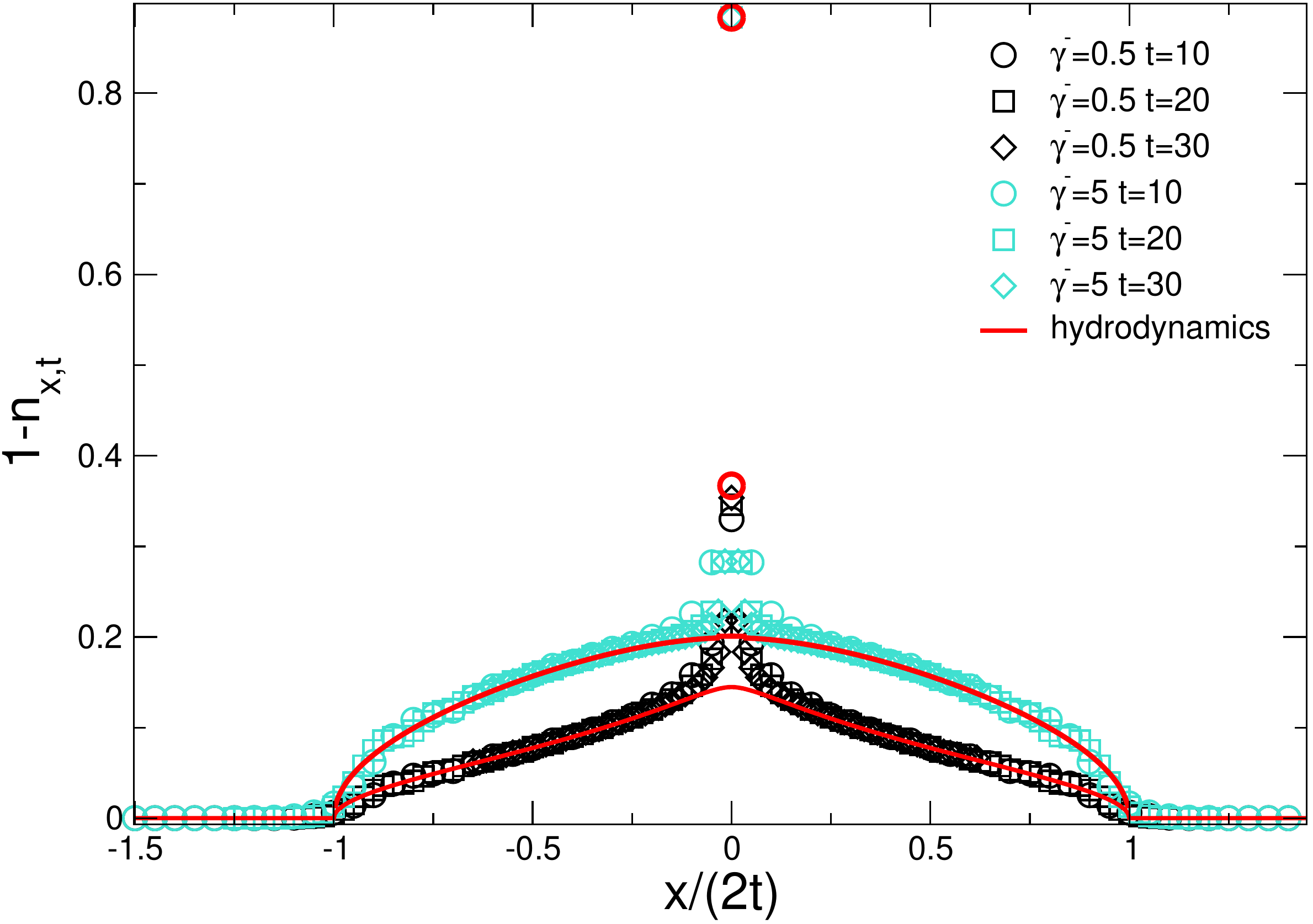}
\caption{ Density profile $n_{x,t}$ in a free fermion chain with localized 
 losses. Here we plot $1-n_{x,t}$ versus $x/(2t)$, 
 with $x$ the position with respect to the center of the chain and 
 $t$ the time. The initial state of the chain is the fully occupied state 
 $|F\rangle$. The symbols are exact numerical data for ``strong'' 
 loss rate $\gamma^-=5$ and ``weak'' loss rate $\gamma^-=0.5$. 
 Lines and the red circle at $x=0$ are the analytic results in 
 the hydrodynamic limit. 
}
\label{fig1:ferro}
\end{figure}
%
The behavior of the density in the space-time scaling limit is 
obtained by using~\eqref{eq:asyJ} and~\eqref{eq:asyK} in~\eqref{eq:den1}. 
Let us assume $\xi_x>0$. We obtain 
\begin{equation}
	\label{eq:1f}
	n_{x,t}=1-\frac{\gamma^-}{\pi}\int_{\arcsin(x/(2t))}^{\pi/2}dq\frac{|v_q|}
	{(\frac{\gamma^-}{2}+|v_q|)^2}\, .
\end{equation}
In deriving~\eqref{eq:1f} we approximated the rapidly oscillating trigonometric 
functions in~\eqref{eq:asyJ} and~\eqref{eq:asyK} with their time average. 
An important remark is that the derivation above is not valid near the origin 
at $x=0$, where the density profile exhibits a singularity. 
For $x=0$ from~\eqref{eq:f-final} we obtain 
\begin{multline}
	S_{k,0}(t)=	
	i^{|k|}J_{|k|}(2t)\\
	-i^{|k|}\frac{\gamma^-}{2}\int_{0}^t dz e^{-\gamma^-z/2}
	\Big(\frac{t-z}{t+z}\Big)^{|k|/2}J_{|k|}(2\sqrt{t^2-z^2}). 
\end{multline}
In the scaling limit $k,t\to\infty$ with $u=k/(2t)$ fixed we have 
\begin{equation}
	\Big(\frac{t-z}{t+z}\Big)^\frac{|k|}{2}\simeq e^{-2|u|z}. 
\end{equation}
This implies 
\begin{equation}
	S_{k,0}(t)\simeq \frac{2i^{|k|}|u|}{\gamma^-/2+2|u|}J_{|k|}(2t). 
\end{equation}
By using the asymptotic expansion of the Bessel function~\eqref{eq:asyJ} we 
obtain 
\begin{equation}
	\label{eq:2f}
	n_{0,t}=\frac{1}{\pi}\int_{-\pi/2}^{\pi/2} 
	dq \frac{|v_q|^2}{\big(\frac{\gamma^-}{2}+|v_q|\big)^2}. 
\end{equation}
We provide an alternative derivation of~\eqref{eq:1f} and~\eqref{eq:2f} 
in section~\ref{sec:ness}. 
It is interesting to observe that from~\eqref{eq:1f} in the limit of large loss 
rate $\gamma^-$ one obtains the Wigner semi-circle law as 
\begin{equation}
	\label{eq:wigner}
	n_{x,t}=1-\frac{1}{\gamma^-}\frac{8}{\pi}\sqrt{1-\frac{x^2}{4t^2}}. 
\end{equation}
The behavior in Eq.~\eqref{eq:wigner} appears also in the case of the 
out-of-equilibrium dynamics from the N\'eel state (see section~\ref{sec:neel}) and 
from inhomogeneous initial states (see section~\ref{sec:ness}). 
Eq.~\eqref{eq:wigner} has a simple physical interpretation. Eq.~\eqref{eq:1f}
in the limit of large $\gamma$ can be rewritten as 
\begin{equation}
	\label{eq:wigner-1}
n_{x,t}=1-\frac{1}{\gamma\pi}\int_{-\pi}^{\pi}
dq v_q\Theta(v_q-x/t). 
\end{equation}
Now the integral in~\eqref{eq:wigner-1} describes the number of holes (equivalently, the absorbed 
fermions) that 
are emitted at the origin and at time $t$ arrive at position $x$. Importantly, 
since $dq v_q=d\epsilon$, this means that the hole is produced at a 
rate $\propto 1/\gamma$  with a uniform distribution in energy, i.e., at 
infinite temperature.

In the limit $\gamma^-\to\infty$ the density remains $n_{x,t}=1$. 
The total number of fermions absorbed at a generic time $t$ in the limit 
$\gamma^-\to\infty$ is given as 
\begin{equation}
	n_{a}:=\int_{-\infty}^\infty dx (1-n_{x,t})=\frac{8}{\gamma^-}t. 
\end{equation}
The number of fermions that are lost at the origin increases linearly 
with time. However, the rate  goes to zero as $\gamma^-\to\infty$, which is consistent with the emergence of a  Zeno effect. These results are checked in Fig.~\ref{fig1:ferro}. 
The symbols are numerical data obtained by using~\eqref{eq:G} for $\gamma^-=0.5$ and 
$\gamma^-=5$. The different symbols correspond to different times. To highlight 
the scaling behavior we plot $1-n_{x,t}$ versus $x/(2t)$. All the data for 
different times collapse on the same curve. 
Note the singularity at $x=0$. Some corrections are visible 
only for very short times. The continuous lines are the 
analytical predictions~\eqref{eq:1f} and~\eqref{eq:2f}, 
and are in perfect agreement with the numerical data.

\subsection{Homogeneous N\'eel quench}
\label{sec:neel}

Let us now discuss the  effect of losses on an out-of-equilibrium 
state arising after the quantum quench from the fermionic N\'eel state. 
The N\'eel state $|N\rangle$ is defined as 
\begin{equation}
	|N\rangle:=\prod_{x\,\mathrm{even}} c_x^\dagger|0\rangle. 
\end{equation}
The initial correlation matrix reads as 
\begin{equation}
	G_{x,y}(0)=\delta_{x,y}, \quad \mathrm{with}\,\,x\,\,\textrm{even}.
\end{equation}
To proceed we impose that the solution of~\eqref{eq:one} is 
factorized as in Eq.~\eqref{eq:ansatz}. 
We obtain the same equation as in~\eqref{eq:S-eq}. 
The initial condition for $S_{k,x}$ is  
\begin{equation}
\label{eq:neel-ini}
S_{k,x}(0)=\delta_{x,k},\quad k\,\,\textrm{even}.
\end{equation}
After performing a Laplace transform with 
respect to time we obtain~\eqref{eq:laplace}. 
Now we can consider separately the cases of $k$ even and $k$ odd. 
Let us first start considering the case with $k$ odd. 
It is straightforward to check that~\eqref{eq:S-eq} together with the 
initial condition~\eqref{eq:neel-ini} implies that $\widehat S_{k,n}=0$ 
for odd $k$. Thus we can restrict ourselves to even $k$. 
For $k$ even we have to distinguish the cases of even $x$ and 
odd $x$. We define $S^{e/o}_{k,x}$, where now $x=0,1,\dots,L/2$ 
labels the ``unit cell'' containing the sites $2x,2x+1$. These 
$S^{e/o}_{k,x}$ satisfy the set of equations 
\begin{align}
	\label{eq:1}
	s\widehat S_{k,x}^{e}-\delta_{k,2x}&=
	i[\widehat S_{k,x}^o+\widehat S_{k,x-1}^o]-
	\frac{\gamma^-}{2}\delta_{x,0}\widehat S^{e}_{k,x}\\
	\label{eq:2}
	s\widehat S_{k,x}^o&=i[\widehat S_{k,x}^e+\widehat S_{k,x+1}^e]. 
\end{align}
We define the Fourier transforms as 
\begin{equation}
\label{eq:ft}
\widehat S_{k,q}^{e/o}=\sum_{x=-\infty}^\infty\widehat S^{e/o}_{k,x}e^{-iqx}. 
\end{equation}
Taking the Fourier transform in~\eqref{eq:1} and~\eqref{eq:2} 
we obtain 
\begin{align}
	\label{eq:lap-ff-1}
	s\widehat S^e_{k,q}-
	e^{-i q k/2}&=i\widehat S_{k,q}^o(1+e^{-iq})-\frac{\gamma^-}{2}\widehat S^e_{k,x=0}\\
	\label{eq:lap-ff-2}
	s\widehat S^e_{k,q}&=i\widehat S_{k,q}^o(1+e^{iq})
\end{align}
Similar manipulations as in section~\ref{sec:ferro} yield 
\begin{multline}
	\label{eq:fin-3}
	\widehat S^e_{k,x}=
	\frac{1}{\sqrt{s^2+4}}\Big(\frac{2i}{s+\sqrt{s^2+4}}\Big)^{|k-2x|}\\
	-\frac{\gamma^-}{2}\frac{1}{(\gamma^-/2+\sqrt{s^2+4})\sqrt{s^2+4}}
	\Big(\frac{2i}{s+\sqrt{s^2+4}}\Big)^{|k|+2|x|}\, ,
\end{multline}
and 
\begin{equation}
	\label{eq:fin-4}
	\widehat S^o_{k,x}=i\int_0^{t}d\tau [\widehat S^e_{k,n}(\tau)+
	\widehat S^e_{k,x+1}(\tau)].
\end{equation}
Eq.~\eqref{eq:fin-3} is the same as for the quench from the 
ferromagnetic state (cf.~\eqref{eq:f-final}) discussed 
in section~\ref{sec:ferro}, after redefining $x\to 2x$, i.e., 
\begin{equation}
	S^e_{k,x}=S_{k,2x}^{|F\rangle}, 
\end{equation}
with $S_{k,x}^{|F\rangle}$ given by~\eqref{eq:f-final}. 
One can use~\eqref{eq:asyJ} and~\eqref{eq:asyK} to obtain the correlators 
$G_{x,y}$ in the space-time scaling limit. We now discuss the dynamics of the 
density profile. We restrict ourselves to even sites $n^e_{n,t}$. 
This is because translation invariance is restored 
by the dynamics at long times and we expect the result for odd sites 
to be the same. 
We now have 
\begin{equation}
	n^e_{x,t}=\sum_{k=-\infty}^\infty |S^e_{2k,x}|^2. 
\end{equation}
One obtains 
\begin{equation}
	\label{eq:ne-den}
	n^e_{x,t}=\frac{1}{2}-\frac{1}{2}\frac{\gamma^-}{\pi}\int_{\arcsin(x/(2t))}^{\pi/2}dq
	\frac{|v_q|}{\big(\frac{\gamma^-}{2}+|v_q|\big)^2}. 
\end{equation}
Note that this is the result obtained for the quench from the 
fully-filled state (see section~\ref{sec:ferro}) divided by two. In addition, once again, the 
density profile is singular in $x=0$. 
The value of the density at $x=0$ is the half of that found in~\eqref{eq:2f}. 
In the absence of dissipation, i.e., for $\gamma^-=0$, 
the fermionic density is uniform and is given as $n_{x,t}=1/2$. 
%
\begin{figure}[t]
\includegraphics[width=0.45\textwidth]{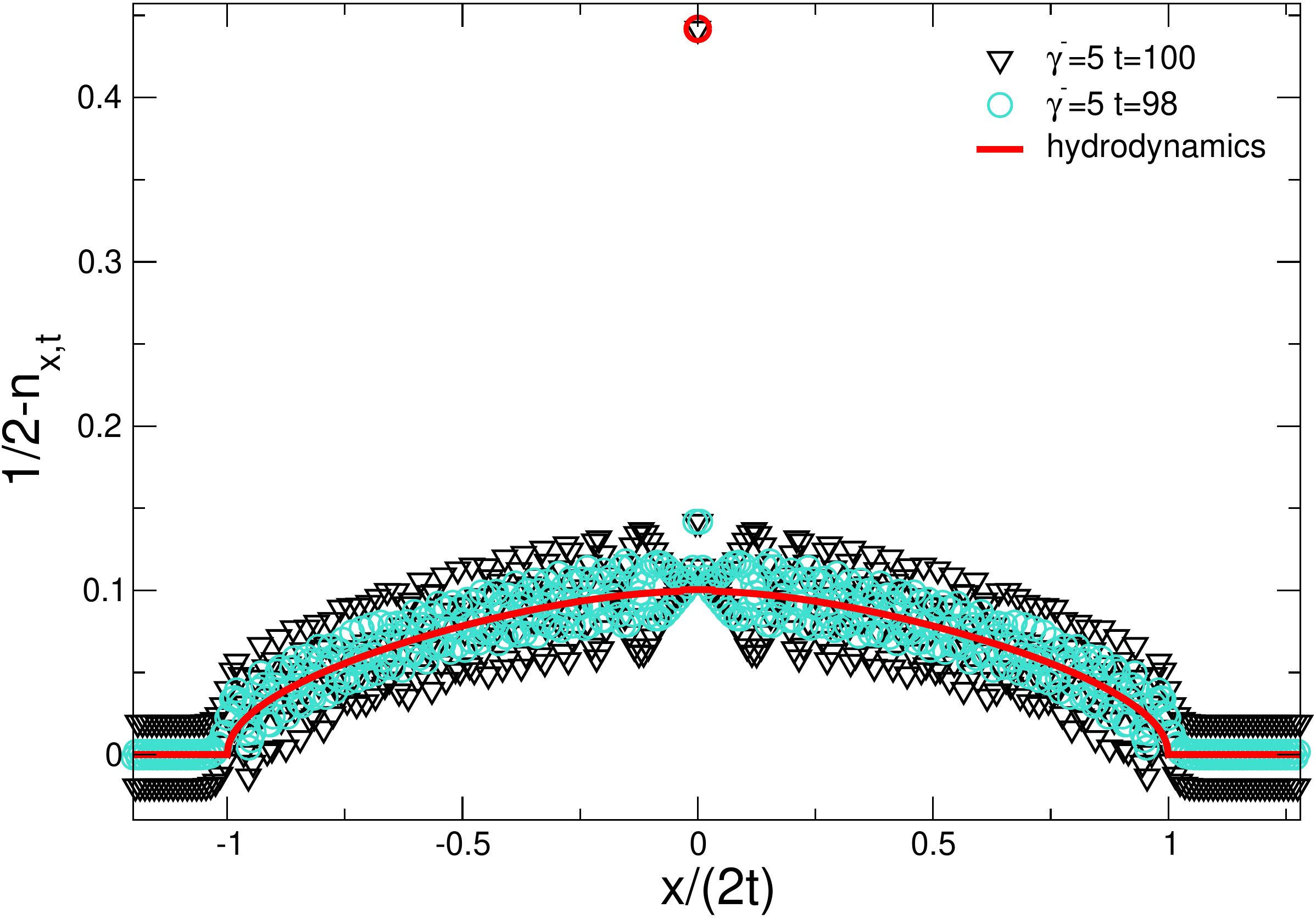}
\caption{ Density profile $n_{x,t}$ in a free fermion chain with local 
 losses: Dynamics starting from the N\'eel state. We plot the shifted 
 density $1/2-n_{x,t}$ versus $x/(2t)$. 
 The symbols are exact numerical data for $\gamma^-=5$ and $t=98$ and 
 $t=100$. Note the strong oscillations with time. 
 Lines are the analytic results in the space-time scaling limit. 
}
\label{fig2:Neel}
\end{figure}
%
For strong dissipation, instead, 
one obtains 
\begin{equation}
	n^e_{x,t}=\frac{1}{2}-\frac{1}{\gamma^-}\frac{4}{\pi}\sqrt{1-
		\frac{x^2}{4t^2}}, 
\end{equation}
which is reminiscent of the Wigner semicircle law in 
Eq.~\eqref{eq:wigner}. It is useful to compare the results in~\eqref{eq:ne-den} 
with numerical data. 
We present some benchmarks in Fig.~\ref{fig2:Neel}. 
In the figure we show $1/2-n_{x,t}$ versus $x/(2t)$. The symbols are 
numerical results obtained by using~\eqref{eq:G}. We only show data for 
$\gamma^-=5$. Strong oscillating corrections are present. They 
disappear in the long time limit $t\to\infty$. Similar corrections are 
also present in the unitary case, i.e., without dissipation. The continuous 
line is the analytic result in the space-time scaling limit. Despite the strong 
oscillations the agreement with the numerical data is satisfactory.

\section{Quenches from inhomogeneous initial states}
\label{sec:ness}

In this section we address the effect of losses in out-of-equilibrium 
dynamics starting from inhomogeneous initial states. We first discuss 
the so-called domain-wall quench in subsection~\ref{sec:DW}. 
This can be straightforwardly treated by using the results of section~\ref{sec:warm-up}. 
We then discuss the generic situation in which the initial state 
is obtained by joining two Fermi seas with different fillings 
in subsection~\ref{sec:ness1}. We show that both the fermionic 
density and the correlation functions admit 
a simple hydrodynamic picture  in the space-time scaling limit.

\subsection{Domain-wall quench}
\label{sec:DW}

Let us consider the domain-wall initial state, in which the left part of the chain 
is fully filled, and the right one is empty. This situation has 
been intensely investigated in the 
past~\cite{antal-1999,bettelheim-2012,gobert-2005,allegra-2016,dubail-2017,collura-2018}. 

The full out-of-equilibrium dynamics ensuing from the domain-wall state 
is obtained by a slight modification of the method employed in section~\ref{sec:warm-up}. 
The same ansatz as in~\eqref{eq:ansatz} holds true, with $S_{k,x}$ 
satisfying~\eqref{eq:S-eq}. The initial condition for $S_{k,x}$ is now 
\begin{equation}
\label{eq:dw-ini}
S_{k,x}(0)=\delta_{k,x}\Theta(-k), 
\end{equation}
where the Heaviside theta function $\Theta(-k)$ takes into account that 
at $t=0$ only the left part of the chain is fully occupied with fermions. 
In taking the Laplace and Fourier transforms of~\eqref{eq:ansatz}, we 
distinguish the case of $k\ge 0$ and $k<0$, obtaining 
\begin{align}
	s\widehat S_{k,q}-e^{-iqk}&=2i\widehat S_{q,k}\cos(q)-\frac{\gamma^-}{2} 
	\widehat S_{k,x=0}\quad\textrm{for}\, k<0\\
s\widehat S_{k,q}&=2i\widehat S_{q,k}\cos(q)-\frac{\gamma^-}{2} \widehat S_{k,x=0}
\quad\textrm{for}\, k\ge0.
\end{align}
The solution of the system above is straightforward and gives 
\begin{align}
	\widehat S_{k,x}=\widehat S_{k,x}^{|F\rangle}\quad&\textrm{for}\,k<0\, ,\\
	\widehat S_{k,x}=0\quad&\textrm{for}\, k\ge 0, 
\end{align}
where $\widehat S_{k,x}^{|F\rangle}$ (cf.~\eqref{eq:f-final}) 
is the same as for the quench from the fully-occupied  state (see section~\ref{sec:ferro}).  
Let us consider the density profile. For $x>0$ we obtain 
\begin{equation}
	\label{eq:e1}
	n_{x,t}=\frac{1}{\pi}\int_{\arcsin(x/(2t))}^{\pi/2}dq
	\frac{|v_q|^2}{\big(\frac{\gamma^-}{2}+|v_q|\big)^2}.
\end{equation}
%
%
\begin{figure}[t]
\includegraphics[width=0.45\textwidth]{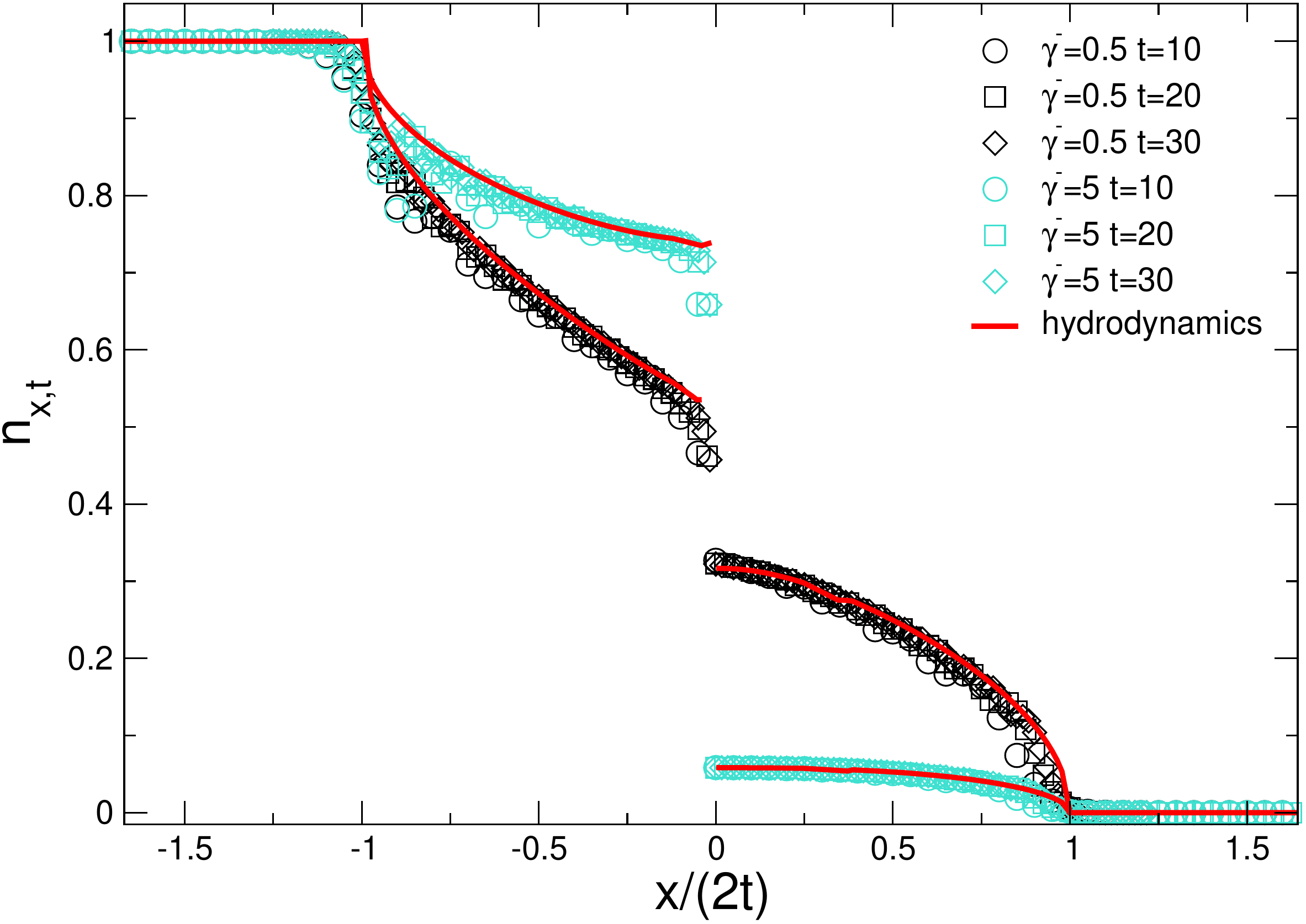}
\caption{ Density profile $n_{x,t}$ in a free fermion chain with localized  
 losses: Dynamics starting from the domain-wall state. We plot $n_{x,t}$ versus 
 $x/(2t)$. The symbols are exact 
 numerical data. Lines are the analytic results in the 
 space-time scaling limit. 
}
\label{fig3:DW}
\end{figure}
%
For $x<0$ the density reads as 
\begin{multline}
	\label{eq:e2}
	n_{x,t}=
	\frac{1}{\pi}\int^{\pi/2}_{\arcsin(x/(2t))}dq\\+
	\frac{(\gamma^-)^2}{4}\frac{1}{\pi}\int^{\pi/2}_{\arcsin(|x|/(2t))}dq\frac{1}
	{(\frac{\gamma^-}{2}+|v_q|)^2}. 
\end{multline}
Here $v_q$ is the fermion group velocity~\eqref{eq:v-k}. 
Clearly, from~\eqref{eq:e1} and~\eqref{eq:e2} 
for $\gamma^-\to0$ one recovers the expected result in the 
absence of dissipation. This corresponds to the first term 
in~\eqref{eq:e2}. Furthermore, in the limit of strong dissipation 
$\gamma^-\to\infty$, Eq.~\eqref{eq:e1} is, again, reminiscent of the Wigner semicircle 
law. In particular, for $x>0$, from~\eqref{eq:e2}, one obtains that  
$n_{x,t}=16/(\gamma^-\pi)\sqrt{1-x^2/(4t^2)}$.  

In Fig.~\ref{fig3:DW} we compare~\eqref{eq:e1} and~\eqref{eq:e2} 
with numerical results obtained from~\eqref{eq:G}. We report data for 
strong dissipation $\gamma^-=5$ and weak one $\gamma^-=0.5$ and several 
times. The data show a perfect agreement when plotting $n_{x,t}$ 
versus $x/(2t)$. The scaling functions are consistent with the 
numerical results in the space-time scaling limit. 
Similarly to the quench from the N\'eel state we should stress that by 
using~\eqref{eq:asyJ} and~\eqref{eq:asyK} it is possible to obtain 
the behavior of the generic fermionic correlator $G_{x,y}$ in the 
space-time scaling limit with arbitrary $\xi_x=x/(2t)$ and $\xi_y=y/(2t)$. 
Finally, we also stress that~\eqref{eq:e2} can be rederived as a particular 
case of the double Fermi seas expansion that we will discuss in the 
next section. 

\subsection{Inhomogeneous Fermi seas}
\label{sec:ness1}

In this section we discuss the situation in which two semi-infinite 
chains [see Fig~\ref{fig0:cartoon}(b)] are prepared in two Fermi 
seas at different fillings $k_F^l$ and $k_F^r$. 
The quench protocol is as follows. 
The two  chains are prepared in the ground state 
of~\eqref{eq:ham} with different fermionic densities, and 
with {\it periodic} boundary conditions. 
At $t=0$ the two chains are joined together. Note that due to 
the initial periodic boundary conditions on the two chains, 
this involves a ``cut and glue'' 
operation. The situation in which the two initial systems 
have open boundary conditions can be treated in a similar way, 
although we expect the out-of-equilibrium dynamics not to be 
dramatically affected by the choice of the boundary conditions. 
In the absence of dissipation, the out-of-equilibrium 
dynamics starting from two open chains that are joined together 
was obtained in Ref.~\onlinecite{viti-2016}, in the space-time 
scaling limit.  
Note that by fixing $k_F^l=\pi$ and $k_F^r=0$, one obtains 
the domain-wall quench (see section~\ref{sec:DW}). Instead, 
for $k_F^r=0$ and $k_F^l=\pi/2$ one has the so-called geometric 
quench~\cite{mossel-2010}, in which the ground state of a 
chain is let to expand in the vacuum. 

It is straightforward to derive the initial correlation matrix 
as 
\begin{multline}
\label{eq:G0}
	G_{x,y}(0)=\frac{\sin(k_F^r(x-y))}{\pi(x-y)}\Theta(x)\Theta(y)\\
	+\frac{\sin(k_F^l(x-y))}{\pi(x-y)}\Theta(-x)\Theta(-y),
\end{multline}
where $\Theta(x)$ is the Heaviside theta function. 
Equation~\eqref{eq:G0} is conveniently rewritten as 
\begin{multline}
	\label{eq:G-par0}
	G_{x,y}(0)=\frac{1}{2\pi}\int_{-k_F^l}^{k_F^l} dk e^{i k(x-y)}\Theta(-x)\Theta(-y)\\
	+\frac{1}{2\pi}\int_{-k_F^r}^{k_F^r}dk e^{i k(x-y)}\Theta(x)\Theta(y). 
\end{multline}
Crucially, Eq.~\eqref{eq:G-par0} suggests that we can 
parametrize $G_{x,y}$ as  
\begin{equation}
\label{eq:G-par}
G_{x,y}=\frac{1}{2\pi}\int_{-k_F^r}^{k_F^r} dk S^l_{k,x}\bar S^l_{k,y}+
\frac{1}{2\pi}\int_{-k_F^l}^{k_F^l} dk S^r_{k,x}\bar S^r_{k,y}, 
\end{equation}
where $S_{k,x}^{l/r}$ have to be determined. Clearly, the 
ansatz~\eqref{eq:G-par} is similar to the one used in section~\eqref{eq:ansatz}. 
After substituting Eq.~\eqref{eq:G-par} in~\eqref{eq:one}, we 
obtain that $S_{k,x}^{\scriptscriptstyle l/r}$ satisfy~\eqref{eq:S-eq}. 
The initial conditions are given as 
\begin{equation}
	\label{eq:lr-ini}
S^{r/l}_{k,x}(0)=e^{i k x}\Theta(\pm x), 
\end{equation}
where the plus and minus signs are for $S^r_{k,x}$ and $S^l_{k,x}$, 
respectively. The Laplace and Fourier transforms of~\eqref{eq:G-par} read 
\begin{equation}
	\label{eq:S-split}
	\widehat S_{k,q}^{l/r}=\widehat S_{k,q}^{l/r,U}
+\widehat S_{k,q}^{l/r,D}, 
\end{equation}
where we separated the unitary part from the 
contribution of the dissipation, as  stressed by 
the superscripts $U$ and $D$ in~\eqref{eq:S-split}. 
Here we defined 
\begin{align}
\label{eq:gq-un-1}
\widehat S_{k,q}^{l,U}&=
\frac{1}{s-2i\cos(q)}\frac{1}{1-e^{i(q-k+i0)}}\\
\label{eq:gq-un-2}
\widehat S_{k,q}^{r,U}&=
-\frac{1}{s-2i\cos(q)}\frac{1}{1-e^{i(q-k-i0)}}\\
\label{eq:gq-dis-1}
\widehat S_{k,q}^{l,D}&=
\int_{-\pi}^{\pi}
\frac{dp}{2\pi}\frac{Z(p)}{1-e^{i(p-k+i0)}}\\
\label{eq:gq-dis-2}
	\widehat S_{k,q}^{r,D}&=
	-\int_{-\pi}^{\pi}
	\frac{dp}{2\pi}\frac{Z(p)}{1-e^{i(p-k-i0)}}. 
\end{align}
The function $Z(p)$ is defined as 
\begin{equation}
\label{eq:Zp}
Z(p)=-\frac{\frac{\gamma^-}{2}}{\frac{\gamma^-}{2}+\sqrt{s^2+4}}
	\frac{\sqrt{s^2+4}}{s-2i\cos(p)}\frac{1}{s-2i\cos(q)}. 
\end{equation}
The terms $\pm i0$ in the equations above are convergence factors, and 
their sign is chosen to impose 
the $\Theta(\pm x)$ in the initial conditions for $S_{k,x}^{l/r}$ (cf.~\eqref{eq:lr-ini}). 
From~\eqref{eq:G-par} it is clear that in order to determine $G_{x,y}$ 
 one has to compute the integrals ${\mathcal I}^l$ and ${\mathcal I}^r$ defined as 
\begin{align}
\label{eq:int-l}
	& {\mathcal I}^l=
	\frac{1}{4\pi^2}\int_{-k^l_F}^{k^l_F}\frac{dk}{(1-e^{i(p-k+i0)})(1-e^{-i(q-k-i0)})}\\
	\label{eq:int-r}
	&{\mathcal I}^r=
	\frac{1}{4\pi^2}\int_{-k^r_F}^{k^r_F}\frac{dk}{(1-e^{i(p-k-i0)})(1-e^{-i(q-k+i0)})}. 
\end{align}
Similar integrals were discussed in Ref.~\onlinecite{viti-2016}. 
%
\begin{figure}[t]
\includegraphics[width=0.45\textwidth]{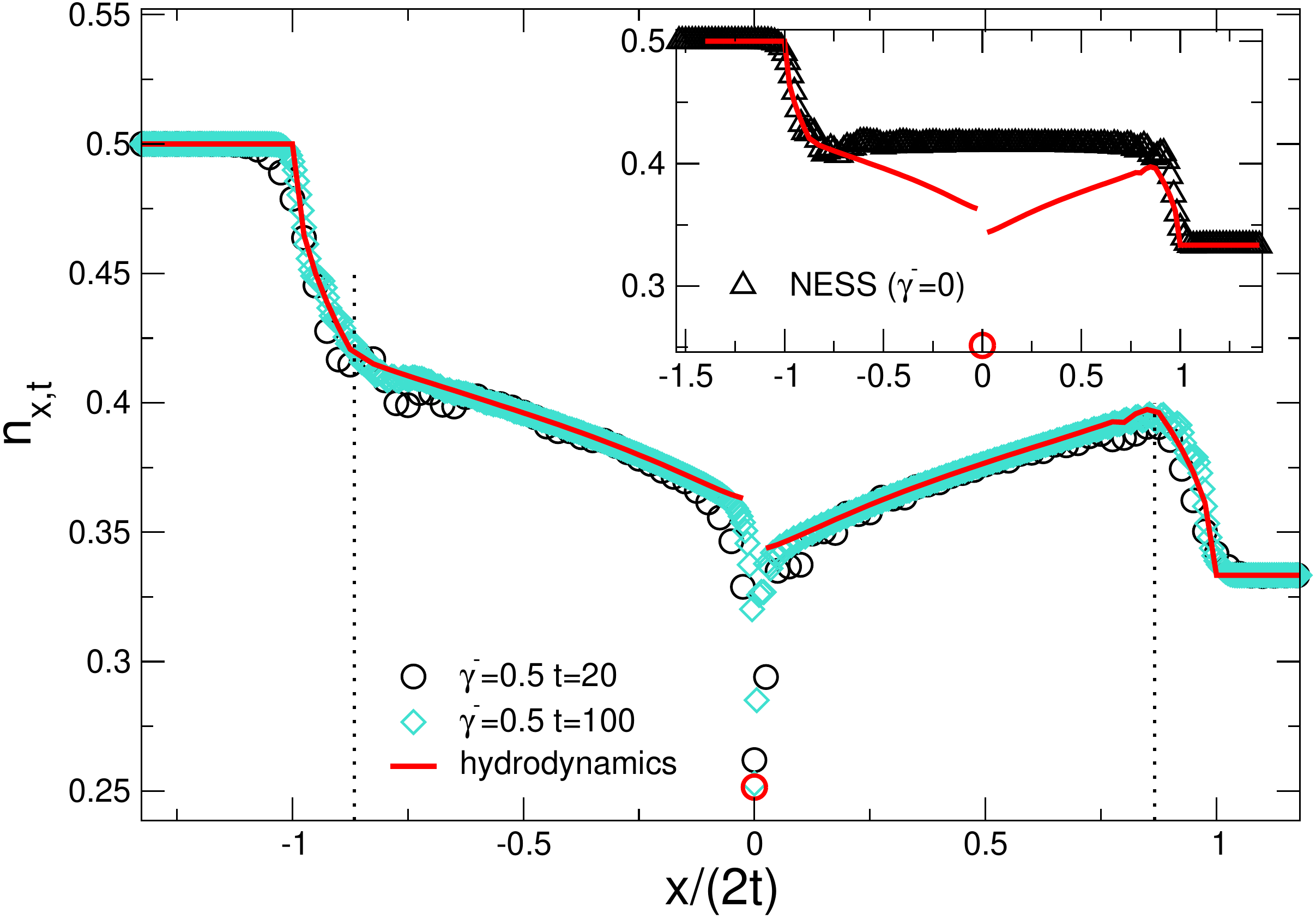}
\caption{ Density profile $n_{x,t}$ in a free fermion chain with localized  
 losses with rate $\gamma^-$: Dynamics starting from the 
 inhomogeneous state (see Fig.~\ref{fig0:cartoon}) 
 obtained by joining two Fermi seas with $k_F^l=\pi/2$ and $k_F^r=\pi/3$. 
 The symbols are exact numerical data for $\gamma^-=0.5$. Lines are the 
 analytic results in the space-time scaling limit. Inset: The case without 
 dissipation. Note the formation of the Non Equilibrium Steady State (NESS) 
 at the interface between the two chains. 
}
\label{fig4:two-FS}
\end{figure}
%
The integration over $k$ in Eqs.~\eqref{eq:int-l}-\eqref{eq:int-r} can be performed easily in the 
complex plane. The derivation is as in Ref.~\onlinecite{viti-2016}, 
and we do not report it here. We obtain 
\begin{multline}
	\label{eq:inte}
	4\pi^2{\mathcal I}^{l/r}=
	\frac{i}{1-e^{i(p-q\pm i0)}}
	\Big[\ln\frac{e^{i k_F^l}-e^{i q}}{e^{-i k_F^l}-e^{iq}}\\
	-\ln\frac{e^{i k_F^l}-e^{i p}}{e^{-i k_F^l}-e^{i p}}\mp 2\pi i\chi^{l/r}(p)\Big], 
\end{multline}
where the terms with $i0$ and $-i0$ in the exponential in~\eqref{eq:inte} 
correspond to ${\mathcal I}^l$ and ${\mathcal I}^r$, respectively. 
Here the function $\chi^{l/r}(p)$ is one if $p$ is in the interval 
$[-k^{l/r}_F,k^{l/r}_F]$, while it is zero otherwise. 

The next step is to determine the large $x$ behavior of 
$S_{k,x}^{\scriptscriptstyle l/r,D}$. 
This requires to calculate the inverse Fourier transform of 
$Z(p)$ with respect to $q$ (cf.~\eqref{eq:Zp}) and its 
inverse Laplace transform with respect to $s$. More precisely, 
one has to determine the asymptotic behavior for large $x,t$ of 
\begin{equation}
	\label{eq:F-def}
	F_x^D:={\mathcal L}^{-1}\Big(\frac{1}{2\pi}\int_{-\pi}^{\pi} dq Z(p)e^{i q x}\Big)\, .
\end{equation}
The derivation is  reported it in Appendix~\ref{sec:asy}. 
Here we quote the final result, which reads  
\begin{equation}
\label{eq:ref-c}
F_x^D(p)=\chi_{x}e^{2it\cos(p)+i |x||p|}r(p), 
\end{equation}
where $v_p$ is the fermions group velocity defined in~\eqref{eq:v-k}. 
Here $\chi_{x}$ is defined as 
\begin{equation}
	\label{eq:chi-def}
	\chi_x(p):=\Theta\Big(|v_p|-\frac{|x|}{t}\Big), 
\end{equation}
with $v_p$ the group velocity of the fermions (cf.~\eqref{eq:v-k}). 
In~\eqref{eq:ref-c} we defined the reflection amplitude 
$r(p)$ as  
\begin{equation}
	\label{eq:r}
	r(p):=-\frac{\gamma^-}{2}\frac{1}{\frac{\gamma^-}{2}+|v_p|}. 
\end{equation}
Note that $r(p)$ appears in the scattering problem of a plane wave 
with a delta potential with imaginary strength~\cite{burke-2020}. 

Finally, we discuss the behavior of $G_{x,y}$ in the space-time 
scaling limit for $t,x,y\to\infty$ with $x/(2t),y/(2t)$ fixed 
and $|x-y|/t\to0$.  
Let us start by discussing the different contributions 
in Eqs.~\eqref{eq:gq-un-1}-\eqref{eq:gq-un-2}-\eqref{eq:gq-dis-1}-\eqref{eq:gq-dis-2}. 
We first consider the unitary contribution 
\begin{multline}
	\label{eq:statio}
	\frac{1}{2\pi}\int_{-k_F^l}^{k_F^l} dk S_{k,x}^{l,U}\bar S_{k,y}^{l,U}=\\
	\frac{1}{2\pi}\int_{-\pi}^\pi dp dq e^{2it\cos(p)-2it\cos{q}+i p x-i q y}\,{\mathcal I}^l(p,q). 
\end{multline}
The analysis is essentially the same as in Ref.~\onlinecite{viti-2016}. 
Let us first consider the case with $x,y>0$. 
We employ the standard stationary phase approximation~\cite{wong}. 
In the large $t,x,y$ limit the stationary points in the 
double integral in~\eqref{eq:statio} satisfy the equations
\begin{align}
	\label{eq:sp-1}
	&-2t\sin(p)+x=0\\
	\label{eq:sp-2}
	&-2t\sin(q)+y=0
\end{align}
As it is clear from~\eqref{eq:sp-1} and~\eqref{eq:sp-2}, 
in the space-time scaling, the integral~\eqref{eq:statio} 
is dominated by the region with $p\to q$. Thus, we 
define $K:=(p+q)/2$ and $Q:=(p-q)$. In the limit $Q\to0$ we have 
\begin{equation}
\label{eq:I-simp}
	{\mathcal I}^{l/r}(p,q)=\mp\frac{1}{2\pi i}\frac{1}{Q\pm i0}\Theta(k_F^{l/r}-|K|), 
\end{equation}
where the term with $Q+i0$ refers to ${\mathcal I}^{l}$, and the one 
with $Q-i0$ to ${\mathcal I}^{r}$. By combining~\eqref{eq:I-simp} 
with the well-known formula 
\begin{equation}
	\label{eq:well-known}
\frac{1}{2\pi i}\int_{-\infty}^\infty dQ \frac{e^{i Q x}}{Q\mp i0}=\pm\Theta(\pm x)
\end{equation}
we obtain the relatively  simple result 
\begin{multline}
\label{eq:b-1}
\frac{1}{2\pi}\int_{-k_F^l}^{k_F^l} dk S_{k,x}^{l,U}\bar S_{k,y}^{l,U}=\\
\int_{-k_F^l}^{k_F^l}\frac{dK}{2\pi}
e^{i K(x-y)}\Theta\Big(2t\sin(K)-\frac{x+y}{2}\Big)\, . 
\end{multline}
This coincides with the result in Ref.~\onlinecite{viti-2016}. 
The derivation of the remaining terms entering in the definition 
of $G_{x,y}$ (cf.~\eqref{eq:G-par} and~\eqref{eq:S-split})  
is similar although more cumbersome due to the presence of the absolute 
values $|p|$ and $|q|$ in the integrands. We illustrate the main steps of 
the derivations in Appendix~\ref{sec:app-tech}. 
We obtain 
\begin{multline}
	\label{eq:FS-final}
G_{x,y}(t)=
	\int_{-k_F^l}^{k_F^l}\frac{dK}{2\pi}
	\Big\{e^{i K(x-y)}\Theta\Big(2t\sin(K)-\frac{x+y}{2}\Big)\\+
	e^{i K(|x|-|y|)}\Theta(K)(\Theta(x)+\Theta(y))
	\chi_{x}\chi_{y}r\\
	+\Theta(K)e^{i K(|x|-|y|)}
	\chi_{x}\chi_{y}r^2
\Big\}+l\leftrightarrow r, 
\end{multline}
with $\chi_x$ as defined in~\eqref{eq:chi-def}. Note that 
$\chi_{x},\chi_{y}$ and $r$ are functions of $K$. 
Here the last term $l\leftrightarrow r$ is obtained by changing 
the integration boundaries as $k_F^l\to k_F^r$,  and by replacing 
$K\to-K$ and $x,y\to-x,-y$ in the integrand in~\eqref{eq:FS-final}. 
Eq.~\eqref{eq:FS-final} holds only in the space-time (hydrodynamic) limit 
$|x|,|y|,t\to\infty$ with fixed $x/(2t)\approx y/(2t)$. 
Note that, similar to the previous sections, the 
correlation matrix $G_{x,y}$ is singular at $x,y\to0$. 
This happens because of fast oscillating terms  
in the limit $x\to\infty$ that cannot be neglected 
at $x\approx 0$. In the region $x/t,y/t\to0$ one obtains 
\begin{multline}
	\label{eq:FS-final-1}
	G_{x,y}=\int_{0}^{k_F^l}\frac{dK}{2\pi}(e^{iKx}+r e^{i K|x|})
(e^{-iK y}+r e^{-i K|y|})\\
\int_{-k_F^r}^0\frac{dK}{2\pi}(e^{iKx}+r e^{i K|x|})
(e^{-iK y}+r e^{-i K|y|}). 
\end{multline}
Before discussing the numerical checks of~\eqref{eq:FS-final} it is 
useful to address its physical interpretation. To this purpose it is 
useful to focus on the dynamics of the fermionic density $G_{x,x}(t)$. 
Equation~\eqref{eq:FS-final} is rewritten as 
\begin{equation}
	G_{x,x}=\int_{-\pi}^\pi \frac{dk}{2\pi}(n_{x,t}^l(k)+n_{x,t}^r(k)), 
\end{equation}
where $n^{(l/r)}_{x,t}(k)$ describe the evolution of the fermions with 
momentum $k$ originated in the initial left and right chains. 
As it is clear from~\eqref{eq:FS-final}, $n_{x,t}^l(k)$ 
is written as  
\begin{multline}
	\label{eq:den-simple}
	n_{x,t}^l(k)=n_0^l(k)\Big[\Theta(-x)\Theta(k)(1+|r|^2\Theta(x+2t\sin(k)))\\
	+\Theta(x)\Theta(k)\Theta(-x+2t\sin(k))|\tau|^2\\
	+\Theta(-x)\Theta(-k)\Theta(-x+2t\sin(k))
\Big].
\end{multline}
In~\eqref{eq:den-simple} we defined the transmission amplitude $\tau(k)$ as 
\begin{equation}
	\label{eq:transmission}
	\tau(k):=\frac{v_k}{\frac{\gamma^-}{2}+|v_k|}, 
\end{equation}
where $v_k$ is the fermion group velocity (cf.~\eqref{eq:v-k}). 
Note that $\tau^2+r^2\ne1$, signaling that the evolution is not unitary. 
Note that $\tau(k)$ coincides with the transmission amplitude for 
the scattering with a delta potential with imaginary 
strength~\cite{burke-2020}. In~\eqref{eq:den-simple}, $n_0^l$ is the 
initial momentum distribution for the left chain $n_0^l=\Theta(k_F^l-|k|)$, 
and $r$ the reflection amplitude defined in~\eqref{eq:r}. 
Now Eq.~\eqref{eq:den-simple} has a simple physical interpretation. The 
first row in~\eqref{eq:den-simple} describes the fermions moving towards 
the dissipative impurity and the scattered ones. The second row describes the 
fermions that are transmitted to the chain on the right at $x>0$. Finally, 
the last row accounts for the fermions that are in the left chain and 
are moving with negative velocity. 

It is useful to check~\eqref{eq:FS-final} and~\eqref{eq:FS-final-1} 
against exact numerical data. 
%
\begin{figure}[t]
\includegraphics[width=0.45\textwidth]{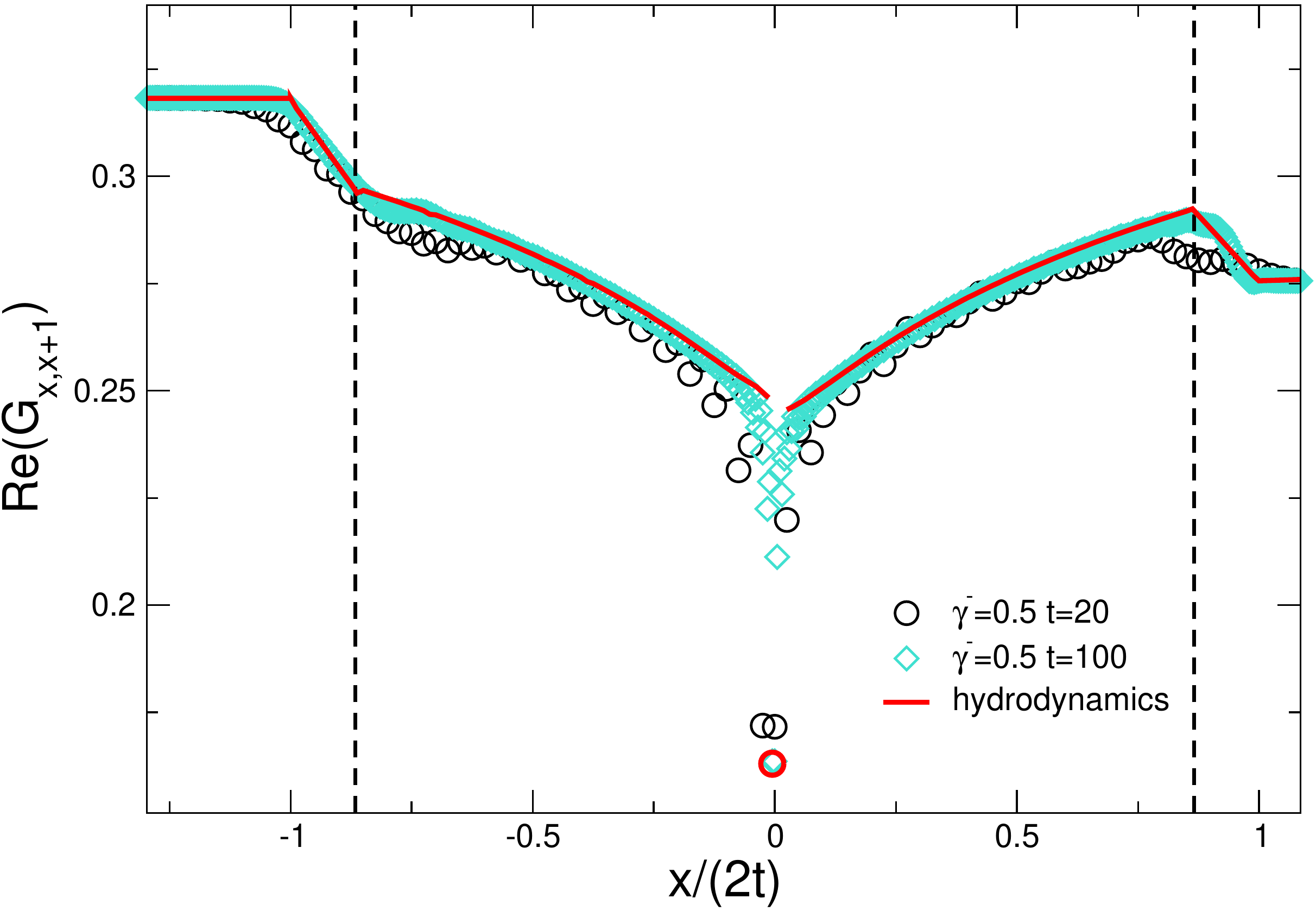}
\caption{Depleted Non-Equilibrium Steady State (NESS). We show the 
 dynamics of the off-diagonal correlator $\mathrm{Re}(G_{x,x+1})$ in the 
 free fermion chain with localized losses. The data are exact numerical 
 results for $\gamma^-=0.5$ for the infinite chain. We show results for the 
 inhomogeneous initial state obtained by joining two Fermi seas with Fermi 
 levels $k_F^l=\pi/2$ and $k_F^r=\pi/3$ (see Fig.~\ref{fig0:cartoon} (b)). 
 The curve is the result in the space-time scaling limit $t,x\to\infty$. 
}
\label{fig5:two-FS}
\end{figure}
%
%
\begin{figure}[t]
\includegraphics[width=0.45\textwidth]{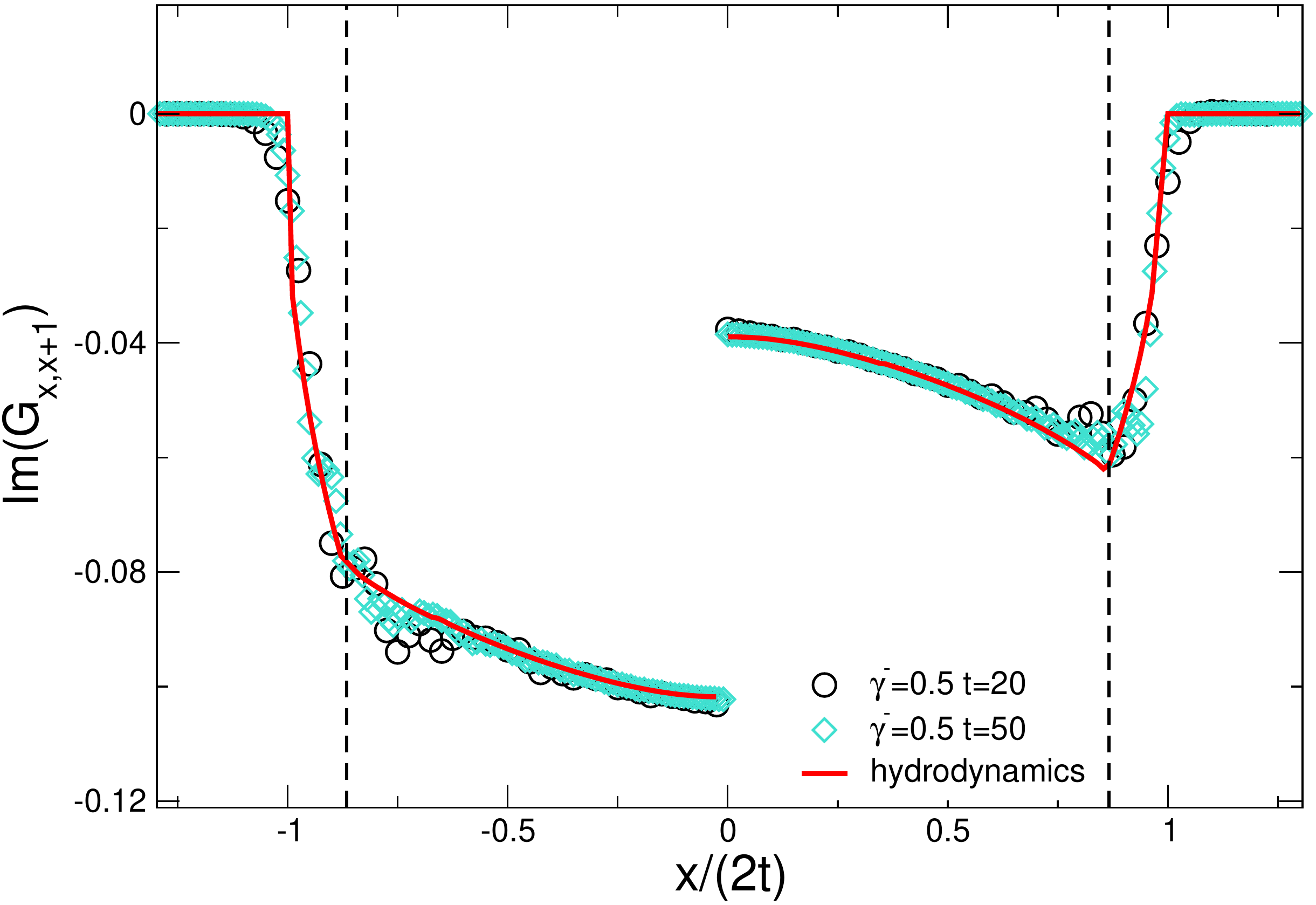}
\caption{Depleted Non-Equilibrium Steady State (NESS). We show the 
 dynamics of the fermion current $\mathrm{Im}(G_{x,x+1})$ in the 
 free fermion chain with localized losses. 
 The initial state and the dissipation are the same as in Fig.~\ref{fig5:two-FS}. 
 The curve is the result in the space-time scaling limit $t,x\to\infty$. 
}
\label{fig5b:two-FS}
\end{figure}
%
%
This is discussed in Fig.~\ref{fig4:two-FS}. We plot the fermionic 
density $n_{x,t}$ versus the scaling variable $x/(2t)$. We consider the 
case with $k_F^l=\pi/2$ and $k_F^r=\pi/3$. Interestingly, 
in the absence of dissipation a Non-Equilibrium Steady State (NESS) emerges 
at the interface between the two chains with 
a  flat density profile for $[-\sin(k_F^r)\le x/(2t)\le 
\sin(k_F^r)]$. The fermionic density in the flat region is the average density 
$(k_F^l+k_F^r)/(2\pi)$. The case without dissipation is shown 
in the inset of Fig.~\ref{fig4:two-FS}. As it is clear from the 
main Figure, in the presence of losses the NESS is depleted. Also the 
density profile exhibits a clear asymmetry under $x\to-x$ with a 
discontinuity at $x=0$. Cusp-like features are present at 
$\pm k_{F}^r$. These are also present in the absence of dissipation~\cite{viti-2016}. 
Finally, we report in the Figure the analytic result in the space-time 
scaling limit~\eqref{eq:FS-final}. This is in perfect agreement with the 
numerical data. Note that the agreement is also good for $x=0$. The 
theoretical prediction for $x=0$ is given by~\eqref{eq:FS-final-1} and it is 
reported as a circle in the Figure. 
Deviations are present near the singularities related to the Fermi momenta, 
similar to the non-dissipative case~\cite{viti-2016}, and are expected to 
vanish in the limit $x,t\to\infty$. Finally, we should mention that 
by imposing $k_F^l=k_F^r=k_F$ in~\eqref{eq:FS-final} one obtains the 
space-time limit behavior of the correlator $G_{x,y}$ for the problem of 
a uniform Fermi sea with Fermi level $k_F$. This is explicitly 
discussed in Appendix~\ref{sec:equal}. 
%
\begin{figure}[t]
\includegraphics[width=0.45\textwidth]{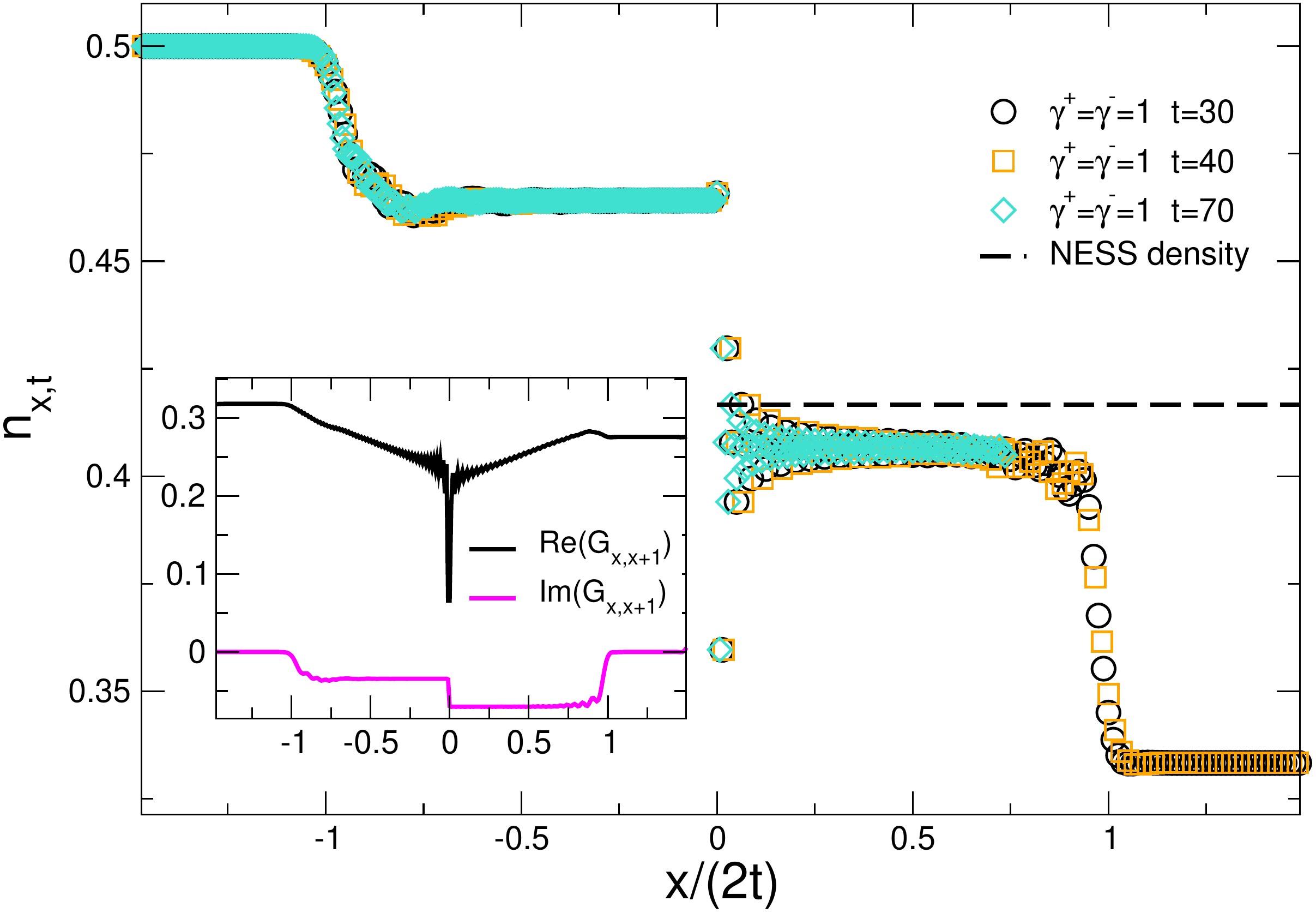}
\caption{Broken NESS with balanced gain and losses. We show the 
 fermionic density $n_{x,t}$ in the free fermion chain with 
 gains and losses. The data are exact numerical 
 results for $\gamma^-=\gamma^+=1$. The initial state is obtained by  
 joining two Fermi seas with Fermi levels $k_F^l=\pi/2$ and 
 $k_F^r=\pi/3$ (see Fig.~\ref{fig0:cartoon} (b)). The dashed line 
 is the NESS density $\pi(k_F^l+k_F^r)/2$ in the absence of dissipation. In the inset we 
 show the off-diagonal correlators $\mathrm{Re}(G_{x,x+1})$ and 
 $\mathrm{Im}(G_{x,x+1})$. 
}
\label{fig6:FS-gl}
\end{figure}
%
In Fig.~\ref{fig5:two-FS} and Fig.~\ref{fig5b:two-FS} we show the behavior of the off-diagonal correlation 
function $G_{x,x+1}$ in the space-time scaling limit. We present $\mathrm{Re}(G_{x,x+1})$ and $\mathrm{Im}(G_{x,x+1})$, which is the fermion current, separately. 
Similar to the density (see Fig.~\ref{fig4:two-FS}), the exact numerical data 
for $G_{x,x+1}$ obtained by numerically solving~\eqref{eq:one} collapse on the 
same curve when plotted as a function of $x/(2t)$, at least for large enough $x,t$. 
The scaling curve is perfectly described by the analytic result~\eqref{eq:FS-final} 
and~\eqref{eq:FS-final-1}. Note that similar to Fig.~\ref{fig4:two-FS} a singularity 
is present at $x=0$ in both Figures and the same cusp-like features at $x/(2t)=\sin(k_F^r)$ can be 
observed. We observe that the current is zero for $|x|/(2t)\ge 1$, 
as expected since for $|x|/(2t)\ge 1$ the system is at equilibrium. Note also that 
$\mathrm{Im}(G_{x,x+1})<0$ for any $x$, and does not change its sign across the singularity. Finally, we should stress that on increasing $\gamma^-$ the transport between 
the two chains is suppressed, which is, again, a manifestation of the quantum Zeno 
effect. This is nicely encoded in the value assumed by the reflection 
and transmission coefficients; for $\gamma_-\to\infty$, we have $r(k)\to-1$ and $\tau(k)\to0$.

To conclude we discuss an interesting effect that arises when one restores the 
gain dissipation. We consider the case of balanced gain/loss dissipation, 
i.e., with $\gamma^+=\gamma^-$. Our results are reported in Fig.~\ref{fig6:FS-gl}. 
We focus on the density profile $n_{x,t}$ plotted versus $x/(2t)$. We fix 
$k_F^l=\pi/2$ and $k_F^r=\pi/3$ and $\gamma^-=\gamma^+=1$. Interestingly, 
as it is clear from Fig.~\ref{fig6:FS-gl} the density profile now exhibits a 
``broken'' NESS structure. Specifically, two flat regions are visible for 
$-\sin(k^r_F)\le x\le0$ and $0<x\le\sin(k^r_F)$, with a step-like discontinuity 
at $x=0$. The dashed line in the figure shows the NESS density in the absence of 
dissipation. In the inset of Fig.~\ref{fig6:FS-gl} we report the behavior of the 
off-diagonal correlation $\mathrm{Re}(G_{x,x+1})$ and $\mathrm{Im}(G_{x,x+1})$, 
which show a nontrivial structure. We should mention that the behavior of the 
correlator in the presence of both gain and loss can be derived analytically 
by using the results in Appendix~\ref{app:gain-losses}, although we do not report  
its explicit expression.


\section{Conclusions}
\label{sec:conc}

We have provided exact results on the out-of-equilibrium dynamics 
of free-fermion systems subject to localized gain/loss dissipation, playing the role of a dissipative defect. 
We considered different setups with both homogeneous and inhomogeneous initial states, and derived general results on the fermionic correlations $G_{x,y}(t)=\mathrm{Tr}(c^\dagger_x c_y\rho(t))$. 
Our findings hold in the space-time scaling limit (hydrodynamic 
limit) with $x,y,t\to\infty$, their ratios $\xi_x=x/(2t),\xi_y=y/(2t)$ being fixed. 
In this limit, we have shown that dissipation acts as an effective delta potential with
momentum-dependent reflection and transmission amplitudes. 
For generic $\xi_x,\xi_y$, the fermionic correlation functions 
depend on the details of the model. On the other hand, in the limit $\xi_x\approx\xi_y$, the 
dynamics of fermionic correlations is completely characterized by the initial fermionic occupations and the 
emergent reflection amplitude of the dissipative impurity. 

Our results pave the way for several further studies. For instance,  it would 
be interesting to extend them to more complicated free-fermion models, 
e.g., the transverse field Ising chain. Another interesting direction 
concerns the investigation of the effect of localized dissipation in free-bosonic systems \cite{krapivsky-2020}. 
An intriguing question is how local dissipation affects the 
entanglement scaling at finite-temperature critical points. An ideal 
setup to explore this is provided by the so-called quantum spherical model, for 
which entanglement properties can be studied effectively~\cite{wald-2020,wald-2020a,alba2020c}. 
Furthermore, it would be of interest to study how localized 
gain/loss dissipations may affect  entanglement spreading, for instance, 
by studying the dynamics of the logarithmic negativity~\cite{vidal-2002,plenio-2005,calabrese-2012,shapourian-2019} 
and comparing with the quasiparticle picture~\cite{alba2019quantum}. In the 
absence of dissipation the entanglement dynamics has been investigated for both the geometric 
quench~\cite{alba-2014,vicari-2012,nespolo-2013} and the domain-wall quench~\cite{sabetta-2013,dubail-2017,collura-2020}. 
Moreover, it is important 
to generalize our findings to the interacting case. Although 
this is a challenging task, the 
results in Ref.~\onlinecite{bouchoule-2020} provide first steps in this direction. It would be 
also interesting to understand whether it is possible to incorporate the effects of 
dissipation in the Conformal Field Theory framework put forward in Refs.~\onlinecite{allegra-2016,dubail-2017} 
or in the quantum GHD~\cite{ruggiero-2020}. Finally, it would be important to clarify the correlation 
structure of the broken NESS discussed in section~\ref{sec:ness1}, and to understand whether it can be 
observed experimentally. 

\section{Acknowledgements}
We would like to thank J\'er\^ome Dubail and Oleksandr Gamayun for useful discussions in 
related projects. 
V.A. acknowledges support from the European Research Council under ERC Advanced grant 743032 DYNAMINT. F.C. acknowledges support from the “Wissenschaftler-R\"uckkehrprogramm GSO/CZS” of the Carl-Zeiss-Stiftung and the German Scholars Organization e.V., as well as through the Deutsche Forschungsgemeinsschaft (DFG, German Research Foundation) under Project No. 435696605.

\appendix

\section{Restoring fermion pumping}
\label{app:gain-losses}

Here we discuss how to obtain the solution of the general equation~\eqref{eq:one} 
for $\gamma^+\ne0$ and $\gamma^-\ne0$ from the solution for the case with 
loss only, i.e., $\gamma^+=0$. The equation for $G_{x,y}$ to be solved 
is~\eqref{eq:one}  
\begin{multline}
\label{eq:app-one}
\frac{d G_{x,y}}{dt}=i(G_{x+1,y}+G_{x-1,y}-G_{x,y+1}-G_{x,y-1})\\
-\frac{\gamma^++\gamma^-}{2}(\delta_{x,0}G_{x,y}+\delta_{y,0}G_{x,y})+
	\gamma^+\delta_{x,0}\delta_{y,0}. 
\end{multline}
Let us define as $\widetilde G_{x,y}$ the solution for the case with pure losses 
with effective loss rate $\gamma^++\gamma^-$, i.e., the solution 
of~\eqref{eq:app-one} where we neglect the last term. We have  
\begin{multline}
\frac{d \widetilde G_{x,y}}{dt}=i(\widetilde G_{x+1,y}+\widetilde G_{x-1,y}-
\widetilde G_{x,y+1}-\widetilde G_{x,y-1})\\
-\frac{\gamma^++\gamma^-}{2}(\delta_{x,0}\widetilde G_{x,y}+\delta_{y,0}
\widetilde G_{x,y}). 
\end{multline}
Let us impose the initial condition as 
\begin{equation}
\label{eq:app-ini}
	\widetilde G_{x,y}(0)=G_{x,y}(0). 
\end{equation}
Now we define the correlator $G_{x,y}'$ as the solution of the 
problem 
\begin{multline}
\frac{d G'_{x,y}}{dt}=i(G'_{x+1,y}+G'_{x-1,y}-
G'_{x,y+1}-G'_{x,y-1})\\
-\frac{\gamma^++\gamma^-}{2}(\delta_{x,0}G'_{x,y}+\delta_{y,0}
G'_{x,y}), 
\end{multline}
with delta initial condition 
\begin{equation}
G'_{x,y}(0)=\delta_{x,0}\delta_{y,0}. 
\end{equation}
Clearly, $G_{x,y}'$ is the solution of the problem with only losses 
with rate $\gamma^++\gamma^-$ for the empty chain with  one fermion 
at $x=0$. We can now write the solution $G_{x,y}$ of~\eqref{eq:app-one} as 
\begin{equation}
\label{eq:gl-sol}
	G_{x,y}(t)=\gamma^+\int_0^t d\tau G'_{x,y}(t-\tau)+\widetilde G_{x,y}(t). 
\end{equation}
By direct substitution, one can verify that Eq.~\eqref{eq:gl-sol} is the 
solution of~\eqref{eq:app-one} with initial condition~\eqref{eq:app-ini}. 

Finally, by using the same strategy as in section~\ref{sec:warm-up} one 
obtains the correlator $G'_{x,t}(t)$ as 
\begin{equation}
	G'_{x,y}=S_x\bar S_{y}, 
\end{equation}
where $S_x$ in the space-time scaling limit is given as~\cite{krapivsky-2019}  
\begin{equation}
	\label{eq:simeq}
S_{x}(t)\simeq 
\left\{\begin{array}{cc}
\frac{2|\xi|}{\frac{\gamma^++\gamma^-}{2}+2|\xi|}J_{|x|}(2t) &|x|>0 \\\\
-\frac{1}{(\gamma^++\gamma^-)^2t}J_1(2t) & x=0
\end{array}\right. 
\end{equation}
where $\xi=x/(2t)$, and $J_x(t)$ are the Bessel functions of the 
first type.

	Finally, we should remark that it is possible to extract from~\eqref{eq:gl-sol} the 
	hydrodynamic behavior in the limit $x,y,t\to\infty$, similar to~\eqref{eq:FS-final-1}. 
	Indeed, the term $\widetilde G_{x,y}$ in~\eqref{eq:gl-sol} is the same as in~\eqref{eq:FS-final-1} except for a redefinition $\gamma^-\to\gamma^++\gamma^-$. Let us now discuss the first term in~\eqref{eq:gl-sol}. By using~\eqref{eq:simeq} in~\eqref{eq:gl-sol}, one obtains 	
	\begin{multline}
\gamma^+\int_0^t d\tau G'_{x,y}(t-\tau)=
\gamma^+ t\int_0^1 d\tau \frac{|x|/(\tau t)}{(\gamma^++\gamma^-)/2+|x|/(t\tau)} \\
		\times\frac{|y|/(\tau t)}{(\gamma^++\gamma^-)/2+|y|/(t\tau)} 
		J_{|x|}(2t\tau)J_{|y|}(2t\tau). 
	\end{multline}
	After using the integral representation for the 
	Bessel function $J_{|x|}(2t)$ (see~\eqref{eq:bessel}) one obtains the three-dimensional 
	integral 
	\begin{multline}
		\label{eq:ref-1}
		\gamma^+\int_0^t d\tau G'_{x,y}(t-\tau)=\gamma^+ t\int_0^1 d\tau r'_x(\tau) r'_y(\tau)\\	
		\times\mathrm{Re}\int_0^\pi \frac{dz}{\pi}		
		e^{i z|x|-2it\tau\sin(z)}		\mathrm{Re}\int_0^\pi\frac{dk}{\pi} 
		e^{i k|y|-2it\tau\sin(k)},
	\end{multline}
	where we defined 
	\begin{equation}
		r'_x(\tau):=\frac{|x|/(\tau t)}{(\gamma^++\gamma^-)/2+|x/(\tau t)|}. 
	\end{equation}
	Now it is clear that in the limit $x,y,t\to\infty$ one can perform two of the integrals in~\eqref{eq:ref-1} (for instance, the integral in $\tau$ and $p$) by using the stationary phase. A similar 
	calculation is reported in Appendix~\ref{sec:asy} (see~\eqref{eq:int-fd}). It is also useful 
	to observe that in the limit $x,y,t\to\infty$ with $x/t,y/t$ fixed and $x/t\approx y/t$ one 
	can replace $r'_x\approx r'_y$ in~\eqref{eq:ref-1}. By performing the integral, one should be 
	able to obtain the hydrodynamic limit of~\eqref{eq:gl-sol} (similar to~\eqref{eq:FS-final-1}).

\section{Effective delta potential of the dissipative impurity}
\label{sec:asy}

In this section we derive the asymptotic behavior of $F_x^D$ 
(cf.~\eqref{eq:F-def}) in the limit $x\to\infty$. This will 
allow us to derive the reflection and transmission amplitudes of 
the effective delta potential associated with the dissipative impurity. 
First, the 
inverse Fourier transform of $Z(p)$ (cf.~\eqref{eq:Zp}) with 
respect to $q$ can be obtained analytically by using~\eqref{eq:ff-nota}. 
Thus, to determine $F_x^D$ (cf.~\eqref{eq:F-def}) 
one has to calculate the inverse Laplace transform 
\begin{equation}
F^D_{x}:={\mathcal L}^{-1}\Big(\frac{-\frac{\gamma^-}{2}}
{\frac{\gamma^-}{2}+\sqrt{s^2+4}}\Big(\frac{2i}{s+\sqrt{s^2+4}}\Big)^{|x|}
\frac{1}{s-2i\cos(p)}\Big)\, .
\end{equation}
This can be obtained by using~\eqref{eq:dn} and that 
\begin{equation}
	{\mathcal L}^{-1}\Big(\frac{1}{s-2i\cos(p)}\Big)=e^{2it\cos(p)}. 
\end{equation}
Now we obtain  
\begin{widetext}
\begin{equation}
	F_{x}^D=-\frac{i^{|x|}\gamma^-}{2}e^{2it\cos(p)}
	\Big[\int_{0}^t d\tau J_{|x|}(2\tau)e^{-2i\tau \cos(p)}
	-\frac{\gamma^-}{2}\int_0^td\tau\\\int_0^\tau dz e^{-\gamma^-z/2}
	\Big(\frac{\tau-z}{\tau+z}\Big)^\frac{|x|}{2}J_{|x|}(2\sqrt{\tau^2-z^2})
e^{-2i\tau\cos(p)} \Big]
\end{equation}
\end{widetext}
We now consider the space-time scaling limit $x,t\to\infty$ with the ratio 
$\xi_x=x/(2t)$ fixed. In the scaling limit we have 
\begin{equation}
	\Big(\frac{\tau-z}{\tau+z}\Big)^{\frac{|x|}{2}}\simeq e^{-2|\xi_x|z}. 
\end{equation}
We can proceed as in section~\ref{sec:ferro} to obtain 
\begin{multline}
\label{eq:ee}
F^D_{x}(p):=-\frac{\gamma^-}{2}I^D_x=\\
-\frac{\gamma^-}{2}i^{|x|}\int_{0}^1 d\tau J_{|x|}(\tau t)e^{2 i t(1-\tau) \cos(p)}
\frac{|x|}{\tau\frac{\gamma^-}{2}+\frac{|x|}{t}}. 
\end{multline}
Now we have to derive the large $x,t$ behavior of the integral $I_x^D$. 
To proceed, we employ the integral representation 
of the Bessel function $J_n(x)$ as 
\begin{equation}
	\label{eq:bessel}
	J_x(x\tau/\xi_x)=\mathrm{Re}\int_0^1 dz e^{i x(\pi z-\tau/\xi_x\sin(\pi z))}. 
\end{equation}
One now has to determine the large $x$ behaviour of the double integral 
\begin{multline}
\label{eq:int-fd}
I_x^D(p)=\frac{1}{2}i^{|x|} e^{2it\cos(p)}\int_0^1d\tau\int_0^1 dz \\
	[e^{i |x|(\pi z-\tau/|\xi_x|\sin(\pi z)-\tau/|\xi_x|\cos(p))}+z\to -z]
	\frac{|x|}{\tau\frac{\gamma^-}{2}+\frac{|x|}{t}}, 
\end{multline}
which is obtained by using~\eqref{eq:bessel} in~\eqref{eq:ee}. 
The integral in~\eqref{eq:int-fd} can be evaluated by using the 
two-dimensional stationary phase approximation~\cite{wong}. 
The stationary point for the first term in the square brackets is at 
\begin{equation}
	\label{eq:s-1}
	\tau^*=\frac{|\xi_x|}{|\sin(p)|},\quad z^*=\frac{|p|}{\pi}-\frac{1}{2}. 
\end{equation}
By imposing that the stationary point is in the integration domain, one 
obtains the condition 
\begin{equation}
	\frac{\pi}{2}\le|p|\le \pi-\arcsin(|\xi_x|). 
\end{equation}
The second term within the square brackets in~\eqref{eq:int-fd} has a 
stationary point at 
\begin{equation}
	\label{eq:s-2}
	\tau^*=\frac{|\xi_x|}{|\sin(p)|},\quad z^*=\frac{1}{2}-\frac{|p|}{\pi}. 
\end{equation}
The condition that the stationary points is in the integration domain 
$[0,1]\times[0,1]$ gives the condition 
\begin{equation}
	\arcsin(|\xi_x|)\le|p|\le\frac{\pi}{2}. 
\end{equation}
The analysis above implies that $F_n^D(p)$ has a contribution 
$\propto 1/|x|$ for $\arcsin(|\xi_x|)\le |p|\le \pi-\arcsin(|\xi_x|)$. 
For values of $p$ outside of this interval the integral in~\eqref{eq:int-fd} 
exhibits a faster decay with increasing $x$ and it does not contribute 
in the space-time scaling limit. 

We now apply the stationary-phase approximation to~\eqref{eq:int-fd}. 
Given two generic functions $g(w)$ and $f(w)$ in two 
dimensions, the stationary phase method states that in the large $x$ limit one 
has~\cite{wong} 
\begin{multline}
	\label{eq:s-phase}
	\int_\Omega g(w)e^{i x f(w)}\\
	\simeq g(w_0)|\det H|^{-1/2}
	\exp\Big(i x f(w_0)+i/4\pi\sigma\Big)\frac{2\pi}{x}. 
\end{multline}
Here $\Omega$ is the integration domain. In our case from~\eqref{eq:int-fd} 
$\Omega$ is the square $\Omega=[0,1]\times [0,1]$. In~\eqref{eq:s-phase} 
$H$ is the Hessian matrix, $\sigma$ denotes the signature of the eigenvalues 
of $H$, and $w_0$ is the stationary point, i.e., 
for which $\nabla_w f(w)=0$.  One can show that in our case $\sigma=0$. 
Moreover, the two stationary points~\eqref{eq:s-1} and~\eqref{eq:s-2} 
give the same contribution. From~\eqref{eq:ee} we obtain 
\begin{equation}
	\label{eq:det}
	|\det H|=\pi^2\frac{\sin^2(p)}{\xi_x^2}. 
\end{equation}
The phase factor in~\eqref{eq:s-phase} reads 
\begin{equation}
	\label{eq:phase}
	f(w_0)=-\frac{\pi}{2}+|p|. 
\end{equation}
After using~\eqref{eq:phase}~\eqref{eq:det} in~\eqref{eq:s-phase} we 
obtain~\eqref{eq:ref-c} 
\begin{equation}
\label{eq:app-res}
	F_x^D(p)=-\frac{\gamma^-}{2}\frac{\chi_{x}}
	{\frac{\gamma^-}{2}+|v_p|}  e^{2it\cos(p)+i |x||p|}, 
\end{equation}
where we used that $v_p=-2\sin(p)$ (cf.~\eqref{eq:v-k}) and $\chi_x$ is 
defined in~\eqref{eq:chi-def}. The prefactor  
multiplying the exponential is the reflection amplitude $r(p)$ in~\eqref{eq:r}.
For $p$ such that $\chi_{x}=0$ the integral in~\eqref{eq:int-fd} does not possess stationary 
points within the integration domain. Then, contributions originate from stationary points at the 
boundary of the domain and are subleading, i.e., they do not contribute in the 
space-time scaling limit. These contributions can be analyzed systematically 
within the stationary-phase approximation. 
Let us now briefly discuss their origin. We use the trivial 
identity~\cite{wong} 
\begin{multline}
	\label{eq:boundary}
	\int_\Omega dw g(w)e^{i x f(w)}=
	-\frac{i}{x}\int_{\partial \Omega}
	ds(\vec u\cdot\hat \nu) e^{ix f(w)}\\
	+\frac{i}{x}\int_\Omega dw (\vec\nabla\cdot\vec u)e^{i x f(w)},
\end{multline}
where $\partial\Omega$ denotes the boundary of $\Omega$ and $\hat\nu$ is the 
unit vector pointing outward normal to $\partial\Omega$. In~\eqref{eq:boundary} 
$\vec u$ is defined as 
\begin{equation}
	\vec u=\frac{\vec\nabla f}{|\vec\nabla f|^2}g. 
\end{equation}
In the presence of a boundary stationary point, 
the first term in the right hand side in~\eqref{eq:boundary} 
gives a contribution $1/x^{3/2}$, whereas the last term is subleading. 

\section{Double Fermi seas expansion: Some technical derivations}
\label{sec:app-tech}

In this section we report the derivation of~\eqref{eq:FS-final}. 
Specifically, we first derive in detail the term  
\begin{multline}
	\label{eq:ixy}
I_{x,y}:=\frac{1}{2\pi}\int_{-k_F^l}^{k_F^l} dk S_{k,x}^{l,U}\bar S_{k,y}^{l,D}=\\
\frac{1}{2\pi}\int_{-\pi}^\pi dp dq e^{2it\cos(p)-2it\cos(q)+i p x-i|q||y|}
\\\times{\mathcal I}^l(p,q)\chi_{y}(q)r(q). 
\end{multline}
This is obtained from~\eqref{eq:G-par}~\eqref{eq:gq-un-1}\eqref{eq:gq-dis-1} 
with~\eqref{eq:Zp}\eqref{eq:int-l}, and the asymptotic expansion~\eqref{eq:app-res}. 
In~\eqref{eq:ixy}, $r(q)$ is the reflection amplitude defined in~\eqref{eq:r}, ${\mathcal I}^l$ is defined 
in~\eqref{eq:int-l}, and $\chi_{y}(q)$ is given by~\eqref{eq:chi-def}.

Here we are interested 
in the space-time scaling limit $x,y,t\to\infty$ with $\xi_x=x/(2t)$ and 
$\xi_y=y/(2t)$ fixed and $|x-y|/t\to0$. In this regime the integral in~\eqref{eq:ixy} 
is dominated by the region $q\approx p$. Thus, it is convenient to define 
$Q:=p-q$ and $K:=(p+q)/2$. By treating carefully the absolute value $|q|$ in~\eqref{eq:ixy}, 
we can rewrite~\eqref{eq:ixy} as   
\begin{multline}
\label{eq:app-fin}
	I_{x,y}=\\
	-\int_{-\pi/2}^{\pi/2}\frac{dK}{2\pi}\int_{-2\pi+2|K|}^{2K} dQ
e^{i K(x-|y|)+i((x+|y|)/2 -2t\sin(K))Q}\\
\frac{1}
{2\pi i(Q+i0)}\chi_{y}(K)r(K)
\\-\int_{\pi/2}^{\pi}\frac{dK}{2\pi}\int_{-2\pi+2K}^{2\pi-2K}
dQ e^{iK(x-|y|)+i((x+|y|)/2 -2t\sin(K))Q}\\
\frac{1}{2\pi i(Q+i0)}\chi_{y}(K) r(K)\\
-
\int_{-\pi/2}^{\pi/2}\frac{dK}{2\pi}\int_{2K}^{2\pi-2|K|} dQ
e^{i K(x+|y|)+i((x-|y|)/2 -2t\sin(K))Q}\\
\frac{1}{2\pi i(Q+i0)}\chi_{y}(K)r(K)\\
-\int_{-\pi}^{-\pi/2}\frac{dK}{2\pi}\int_{2|K|-2\pi}^{2\pi-2|K|}
dQ e^{iK(x+|y|)+i((x-|y|)/2 -2t\sin(K))Q}\\
\frac{1}{2\pi i(Q+i0)}\chi_{y}(K)r(K). 
\end{multline}
The first two rows correspond to $q>0$, the other two 
to $q<0$. In~\eqref{eq:app-fin} we used that in the limit 
$q\approx p$ Eq.~\eqref{eq:I-simp} holds. 
We now observe that in the first two integrals in~\eqref{eq:app-fin} 
only the region with $K>0$ contribute, whereas the remaining 
two integrals get contributions from the region with $K<0$. 
The final result reads as  
\begin{multline}
\label{eq:ug-1-app}
\frac{1}{2\pi}\int_{-k_F^l}^{k_F^l} dk S_{k,x}^{l,U}\bar S_{k,y}^{l,D}=\\
\int_{0}^{k_F^l}\frac{dK}{2\pi}
e^{i K(x-|y|)}\Theta\Big(2t\sin(K)-\frac{x+|y|}{2}\Big)\chi_{y}r 
\\+\int_{-k_F^l}^{0}\frac{dK}{2\pi} e^{iK(x+|y|)}\Theta\Big(2t\sin(K)-\frac{x-|y|)}{2}\Big)
\chi_{y}r, 
\end{multline}
where we used~\eqref{eq:well-known}. 
A similar calculation allows us to obtain  
\begin{multline}
\label{eq:ug-2-app}
\frac{1}{2\pi}\int_{-k_F^l}^{k_F^l} dk S_{k,x}^{l,D}\bar S_{k,y}^{l,U}=\\
\int_{0}^{k_F^l}\frac{dK}{2\pi}
e^{i K(|x|-y)}\Theta\Big(-\frac{|x|+y}{2} +2t\sin(K)\Big)\chi_{x}r
\\+\int_{-k_F^l}^{0}\frac{dK}{2\pi} 
e^{iK(-|x|-y)}\Theta\Big(-\frac{-|x|+y}{2} +2t\sin(K)\Big)
\chi_{x}r.
\end{multline}
We also have 
\begin{multline}
\label{eq:ug-3-app}
\frac{1}{2\pi}\int_{-k_F^l}^{k_F^l} dk S_{k,x}^{l,D}\bar S_{k,y}^{l,D}=\\
\int_{0}^{k_F^l}\frac{dK}{2\pi}e^{iK(|x|-|y|)}\Theta\Big(2t\sin(K)-\frac{|x|+|y|}{2}\Big)
\chi_{x}\chi_{y|}r^2
\\+\int_{-k_F^l}^{0}\frac{dK}{2\pi}e^{iK(-|x|+|y|)}\Theta\Big(2t\sin(K)-
\frac{-|x|-|y|}{2}\Big)
\chi_{x}\chi_{y}r^2
\end{multline}
The expressions above can be simplified 
as follows. Let us consider the case with $x,y\ne0$. We observe that in 
Eq.~\eqref{eq:ug-1-app} we can neglect oscillating contributions in the 
limit $x,y\to\infty$. Eq.~\eqref{eq:ug-1-app} becomes 
\begin{multline}
\label{eq:ug-1-ap}
\frac{1}{2\pi}\int_{-k_F^l}^{k_F^l} dk S_{k,x}^{l,U}
\bar S_{k,y}^{l,D}=\int_{0}^{k_F^l}\frac{dK}{2\pi}
e^{i K(|x|-|y|)}\\
\times\Theta\big[\mathrm{sign}(x)(v_K-|\xi_x|-|\xi_y|)\big]\chi_{y}r.  
\end{multline}
In a similar way we obtain that 
\begin{multline}
\label{eq:ug-2-ap}
\frac{1}{2\pi}\int_{-k_F^l}^{k_F^l} dk S_{k,x}^{l,D}\bar S_{k,y}^{l,U}=
\int_{0}^{k_F^l}\frac{dK}{2\pi}
e^{i K(|x|-|y|)}\\
\times\Theta\big[\mathrm{sign}(y)(v_K-|\xi_x|-|\xi_y|
)\big]\chi_{x}r. 
\end{multline}
One should observe that~\eqref{eq:ug-1-ap} and~\eqref{eq:ug-2-ap} 
are nonzero only for $x>0$ and $y>0$, respectively. 
Finally, we observe that in~\eqref{eq:ug-3-app} 
only the first term contributes. We obtain 
\begin{multline}
\label{eq:ug-3-ap}
\frac{1}{2\pi}\int_{-k_F^l}^{k_F^l} dk S_{k,x}^{l,D}\bar S_{k,y}^{l,D}=
\int_{0}^{k_F^l}\frac{dK}{2\pi}e^{iK(|x|-|y|)}\\
\times\Theta\big[v_K-|\xi_x|-|\xi_y|\big]
\chi_{x}\chi_{y}r^2. 
\end{multline}
By using that in the space-time scaling limit $|x-y|/t\to0$, 
we can rewrite~\eqref{eq:ug-1-ap}  as 
\begin{multline}
	\label{eq:ud1}
\frac{1}{2\pi}\int_{-k_F^l}^{k_F^l} dk S_{k,x}^{l,U}
\bar S_{k,y}^{l,D}=\\
\int_{0}^{k_F^l}\frac{dK}{2\pi}
e^{i K(x-|y|)}\Theta(x)\chi_{x}\chi_{y}r.  
\end{multline}
Eq.~\eqref{eq:ug-2-ap} becomes 
\begin{multline}
	\label{eq:du1}
\frac{1}{2\pi}\int_{-k_F^l}^{k_F^l} dk S_{k,x}^{l,D}\bar S_{k,y}^{l,U}=\\
\int_{0}^{k_F^l}\frac{dK}{2\pi}
e^{i K(|x|-y)}
\Theta(y)\chi_{y}\chi_{x}r. 
\end{multline}
Finally, we have that~\eqref{eq:ug-3-ap} is rewritten as 
\begin{multline}
	\label{eq:dd1}
\frac{1}{2\pi}\int_{-k_F^l}^{k_F^l} dk S_{k,x}^{l,D}\bar S_{k,y}^{l,D}=\\
\int_{0}^{k_F^l}\frac{dK}{2\pi}e^{iK(|x|-|y|)}
\chi_{x}\chi_{y}r^2. 
\end{multline}
Let us comment on the terms with $S_{k,x}^r$ (see~\eqref{eq:G-par}).  
It is clear from the symmetry of the problem (see Fig.~\ref{fig0:cartoon}) 
that these coincide with~\eqref{eq:b-1}~\eqref{eq:ud1}\eqref{eq:du1} 
and~\eqref{eq:dd1} after replacing $k_F^l\to k_F^r$ and after changing 
$K\to-K$ and $x,y\to-x,-y$ in the integrands. 

%
\section{Two equal Fermi seas: Direct derivation}
\label{sec:equal}

In this section we present the direct derivation of the fermionic 
correlator $G_{x,y}$ for the free-fermion chain with only losses 
and $k_F^l=k_F^r=k_F$. This corresponds to the uniform Fermi sea as 
initial state. For a uniform state the calculations are somewhat 
easier than in section~\ref{sec:ness1} because only the asymptotic 
behavior of $F_x^D(p)$ (cf.~\eqref{eq:F-def}) in the limit $x,t\to\infty$ 
is required, whereas the stationary phase approximation discussed in 
section~\ref{sec:ness1} is not necessary. The 
result coincides with  Eq.~\eqref{eq:FS-final} after fixing 
$k_F^l=k_F^r=k_F$, confirming the correctness of the derivation in 
section~\ref{sec:ness1}. 

%
\begin{figure}[t]
\includegraphics[width=0.45\textwidth]{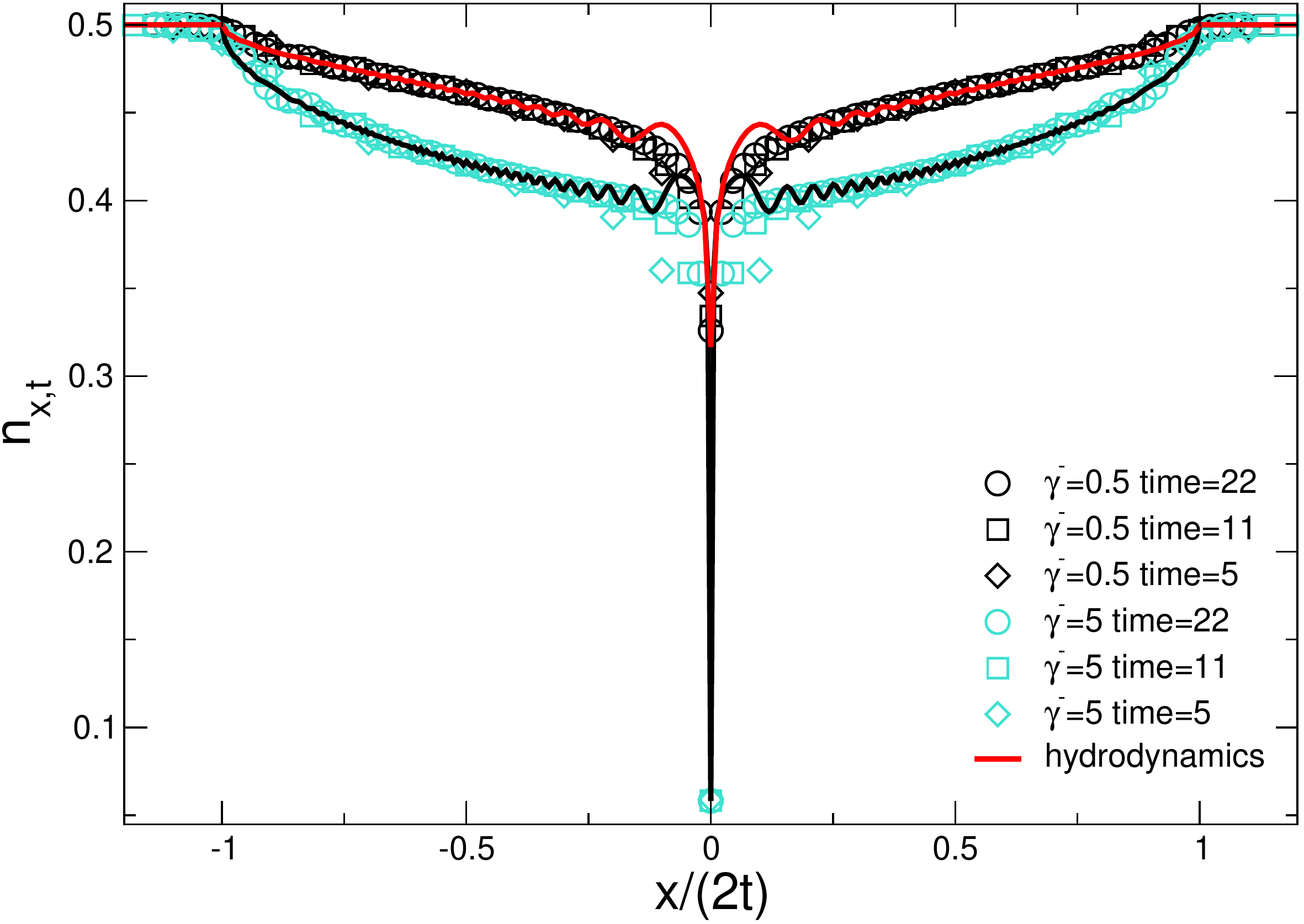}
\caption{ Density profile $n_{x,t}$ in a free fermion chain with local 
 losses. The chain is prepared in a uniform Fermi sea 
 at half filling, i.e., with $k_F=\pi/2$. The symbols are exact 
 numerical data for $\gamma^-=0.5$ and $\gamma^-=5$. Lines are exact 
 results in the space-time scaling limit. The oscillating correction 
 are an artifact of the approximations and vanish in the space-time 
 scaling limit. 
}
\label{fig7:one-FS}
\end{figure}
%

For a Fermi sea with Fermi level $k_F$ the initial correlation matrix reads 
\begin{equation}
G_{x,y}=\frac{\sin(k_F(x-y))}{\pi(x-y)}. 
\end{equation}
We now use the parametrization 
\begin{equation}
	G_{x,y}=\frac{1}{2\pi}\int_{-k_F}^{k_F}dk S_{k,x}\bar S_{k,y}
\end{equation}
As in section~\ref{sec:ness1}, the equation for $S_{k,x}$ is 
\begin{equation}
	\label{eq:app-lap}
	\frac{S_{k,x}}{dt}=i[S_{k,x+1}+S_{k,x-1}]-\frac{\gamma^-}{2}\delta_{x,0}S_{k,x}. 
\end{equation}
The initial condition reads as 
\begin{equation}
	S_{k,x}(0)=e^{ik x}. 
\end{equation}
We now use that 
\begin{equation}
	\label{eq:comb}
	\sum_{x=-\infty}^\infty e^{i x z}=W(z):=2\pi\sum_{p=-\infty}^\infty  
	\delta(z-2\pi p).
\end{equation}
The Laplace/Fourier transforms of~\eqref{eq:app-lap} read  as 
\begin{equation}
	\widehat S_{k,q}=\widehat S^U_{k,q}+\widehat S_{k,q}^{D}. 
\end{equation}
Here we defined 
\begin{equation}
	\widehat S_{k,q}^U=\frac{1}{s-2i\cos(q)}W(k-q)
\end{equation}
and 
\begin{multline}
	\widehat S_{k,q}^D=
	-\frac{\gamma^-}{4\pi}\int_{-\pi}^{\pi} dp\frac{W(k-p)}{\gamma^-/2+\sqrt{s^2+4}}
	\\\times\frac{\sqrt{s^2+4}}{s-2i\cos(p)}\frac{1}{s-2i\cos(q)}. 
\end{multline}
Following the same steps as in section~\ref{sec:ness1}, we can rewrite 
$S_{k,n}^D$ and $S^{U}_{k,n}$ 
as 
\begin{align}
	\label{eq:two-1}
	S_{k,x}^U &=\frac{1}{2\pi}\int_{-\pi}^\pi dq e^{2i t\cos(q)+iq x}W(k-q)\\
	\label{eq:two-2}
	S_{k,y}^D&=\frac{1}{2\pi}\int_{-\pi}^\pi dp W(k-p) e^{2 i t\cos(p)+i |x||p|}\chi_{x}r.
\end{align}
The integrations over $p$ and $q$ in~\eqref{eq:two-1} and~\eqref{eq:two-2} are 
straightforward, in contrast with section~\ref{sec:ness1}, because of the simple structure of 
$W(z)$ (cf.~\eqref{eq:comb}). The net effect of the integration is to fix  
$q=k$ and $p=k$. The final result is given as 
\begin{multline}
	\label{eq:g-ofs}
	G_{x,y}(t)=\frac{1}{2\pi}\int_{-k_F}^{k_F} dk \Big[
		e^{i k x}+\chi_{x}r e^{i |k||x|}
	\Big]\\
\times \Big[
	 e^{-i k y}+\chi_{y}r e^{-i |k||y|}
	\Big]
\end{multline}
It is interesting to observe that in~\eqref{eq:g-ofs} the time-dependent 
terms $e^{2 i t\cos(k)}$ drop out. The only time dependence is in the 
term $\chi_{|\xi_x|}$ and $\chi_{|\xi_y|}$. 
Finally, it is straightforward to check that in the space-time 
scaling limit $x,y,t\to\infty$ after neglecting oscillating 
terms Eq.~\eqref{eq:g-ofs} coincides with~\eqref{eq:FS-final} 
with $k_F^l=k_F^r=k_F$. 

In Fig.~\ref{fig7:one-FS} we discuss exact numerical data 
for the fermionic density $n_{x,t}=G_{xx}$ obtained by solving 
numerically Eq.~\eqref{eq:one} with $\gamma^+=0$. We fix $k_F=\pi/2$. 
In the Figure we show results for both ``strong'' dissipation ($\gamma^-=5$) and 
``weak'' dissipation ($\gamma^-=0.5$). Note that a singularity is present at 
$x=0$, as expected. Also, for $|x|/(2t)>1$ one has $n_{x,t}=1/2$. 
The continuous lines in Fig.~\ref{fig7:one-FS} are~\eqref{eq:g-ofs}. 
Note that oscillating corrections are present. These are an artifact of the 
derivation of~\eqref{eq:g-ofs}. The corrections vanish in the 
limit $x,t\to\infty$ and Eq.~\eqref{eq:g-ofs} fully describes 
the numerical data. 


%
\bibliography{bibliography}

\begin{thebibliography}{95}%
\makeatletter
\providecommand \@ifxundefined [1]{%
 \@ifx{#1\undefined}
}%
\providecommand \@ifnum [1]{%
 \ifnum #1\expandafter \@firstoftwo
 \else \expandafter \@secondoftwo
 \fi
}%
\providecommand \@ifx [1]{%
 \ifx #1\expandafter \@firstoftwo
 \else \expandafter \@secondoftwo
 \fi
}%
\providecommand \natexlab [1]{#1}%
\providecommand \enquote  [1]{``#1''}%
\providecommand \bibnamefont  [1]{#1}%
\providecommand \bibfnamefont [1]{#1}%
\providecommand \citenamefont [1]{#1}%
\providecommand \href@noop [0]{\@secondoftwo}%
\providecommand \href [0]{\begingroup \@sanitize@url \@href}%
\providecommand \@href[1]{\@@startlink{#1}\@@href}%
\providecommand \@@href[1]{\endgroup#1\@@endlink}%
\providecommand \@sanitize@url [0]{\catcode `\\12\catcode `\$12\catcode
  `\&12\catcode `\#12\catcode `\^12\catcode `\_12\catcode `\%12\relax}%
\providecommand \@@startlink[1]{}%
\providecommand \@@endlink[0]{}%
\providecommand \url  [0]{\begingroup\@sanitize@url \@url }%
\providecommand \@url [1]{\endgroup\@href {#1}{\urlprefix }}%
\providecommand \urlprefix  [0]{URL }%
\providecommand \Eprint [0]{\href }%
\providecommand \doibase [0]{http://dx.doi.org/}%
\providecommand \selectlanguage [0]{\@gobble}%
\providecommand \bibinfo  [0]{\@secondoftwo}%
\providecommand \bibfield  [0]{\@secondoftwo}%
\providecommand \translation [1]{[#1]}%
\providecommand \BibitemOpen [0]{}%
\providecommand \bibitemStop [0]{}%
\providecommand \bibitemNoStop [0]{.\EOS\space}%
\providecommand \EOS [0]{\spacefactor3000\relax}%
\providecommand \BibitemShut  [1]{\csname bibitem#1\endcsname}%
\let\auto@bib@innerbib\@empty
\bibitem [{\citenamefont {Degasperis}\ \emph {et~al.}(1974)\citenamefont
  {Degasperis}, \citenamefont {Fonda},\ and\ \citenamefont
  {Ghirardi}}]{degasperis-1974}%
  \BibitemOpen
  \bibfield  {author} {\bibinfo {author} {\bibfnamefont {A.}~\bibnamefont
  {Degasperis}}, \bibinfo {author} {\bibfnamefont {L.}~\bibnamefont {Fonda}}, \
  and\ \bibinfo {author} {\bibfnamefont {G.~C.}\ \bibnamefont {Ghirardi}},\
  }\href {\doibase 10.1007/BF02731351} {\bibfield  {journal} {\bibinfo
  {journal} {Il Nuovo Cimento A (1965-1970)}\ }\textbf {\bibinfo {volume}
  {21}},\ \bibinfo {pages} {471} (\bibinfo {year} {1974})}\BibitemShut
  {NoStop}%
\bibitem [{\citenamefont {Misra}\ and\ \citenamefont
  {Sudarshan}(1977)}]{misra-1977}%
  \BibitemOpen
  \bibfield  {author} {\bibinfo {author} {\bibfnamefont {B.}~\bibnamefont
  {Misra}}\ and\ \bibinfo {author} {\bibfnamefont {E.~C.~G.}\ \bibnamefont
  {Sudarshan}},\ }\href {\doibase 10.1063/1.523304} {\bibfield  {journal}
  {\bibinfo  {journal} {Journal of Mathematical Physics}\ }\textbf {\bibinfo
  {volume} {18}},\ \bibinfo {pages} {756} (\bibinfo {year} {1977})}\BibitemShut
  {NoStop}%
\bibitem [{\citenamefont {Facchi}\ and\ \citenamefont
  {Pascazio}(2002)}]{facchi-2002}%
  \BibitemOpen
  \bibfield  {author} {\bibinfo {author} {\bibfnamefont {P.}~\bibnamefont
  {Facchi}}\ and\ \bibinfo {author} {\bibfnamefont {S.}~\bibnamefont
  {Pascazio}},\ }\href {\doibase 10.1103/PhysRevLett.89.080401} {\bibfield
  {journal} {\bibinfo  {journal} {Phys. Rev. Lett.}\ }\textbf {\bibinfo
  {volume} {89}},\ \bibinfo {pages} {080401} (\bibinfo {year}
  {2002})}\BibitemShut {NoStop}%
\bibitem [{\citenamefont {Bernard}\ \emph {et~al.}(2018)\citenamefont
  {Bernard}, \citenamefont {Jin},\ and\ \citenamefont
  {Shpielberg}}]{bernard2018}%
  \BibitemOpen
  \bibfield  {author} {\bibinfo {author} {\bibfnamefont {D.}~\bibnamefont
  {Bernard}}, \bibinfo {author} {\bibfnamefont {T.}~\bibnamefont {Jin}}, \ and\
  \bibinfo {author} {\bibfnamefont {O.}~\bibnamefont {Shpielberg}},\ }\href
  {\doibase 10.1209/0295-5075/121/60006} {\bibfield  {journal} {\bibinfo
  {journal} {{EPL} (Europhysics Letters)}\ }\textbf {\bibinfo {volume} {121}},\
  \bibinfo {pages} {60006} (\bibinfo {year} {2018})}\BibitemShut {NoStop}%
\bibitem [{\citenamefont {Carollo}\ \emph {et~al.}(2018)\citenamefont
  {Carollo}, \citenamefont {Garrahan},\ and\ \citenamefont
  {Lesanovsky}}]{carollo2018}%
  \BibitemOpen
  \bibfield  {author} {\bibinfo {author} {\bibfnamefont {F.}~\bibnamefont
  {Carollo}}, \bibinfo {author} {\bibfnamefont {J.~P.}\ \bibnamefont
  {Garrahan}}, \ and\ \bibinfo {author} {\bibfnamefont {I.}~\bibnamefont
  {Lesanovsky}},\ }\href {\doibase 10.1103/PhysRevB.98.094301} {\bibfield
  {journal} {\bibinfo  {journal} {Phys. Rev. B}\ }\textbf {\bibinfo {volume}
  {98}},\ \bibinfo {pages} {094301} (\bibinfo {year} {2018})}\BibitemShut
  {NoStop}%
\bibitem [{\citenamefont {Popkov}\ \emph {et~al.}(2018)\citenamefont {Popkov},
  \citenamefont {Essink}, \citenamefont {Presilla},\ and\ \citenamefont
  {Sch\"utz}}]{popkov2018}%
  \BibitemOpen
  \bibfield  {author} {\bibinfo {author} {\bibfnamefont {V.}~\bibnamefont
  {Popkov}}, \bibinfo {author} {\bibfnamefont {S.}~\bibnamefont {Essink}},
  \bibinfo {author} {\bibfnamefont {C.}~\bibnamefont {Presilla}}, \ and\
  \bibinfo {author} {\bibfnamefont {G.}~\bibnamefont {Sch\"utz}},\ }\href
  {\doibase 10.1103/PhysRevA.98.052110} {\bibfield  {journal} {\bibinfo
  {journal} {Phys. Rev. A}\ }\textbf {\bibinfo {volume} {98}},\ \bibinfo
  {pages} {052110} (\bibinfo {year} {2018})}\BibitemShut {NoStop}%
\bibitem [{\citenamefont {Lin}\ \emph {et~al.}(2013)\citenamefont {Lin},
  \citenamefont {Gaebler}, \citenamefont {Reiter}, \citenamefont {Tan},
  \citenamefont {Bowler}, \citenamefont {S{\o}rensen}, \citenamefont
  {Leibfried},\ and\ \citenamefont {Wineland}}]{lin2013}%
  \BibitemOpen
  \bibfield  {author} {\bibinfo {author} {\bibfnamefont {Y.}~\bibnamefont
  {Lin}}, \bibinfo {author} {\bibfnamefont {J.~P.}\ \bibnamefont {Gaebler}},
  \bibinfo {author} {\bibfnamefont {F.}~\bibnamefont {Reiter}}, \bibinfo
  {author} {\bibfnamefont {T.~R.}\ \bibnamefont {Tan}}, \bibinfo {author}
  {\bibfnamefont {R.}~\bibnamefont {Bowler}}, \bibinfo {author} {\bibfnamefont
  {A.~S.}\ \bibnamefont {S{\o}rensen}}, \bibinfo {author} {\bibfnamefont
  {D.}~\bibnamefont {Leibfried}}, \ and\ \bibinfo {author} {\bibfnamefont
  {D.~J.}\ \bibnamefont {Wineland}},\ }\href {\doibase 10.1038/nature12801}
  {\bibfield  {journal} {\bibinfo  {journal} {Nature}\ }\textbf {\bibinfo
  {volume} {504}},\ \bibinfo {pages} {415} (\bibinfo {year}
  {2013})}\BibitemShut {NoStop}%
\bibitem [{\citenamefont {Verstraete}\ \emph {et~al.}(2009)\citenamefont
  {Verstraete}, \citenamefont {Wolf},\ and\ \citenamefont
  {Ignacio~Cirac}}]{verstraete-2009}%
  \BibitemOpen
  \bibfield  {author} {\bibinfo {author} {\bibfnamefont {F.}~\bibnamefont
  {Verstraete}}, \bibinfo {author} {\bibfnamefont {M.~M.}\ \bibnamefont
  {Wolf}}, \ and\ \bibinfo {author} {\bibfnamefont {J.}~\bibnamefont
  {Ignacio~Cirac}},\ }\href {\doibase 10.1038/nphys1342} {\bibfield  {journal}
  {\bibinfo  {journal} {Nature Physics}\ }\textbf {\bibinfo {volume} {5}},\
  \bibinfo {pages} {633} (\bibinfo {year} {2009})}\BibitemShut {NoStop}%
\bibitem [{\citenamefont {Diehl}\ \emph {et~al.}(2011)\citenamefont {Diehl},
  \citenamefont {Rico}, \citenamefont {Baranov},\ and\ \citenamefont
  {Zoller}}]{diehl-2011}%
  \BibitemOpen
  \bibfield  {author} {\bibinfo {author} {\bibfnamefont {S.}~\bibnamefont
  {Diehl}}, \bibinfo {author} {\bibfnamefont {E.}~\bibnamefont {Rico}},
  \bibinfo {author} {\bibfnamefont {M.~A.}\ \bibnamefont {Baranov}}, \ and\
  \bibinfo {author} {\bibfnamefont {P.}~\bibnamefont {Zoller}},\ }\href
  {\doibase 10.1038/nphys2106} {\bibfield  {journal} {\bibinfo  {journal}
  {Nature Physics}\ }\textbf {\bibinfo {volume} {7}},\ \bibinfo {pages} {971}
  (\bibinfo {year} {2011})}\BibitemShut {NoStop}%
\bibitem [{\citenamefont {Vicari}(2018)}]{vicari-2018}%
  \BibitemOpen
  \bibfield  {author} {\bibinfo {author} {\bibfnamefont {E.}~\bibnamefont
  {Vicari}},\ }\href {\doibase 10.1103/PhysRevA.98.052127} {\bibfield
  {journal} {\bibinfo  {journal} {Phys. Rev. A}\ }\textbf {\bibinfo {volume}
  {98}},\ \bibinfo {pages} {052127} (\bibinfo {year} {2018})}\BibitemShut
  {NoStop}%
\bibitem [{\citenamefont {Rossini}\ and\ \citenamefont
  {Vicari}(2019{\natexlab{a}})}]{rossini-2019a}%
  \BibitemOpen
  \bibfield  {author} {\bibinfo {author} {\bibfnamefont {D.}~\bibnamefont
  {Rossini}}\ and\ \bibinfo {author} {\bibfnamefont {E.}~\bibnamefont
  {Vicari}},\ }\href {\doibase 10.1103/PhysRevA.99.052113} {\bibfield
  {journal} {\bibinfo  {journal} {Phys. Rev. A}\ }\textbf {\bibinfo {volume}
  {99}},\ \bibinfo {pages} {052113} (\bibinfo {year}
  {2019}{\natexlab{a}})}\BibitemShut {NoStop}%
\bibitem [{\citenamefont {Nigro}\ \emph {et~al.}(2019)\citenamefont {Nigro},
  \citenamefont {Rossini},\ and\ \citenamefont {Vicari}}]{nigro-2019}%
  \BibitemOpen
  \bibfield  {author} {\bibinfo {author} {\bibfnamefont {D.}~\bibnamefont
  {Nigro}}, \bibinfo {author} {\bibfnamefont {D.}~\bibnamefont {Rossini}}, \
  and\ \bibinfo {author} {\bibfnamefont {E.}~\bibnamefont {Vicari}},\ }\href
  {\doibase 10.1103/PhysRevA.100.052108} {\bibfield  {journal} {\bibinfo
  {journal} {Phys. Rev. A}\ }\textbf {\bibinfo {volume} {100}},\ \bibinfo
  {pages} {052108} (\bibinfo {year} {2019})}\BibitemShut {NoStop}%
\bibitem [{\citenamefont {Rossini}\ and\ \citenamefont
  {Vicari}(2019{\natexlab{b}})}]{rossini-2019}%
  \BibitemOpen
  \bibfield  {author} {\bibinfo {author} {\bibfnamefont {D.}~\bibnamefont
  {Rossini}}\ and\ \bibinfo {author} {\bibfnamefont {E.}~\bibnamefont
  {Vicari}},\ }\href {\doibase 10.1103/PhysRevB.100.174303} {\bibfield
  {journal} {\bibinfo  {journal} {Phys. Rev. B}\ }\textbf {\bibinfo {volume}
  {100}},\ \bibinfo {pages} {174303} (\bibinfo {year}
  {2019}{\natexlab{b}})}\BibitemShut {NoStop}%
\bibitem [{\citenamefont {Di~Meglio}\ \emph {et~al.}(2020)\citenamefont
  {Di~Meglio}, \citenamefont {Rossini},\ and\ \citenamefont
  {Vicari}}]{di-meglio-2020}%
  \BibitemOpen
  \bibfield  {author} {\bibinfo {author} {\bibfnamefont {G.}~\bibnamefont
  {Di~Meglio}}, \bibinfo {author} {\bibfnamefont {D.}~\bibnamefont {Rossini}},
  \ and\ \bibinfo {author} {\bibfnamefont {E.}~\bibnamefont {Vicari}},\ }\href
  {\doibase 10.1103/PhysRevB.102.224302} {\bibfield  {journal} {\bibinfo
  {journal} {Phys. Rev. B}\ }\textbf {\bibinfo {volume} {102}},\ \bibinfo
  {pages} {224302} (\bibinfo {year} {2020})}\BibitemShut {NoStop}%
\bibitem [{\citenamefont {Rossini}\ and\ \citenamefont
  {Vicari}(2021)}]{rossini2021coherent}%
  \BibitemOpen
  \bibfield  {author} {\bibinfo {author} {\bibfnamefont {D.}~\bibnamefont
  {Rossini}}\ and\ \bibinfo {author} {\bibfnamefont {E.}~\bibnamefont
  {Vicari}},\ }\href@noop {} {\enquote {\bibinfo {title} {Coherent and
  dissipative dynamics at quantum phase transitions},}\ } (\bibinfo {year}
  {2021}),\ \Eprint {http://arxiv.org/abs/2103.02626} {arXiv:2103.02626
  [cond-mat.stat-mech]} \BibitemShut {NoStop}%
\bibitem [{\citenamefont {Breuer}\ and\ \citenamefont
  {Petruccione}(2002)}]{petruccione}%
  \BibitemOpen
  \bibfield  {author} {\bibinfo {author} {\bibfnamefont {H.~P.}\ \bibnamefont
  {Breuer}}\ and\ \bibinfo {author} {\bibfnamefont {F.}~\bibnamefont
  {Petruccione}},\ }\href@noop {} {\emph {\bibinfo {title} {The theory of open
  quantum systems}}}\ (\bibinfo  {publisher} {Oxford University Press},\
  \bibinfo {address} {Great Clarendon Street},\ \bibinfo {year}
  {2002})\BibitemShut {NoStop}%
\bibitem [{\citenamefont {Prosen}(2008)}]{prosen-2008}%
  \BibitemOpen
  \bibfield  {author} {\bibinfo {author} {\bibfnamefont {T.}~\bibnamefont
  {Prosen}},\ }\href {\doibase 10.1088/1367-2630/10/4/043026} {\bibfield
  {journal} {\bibinfo  {journal} {New Journal of Physics}\ }\textbf {\bibinfo
  {volume} {10}},\ \bibinfo {pages} {43026} (\bibinfo {year}
  {2008})}\BibitemShut {NoStop}%
\bibitem [{\citenamefont {Prosen}(2011)}]{prosen-2011}%
  \BibitemOpen
  \bibfield  {author} {\bibinfo {author} {\bibfnamefont {T.}~\bibnamefont
  {Prosen}},\ }\href {\doibase 10.1103/PhysRevLett.107.137201} {\bibfield
  {journal} {\bibinfo  {journal} {Phys. Rev. Lett.}\ }\textbf {\bibinfo
  {volume} {107}},\ \bibinfo {pages} {137201} (\bibinfo {year}
  {2011})}\BibitemShut {NoStop}%
\bibitem [{\citenamefont {Prosen}(2014)}]{prosen-2014}%
  \BibitemOpen
  \bibfield  {author} {\bibinfo {author} {\bibfnamefont {T.}~\bibnamefont
  {Prosen}},\ }\href {\doibase 10.1103/PhysRevLett.112.030603} {\bibfield
  {journal} {\bibinfo  {journal} {Phys. Rev. Lett.}\ }\textbf {\bibinfo
  {volume} {112}},\ \bibinfo {pages} {030603} (\bibinfo {year}
  {2014})}\BibitemShut {NoStop}%
\bibitem [{\citenamefont {Prosen}(2015)}]{prosen-2015}%
  \BibitemOpen
  \bibfield  {author} {\bibinfo {author} {\bibfnamefont {T.}~\bibnamefont
  {Prosen}},\ }\href {\doibase 10.1088/1751-8113/48/37/373001} {\bibfield
  {journal} {\bibinfo  {journal} {Journal of Physics A: Mathematical and
  Theoretical}\ }\textbf {\bibinfo {volume} {48}},\ \bibinfo {pages} {373001}
  (\bibinfo {year} {2015})}\BibitemShut {NoStop}%
\bibitem [{\citenamefont {{\v{Z}}nidari{\v{c}}}(2010)}]{znidaric-2010}%
  \BibitemOpen
  \bibfield  {author} {\bibinfo {author} {\bibfnamefont {M.}~\bibnamefont
  {{\v{Z}}nidari{\v{c}}}},\ }\href {\doibase 10.1088/1742-5468/2010/05/l05002}
  {\bibfield  {journal} {\bibinfo  {journal} {Journal of Statistical Mechanics:
  Theory and Experiment}\ }\textbf {\bibinfo {volume} {2010}},\ \bibinfo
  {pages} {L05002} (\bibinfo {year} {2010})}\BibitemShut {NoStop}%
\bibitem [{\citenamefont {\ifmmode \check{Z}\else
  \v{Z}\fi{}nidari\ifmmode~\check{c}\else \v{c}\fi{}}(2011)}]{znidaric-2011}%
  \BibitemOpen
  \bibfield  {author} {\bibinfo {author} {\bibfnamefont {M.}~\bibnamefont
  {\ifmmode \check{Z}\else \v{Z}\fi{}nidari\ifmmode~\check{c}\else
  \v{c}\fi{}}},\ }\href {\doibase 10.1103/PhysRevE.83.011108} {\bibfield
  {journal} {\bibinfo  {journal} {Phys. Rev. E}\ }\textbf {\bibinfo {volume}
  {83}},\ \bibinfo {pages} {011108} (\bibinfo {year} {2011})}\BibitemShut
  {NoStop}%
\bibitem [{\citenamefont {Medvedyeva}\ \emph {et~al.}(2016)\citenamefont
  {Medvedyeva}, \citenamefont {Essler},\ and\ \citenamefont
  {Prosen}}]{medvedyeva-2016}%
  \BibitemOpen
  \bibfield  {author} {\bibinfo {author} {\bibfnamefont {M.~V.}\ \bibnamefont
  {Medvedyeva}}, \bibinfo {author} {\bibfnamefont {F.~H.~L.}\ \bibnamefont
  {Essler}}, \ and\ \bibinfo {author} {\bibfnamefont {T.}~\bibnamefont
  {Prosen}},\ }\href {\doibase 10.1103/PhysRevLett.117.137202} {\bibfield
  {journal} {\bibinfo  {journal} {Phys. Rev. Lett.}\ }\textbf {\bibinfo
  {volume} {117}},\ \bibinfo {pages} {137202} (\bibinfo {year}
  {2016})}\BibitemShut {NoStop}%
\bibitem [{\citenamefont {Ilievski}(2017)}]{ilievski2017dissipation}%
  \BibitemOpen
  \bibfield  {author} {\bibinfo {author} {\bibfnamefont {E.}~\bibnamefont
  {Ilievski}},\ }\href {\doibase 10.21468/SciPostPhys.3.4.031} {\bibfield
  {journal} {\bibinfo  {journal} {SciPost Phys.}\ }\textbf {\bibinfo {volume}
  {3}},\ \bibinfo {pages} {031} (\bibinfo {year} {2017})}\BibitemShut {NoStop}%
\bibitem [{\citenamefont {Buca}\ \emph {et~al.}(2020)\citenamefont {Buca},
  \citenamefont {Booker}, \citenamefont {Medenjak},\ and\ \citenamefont
  {Jaksch}}]{buca-2020}%
  \BibitemOpen
  \bibfield  {author} {\bibinfo {author} {\bibfnamefont {B.}~\bibnamefont
  {Buca}}, \bibinfo {author} {\bibfnamefont {C.}~\bibnamefont {Booker}},
  \bibinfo {author} {\bibfnamefont {M.}~\bibnamefont {Medenjak}}, \ and\
  \bibinfo {author} {\bibfnamefont {D.}~\bibnamefont {Jaksch}},\ }\href@noop {}
  {\enquote {\bibinfo {title} {Dissipative bethe ansatz: Exact solutions of
  quantum many-body dynamics under loss},}\ } (\bibinfo {year} {2020}),\
  \Eprint {http://arxiv.org/abs/2004.05955} {arXiv:2004.05955
  [cond-mat.stat-mech]} \BibitemShut {NoStop}%
\bibitem [{\citenamefont {Bastianello}\ \emph {et~al.}(2020)\citenamefont
  {Bastianello}, \citenamefont {De~Nardis},\ and\ \citenamefont
  {De~Luca}}]{bastianello-2020}%
  \BibitemOpen
  \bibfield  {author} {\bibinfo {author} {\bibfnamefont {A.}~\bibnamefont
  {Bastianello}}, \bibinfo {author} {\bibfnamefont {J.}~\bibnamefont
  {De~Nardis}}, \ and\ \bibinfo {author} {\bibfnamefont {A.}~\bibnamefont
  {De~Luca}},\ }\href {\doibase 10.1103/PhysRevB.102.161110} {\bibfield
  {journal} {\bibinfo  {journal} {Phys. Rev. B}\ }\textbf {\bibinfo {volume}
  {102}},\ \bibinfo {pages} {161110} (\bibinfo {year} {2020})}\BibitemShut
  {NoStop}%
\bibitem [{\citenamefont {Essler}\ and\ \citenamefont
  {Piroli}(2020)}]{essler-2020}%
  \BibitemOpen
  \bibfield  {author} {\bibinfo {author} {\bibfnamefont {F.~H.~L.}\
  \bibnamefont {Essler}}\ and\ \bibinfo {author} {\bibfnamefont
  {L.}~\bibnamefont {Piroli}},\ }\href {\doibase 10.1103/PhysRevE.102.062210}
  {\bibfield  {journal} {\bibinfo  {journal} {Phys. Rev. E}\ }\textbf {\bibinfo
  {volume} {102}},\ \bibinfo {pages} {062210} (\bibinfo {year}
  {2020})}\BibitemShut {NoStop}%
\bibitem [{\citenamefont {Ziolkowska}\ and\ \citenamefont
  {Essler}(2020)}]{ziolkowska-2020}%
  \BibitemOpen
  \bibfield  {author} {\bibinfo {author} {\bibfnamefont {A.~A.}\ \bibnamefont
  {Ziolkowska}}\ and\ \bibinfo {author} {\bibfnamefont {F.~H.}\ \bibnamefont
  {Essler}},\ }\href {\doibase 10.21468/SciPostPhys.8.3.044} {\bibfield
  {journal} {\bibinfo  {journal} {SciPost Phys.}\ }\textbf {\bibinfo {volume}
  {8}},\ \bibinfo {pages} {44} (\bibinfo {year} {2020})}\BibitemShut {NoStop}%
\bibitem [{\citenamefont {Sieberer}\ \emph {et~al.}(2016)\citenamefont
  {Sieberer}, \citenamefont {Buchhold},\ and\ \citenamefont
  {Diehl}}]{sieberer-2016}%
  \BibitemOpen
  \bibfield  {author} {\bibinfo {author} {\bibfnamefont {L.~M.}\ \bibnamefont
  {Sieberer}}, \bibinfo {author} {\bibfnamefont {M.}~\bibnamefont {Buchhold}},
  \ and\ \bibinfo {author} {\bibfnamefont {S.}~\bibnamefont {Diehl}},\ }\href
  {\doibase 10.1088/0034-4885/79/9/096001} {\bibfield  {journal} {\bibinfo
  {journal} {Reports on Progress in Physics}\ }\textbf {\bibinfo {volume}
  {79}},\ \bibinfo {pages} {96001} (\bibinfo {year} {2016})}\BibitemShut
  {NoStop}%
\bibitem [{\citenamefont {Bertini}\ \emph {et~al.}(2016)\citenamefont
  {Bertini}, \citenamefont {Collura}, \citenamefont {De~Nardis},\ and\
  \citenamefont {Fagotti}}]{bertini-2016}%
  \BibitemOpen
  \bibfield  {author} {\bibinfo {author} {\bibfnamefont {B.}~\bibnamefont
  {Bertini}}, \bibinfo {author} {\bibfnamefont {M.}~\bibnamefont {Collura}},
  \bibinfo {author} {\bibfnamefont {J.}~\bibnamefont {De~Nardis}}, \ and\
  \bibinfo {author} {\bibfnamefont {M.}~\bibnamefont {Fagotti}},\ }\href
  {\doibase 10.1103/PhysRevLett.117.207201} {\bibfield  {journal} {\bibinfo
  {journal} {Phys. Rev. Lett.}\ }\textbf {\bibinfo {volume} {117}},\ \bibinfo
  {pages} {207201} (\bibinfo {year} {2016})}\BibitemShut {NoStop}%
\bibitem [{\citenamefont {Castro-Alvaredo}\ \emph {et~al.}(2016)\citenamefont
  {Castro-Alvaredo}, \citenamefont {Doyon},\ and\ \citenamefont
  {Yoshimura}}]{olalla-2016}%
  \BibitemOpen
  \bibfield  {author} {\bibinfo {author} {\bibfnamefont {O.~A.}\ \bibnamefont
  {Castro-Alvaredo}}, \bibinfo {author} {\bibfnamefont {B.}~\bibnamefont
  {Doyon}}, \ and\ \bibinfo {author} {\bibfnamefont {T.}~\bibnamefont
  {Yoshimura}},\ }\href {\doibase 10.1103/PhysRevX.6.041065} {\bibfield
  {journal} {\bibinfo  {journal} {Phys. Rev. X}\ }\textbf {\bibinfo {volume}
  {6}},\ \bibinfo {pages} {041065} (\bibinfo {year} {2016})}\BibitemShut
  {NoStop}%
\bibitem [{\citenamefont {Bouchoule}\ \emph {et~al.}(2020)\citenamefont
  {Bouchoule}, \citenamefont {Doyon},\ and\ \citenamefont
  {Dubail}}]{bouchoule-2020}%
  \BibitemOpen
  \bibfield  {author} {\bibinfo {author} {\bibfnamefont {I.}~\bibnamefont
  {Bouchoule}}, \bibinfo {author} {\bibfnamefont {B.}~\bibnamefont {Doyon}}, \
  and\ \bibinfo {author} {\bibfnamefont {J.}~\bibnamefont {Dubail}},\ }\href
  {\doibase 10.21468/SciPostPhys.9.4.044} {\bibfield  {journal} {\bibinfo
  {journal} {SciPost Phys.}\ }\textbf {\bibinfo {volume} {9}},\ \bibinfo
  {pages} {44} (\bibinfo {year} {2020})}\BibitemShut {NoStop}%
\bibitem [{\citenamefont {Friedman}\ \emph {et~al.}(2020)\citenamefont
  {Friedman}, \citenamefont {Gopalakrishnan},\ and\ \citenamefont
  {Vasseur}}]{Friedman_2020}%
  \BibitemOpen
  \bibfield  {author} {\bibinfo {author} {\bibfnamefont {A.~J.}\ \bibnamefont
  {Friedman}}, \bibinfo {author} {\bibfnamefont {S.}~\bibnamefont
  {Gopalakrishnan}}, \ and\ \bibinfo {author} {\bibfnamefont {R.}~\bibnamefont
  {Vasseur}},\ }\href {http://dx.doi.org/10.1103/PhysRevB.101.180302}
  {\bibfield  {journal} {\bibinfo  {journal} {Physical Review B}\ }\textbf
  {\bibinfo {volume} {101}} (\bibinfo {year} {2020})}\BibitemShut {NoStop}%
\bibitem [{\citenamefont {de~Leeuw}\ \emph {et~al.}(2021)\citenamefont
  {de~Leeuw}, \citenamefont {Paletta},\ and\ \citenamefont
  {Pozsgay}}]{deleeuw-2021}%
  \BibitemOpen
  \bibfield  {author} {\bibinfo {author} {\bibfnamefont {M.}~\bibnamefont
  {de~Leeuw}}, \bibinfo {author} {\bibfnamefont {C.}~\bibnamefont {Paletta}}, \
  and\ \bibinfo {author} {\bibfnamefont {B.}~\bibnamefont {Pozsgay}},\
  }\href@noop {} {\enquote {\bibinfo {title} {Constructing integrable lindblad
  superoperators},}\ } (\bibinfo {year} {2021}),\ \Eprint
  {http://arxiv.org/abs/2101.08279} {arXiv:2101.08279 [cond-mat.stat-mech]}
  \BibitemShut {NoStop}%
\bibitem [{\citenamefont {Nardis}\ \emph {et~al.}(2021)\citenamefont {Nardis},
  \citenamefont {Gopalakrishnan}, \citenamefont {Vasseur},\ and\ \citenamefont
  {Ware}}]{denardis2021}%
  \BibitemOpen
  \bibfield  {author} {\bibinfo {author} {\bibfnamefont {J.~D.}\ \bibnamefont
  {Nardis}}, \bibinfo {author} {\bibfnamefont {S.}~\bibnamefont
  {Gopalakrishnan}}, \bibinfo {author} {\bibfnamefont {R.}~\bibnamefont
  {Vasseur}}, \ and\ \bibinfo {author} {\bibfnamefont {B.}~\bibnamefont
  {Ware}},\ }\href@noop {} {\enquote {\bibinfo {title} {Stability of
  superdiffusion in nearly integrable spin chains},}\ } (\bibinfo {year}
  {2021}),\ \Eprint {http://arxiv.org/abs/2102.02219} {arXiv:2102.02219
  [cond-mat.stat-mech]} \BibitemShut {NoStop}%
\bibitem [{\citenamefont {Calabrese}\ and\ \citenamefont
  {Cardy}(2005)}]{calabrese-2005}%
  \BibitemOpen
  \bibfield  {author} {\bibinfo {author} {\bibfnamefont {P.}~\bibnamefont
  {Calabrese}}\ and\ \bibinfo {author} {\bibfnamefont {J.}~\bibnamefont
  {Cardy}},\ }\href {\doibase 10.1088/1742-5468/2005/04/p04010} {\bibfield
  {journal} {\bibinfo  {journal} {Journal of Statistical Mechanics: Theory and
  Experiment}\ }\textbf {\bibinfo {volume} {2005}},\ \bibinfo {pages} {P04010}
  (\bibinfo {year} {2005})}\BibitemShut {NoStop}%
\bibitem [{\citenamefont {Fagotti}\ and\ \citenamefont
  {Calabrese}(2008)}]{fagotti-2008}%
  \BibitemOpen
  \bibfield  {author} {\bibinfo {author} {\bibfnamefont {M.}~\bibnamefont
  {Fagotti}}\ and\ \bibinfo {author} {\bibfnamefont {P.}~\bibnamefont
  {Calabrese}},\ }\href {\doibase 10.1103/PhysRevA.78.010306} {\bibfield
  {journal} {\bibinfo  {journal} {Phys. Rev. A}\ }\textbf {\bibinfo {volume}
  {78}},\ \bibinfo {pages} {010306} (\bibinfo {year} {2008})}\BibitemShut
  {NoStop}%
\bibitem [{\citenamefont {Alba}\ and\ \citenamefont
  {Calabrese}(2017)}]{alba-2017}%
  \BibitemOpen
  \bibfield  {author} {\bibinfo {author} {\bibfnamefont {V.}~\bibnamefont
  {Alba}}\ and\ \bibinfo {author} {\bibfnamefont {P.}~\bibnamefont
  {Calabrese}},\ }\href {\doibase 10.1073/pnas.1703516114} {\bibfield
  {journal} {\bibinfo  {journal} {Proceedings of the National Academy of
  Sciences of the United States of America}\ }\textbf {\bibinfo {volume} {114}}
  (\bibinfo {year} {2017}),\ 10.1073/pnas.1703516114}\BibitemShut {NoStop}%
\bibitem [{\citenamefont {Alba}\ and\ \citenamefont
  {Calabrese}(2018)}]{alba-2018}%
  \BibitemOpen
  \bibfield  {author} {\bibinfo {author} {\bibfnamefont {V.}~\bibnamefont
  {Alba}}\ and\ \bibinfo {author} {\bibfnamefont {P.}~\bibnamefont
  {Calabrese}},\ }\href {\doibase 10.21468/SciPostPhys.4.3.017} {\bibfield
  {journal} {\bibinfo  {journal} {SciPost Phys.}\ }\textbf {\bibinfo {volume}
  {4}},\ \bibinfo {pages} {17} (\bibinfo {year} {2018})}\BibitemShut {NoStop}%
\bibitem [{\citenamefont {Alba}\ and\ \citenamefont
  {Carollo}(2021)}]{alba-2021}%
  \BibitemOpen
  \bibfield  {author} {\bibinfo {author} {\bibfnamefont {V.}~\bibnamefont
  {Alba}}\ and\ \bibinfo {author} {\bibfnamefont {F.}~\bibnamefont {Carollo}},\
  }\href {\doibase 10.1103/PhysRevB.103.L020302} {\bibfield  {journal}
  {\bibinfo  {journal} {Phys. Rev. B}\ }\textbf {\bibinfo {volume} {103}},\
  \bibinfo {pages} {L020302} (\bibinfo {year} {2021})}\BibitemShut {NoStop}%
\bibitem [{\citenamefont {Maity}\ \emph {et~al.}(2020)\citenamefont {Maity},
  \citenamefont {Bandyopadhyay}, \citenamefont {Bhattacharjee},\ and\
  \citenamefont {Dutta}}]{maity-2020}%
  \BibitemOpen
  \bibfield  {author} {\bibinfo {author} {\bibfnamefont {S.}~\bibnamefont
  {Maity}}, \bibinfo {author} {\bibfnamefont {S.}~\bibnamefont
  {Bandyopadhyay}}, \bibinfo {author} {\bibfnamefont {S.}~\bibnamefont
  {Bhattacharjee}}, \ and\ \bibinfo {author} {\bibfnamefont {A.}~\bibnamefont
  {Dutta}},\ }\href {\doibase 10.1103/PhysRevB.101.180301} {\bibfield
  {journal} {\bibinfo  {journal} {Phys. Rev. B}\ }\textbf {\bibinfo {volume}
  {101}},\ \bibinfo {pages} {180301} (\bibinfo {year} {2020})}\BibitemShut
  {NoStop}%
\bibitem [{\citenamefont {Dolgirev}\ \emph {et~al.}(2020)\citenamefont
  {Dolgirev}, \citenamefont {Marino}, \citenamefont {Sels},\ and\ \citenamefont
  {Demler}}]{dolgirev-2020}%
  \BibitemOpen
  \bibfield  {author} {\bibinfo {author} {\bibfnamefont {P.~E.}\ \bibnamefont
  {Dolgirev}}, \bibinfo {author} {\bibfnamefont {J.}~\bibnamefont {Marino}},
  \bibinfo {author} {\bibfnamefont {D.}~\bibnamefont {Sels}}, \ and\ \bibinfo
  {author} {\bibfnamefont {E.}~\bibnamefont {Demler}},\ }\href {\doibase
  10.1103/PhysRevB.102.100301} {\bibfield  {journal} {\bibinfo  {journal}
  {Phys. Rev. B}\ }\textbf {\bibinfo {volume} {102}},\ \bibinfo {pages}
  {100301} (\bibinfo {year} {2020})}\BibitemShut {NoStop}%
\bibitem [{\citenamefont {Jin}\ \emph {et~al.}(2020)\citenamefont {Jin},
  \citenamefont {Filippone},\ and\ \citenamefont {Giamarchi}}]{jin-2020}%
  \BibitemOpen
  \bibfield  {author} {\bibinfo {author} {\bibfnamefont {T.}~\bibnamefont
  {Jin}}, \bibinfo {author} {\bibfnamefont {M.}~\bibnamefont {Filippone}}, \
  and\ \bibinfo {author} {\bibfnamefont {T.}~\bibnamefont {Giamarchi}},\ }\href
  {\doibase 10.1103/PhysRevB.102.205131} {\bibfield  {journal} {\bibinfo
  {journal} {Phys. Rev. B}\ }\textbf {\bibinfo {volume} {102}},\ \bibinfo
  {pages} {205131} (\bibinfo {year} {2020})}\BibitemShut {NoStop}%
\bibitem [{\citenamefont {Maimbourg}\ \emph {et~al.}(2020)\citenamefont
  {Maimbourg}, \citenamefont {Basko}, \citenamefont {Holzmann},\ and\
  \citenamefont {Rosso}}]{maimbourg-2020}%
  \BibitemOpen
  \bibfield  {author} {\bibinfo {author} {\bibfnamefont {T.}~\bibnamefont
  {Maimbourg}}, \bibinfo {author} {\bibfnamefont {D.~M.}\ \bibnamefont
  {Basko}}, \bibinfo {author} {\bibfnamefont {M.}~\bibnamefont {Holzmann}}, \
  and\ \bibinfo {author} {\bibfnamefont {A.}~\bibnamefont {Rosso}},\
  }\href@noop {} {\enquote {\bibinfo {title} {Bath-induced zeno localization in
  driven many-body quantum systems},}\ } (\bibinfo {year} {2020}),\ \Eprint
  {http://arxiv.org/abs/2009.11784} {arXiv:2009.11784 [cond-mat.stat-mech]}
  \BibitemShut {NoStop}%
\bibitem [{\citenamefont {Fr\"oml}\ \emph {et~al.}(2019)\citenamefont
  {Fr\"oml}, \citenamefont {Chiocchetta}, \citenamefont {Kollath},\ and\
  \citenamefont {Diehl}}]{froml-2019}%
  \BibitemOpen
  \bibfield  {author} {\bibinfo {author} {\bibfnamefont {H.}~\bibnamefont
  {Fr\"oml}}, \bibinfo {author} {\bibfnamefont {A.}~\bibnamefont
  {Chiocchetta}}, \bibinfo {author} {\bibfnamefont {C.}~\bibnamefont
  {Kollath}}, \ and\ \bibinfo {author} {\bibfnamefont {S.}~\bibnamefont
  {Diehl}},\ }\href {\doibase 10.1103/PhysRevLett.122.040402} {\bibfield
  {journal} {\bibinfo  {journal} {Phys. Rev. Lett.}\ }\textbf {\bibinfo
  {volume} {122}},\ \bibinfo {pages} {040402} (\bibinfo {year}
  {2019})}\BibitemShut {NoStop}%
\bibitem [{\citenamefont {Tonielli}\ \emph {et~al.}(2019)\citenamefont
  {Tonielli}, \citenamefont {Fazio}, \citenamefont {Diehl},\ and\ \citenamefont
  {Marino}}]{tonielli-2019}%
  \BibitemOpen
  \bibfield  {author} {\bibinfo {author} {\bibfnamefont {F.}~\bibnamefont
  {Tonielli}}, \bibinfo {author} {\bibfnamefont {R.}~\bibnamefont {Fazio}},
  \bibinfo {author} {\bibfnamefont {S.}~\bibnamefont {Diehl}}, \ and\ \bibinfo
  {author} {\bibfnamefont {J.}~\bibnamefont {Marino}},\ }\href {\doibase
  10.1103/PhysRevLett.122.040604} {\bibfield  {journal} {\bibinfo  {journal}
  {Phys. Rev. Lett.}\ }\textbf {\bibinfo {volume} {122}},\ \bibinfo {pages}
  {040604} (\bibinfo {year} {2019})}\BibitemShut {NoStop}%
\bibitem [{\citenamefont {Fr\"oml}\ \emph {et~al.}(2020)\citenamefont
  {Fr\"oml}, \citenamefont {Muckel}, \citenamefont {Kollath}, \citenamefont
  {Chiocchetta},\ and\ \citenamefont {Diehl}}]{froml-2020}%
  \BibitemOpen
  \bibfield  {author} {\bibinfo {author} {\bibfnamefont {H.}~\bibnamefont
  {Fr\"oml}}, \bibinfo {author} {\bibfnamefont {C.}~\bibnamefont {Muckel}},
  \bibinfo {author} {\bibfnamefont {C.}~\bibnamefont {Kollath}}, \bibinfo
  {author} {\bibfnamefont {A.}~\bibnamefont {Chiocchetta}}, \ and\ \bibinfo
  {author} {\bibfnamefont {S.}~\bibnamefont {Diehl}},\ }\href {\doibase
  10.1103/PhysRevB.101.144301} {\bibfield  {journal} {\bibinfo  {journal}
  {Phys. Rev. B}\ }\textbf {\bibinfo {volume} {101}},\ \bibinfo {pages}
  {144301} (\bibinfo {year} {2020})}\BibitemShut {NoStop}%
\bibitem [{\citenamefont {Krapivsky}\ \emph {et~al.}(2019)\citenamefont
  {Krapivsky}, \citenamefont {Mallick},\ and\ \citenamefont
  {Sels}}]{krapivsky-2019}%
  \BibitemOpen
  \bibfield  {author} {\bibinfo {author} {\bibfnamefont {P.~L.}\ \bibnamefont
  {Krapivsky}}, \bibinfo {author} {\bibfnamefont {K.}~\bibnamefont {Mallick}},
  \ and\ \bibinfo {author} {\bibfnamefont {D.}~\bibnamefont {Sels}},\ }\href
  {\doibase 10.1088/1742-5468/ab4e8e} {\bibfield  {journal} {\bibinfo
  {journal} {Journal of Statistical Mechanics: Theory and Experiment}\ }\textbf
  {\bibinfo {volume} {2019}},\ \bibinfo {pages} {113108} (\bibinfo {year}
  {2019})}\BibitemShut {NoStop}%
\bibitem [{\citenamefont {Krapivsky}\ \emph {et~al.}(2020)\citenamefont
  {Krapivsky}, \citenamefont {Mallick},\ and\ \citenamefont
  {Sels}}]{krapivsky-2020}%
  \BibitemOpen
  \bibfield  {author} {\bibinfo {author} {\bibfnamefont {P.~L.}\ \bibnamefont
  {Krapivsky}}, \bibinfo {author} {\bibfnamefont {K.}~\bibnamefont {Mallick}},
  \ and\ \bibinfo {author} {\bibfnamefont {D.}~\bibnamefont {Sels}},\ }\href
  {\doibase 10.1088/1742-5468/ab8118} {\bibfield  {journal} {\bibinfo
  {journal} {Journal of Statistical Mechanics: Theory and Experiment}\ }\textbf
  {\bibinfo {volume} {2020}},\ \bibinfo {pages} {63101} (\bibinfo {year}
  {2020})}\BibitemShut {NoStop}%
\bibitem [{\citenamefont {Rosso}\ \emph {et~al.}(2020)\citenamefont {Rosso},
  \citenamefont {Iemini}, \citenamefont {Schir{\`o}},\ and\ \citenamefont
  {Mazza}}]{rosso-2020}%
  \BibitemOpen
  \bibfield  {author} {\bibinfo {author} {\bibfnamefont {L.}~\bibnamefont
  {Rosso}}, \bibinfo {author} {\bibfnamefont {F.}~\bibnamefont {Iemini}},
  \bibinfo {author} {\bibfnamefont {M.}~\bibnamefont {Schir{\`o}}}, \ and\
  \bibinfo {author} {\bibfnamefont {L.}~\bibnamefont {Mazza}},\ }\href
  {\doibase 10.21468/SciPostPhys.9.6.091} {\bibfield  {journal} {\bibinfo
  {journal} {SciPost Phys.}\ }\textbf {\bibinfo {volume} {9}},\ \bibinfo
  {pages} {91} (\bibinfo {year} {2020})}\BibitemShut {NoStop}%
\bibitem [{\citenamefont {Vernier}(2020)}]{vernier-2020}%
  \BibitemOpen
  \bibfield  {author} {\bibinfo {author} {\bibfnamefont {E.}~\bibnamefont
  {Vernier}},\ }\href {\doibase 10.21468/SciPostPhys.9.4.049} {\bibfield
  {journal} {\bibinfo  {journal} {SciPost Phys.}\ }\textbf {\bibinfo {volume}
  {9}},\ \bibinfo {pages} {49} (\bibinfo {year} {2020})}\BibitemShut {NoStop}%
\bibitem [{\citenamefont {Gericke}\ \emph {et~al.}(2008)\citenamefont
  {Gericke}, \citenamefont {W{\"u}rtz}, \citenamefont {Reitz}, \citenamefont
  {Langen},\ and\ \citenamefont {Ott}}]{gericke-2008}%
  \BibitemOpen
  \bibfield  {author} {\bibinfo {author} {\bibfnamefont {T.}~\bibnamefont
  {Gericke}}, \bibinfo {author} {\bibfnamefont {P.}~\bibnamefont {W{\"u}rtz}},
  \bibinfo {author} {\bibfnamefont {D.}~\bibnamefont {Reitz}}, \bibinfo
  {author} {\bibfnamefont {T.}~\bibnamefont {Langen}}, \ and\ \bibinfo {author}
  {\bibfnamefont {H.}~\bibnamefont {Ott}},\ }\href {\doibase 10.1038/nphys1102}
  {\bibfield  {journal} {\bibinfo  {journal} {Nature Physics}\ }\textbf
  {\bibinfo {volume} {4}},\ \bibinfo {pages} {949} (\bibinfo {year}
  {2008})}\BibitemShut {NoStop}%
\bibitem [{\citenamefont {Brazhnyi}\ \emph {et~al.}(2009)\citenamefont
  {Brazhnyi}, \citenamefont {Konotop}, \citenamefont {P\'erez-Garc\'{\i}a},\
  and\ \citenamefont {Ott}}]{brazhnyi-2009}%
  \BibitemOpen
  \bibfield  {author} {\bibinfo {author} {\bibfnamefont {V.~A.}\ \bibnamefont
  {Brazhnyi}}, \bibinfo {author} {\bibfnamefont {V.~V.}\ \bibnamefont
  {Konotop}}, \bibinfo {author} {\bibfnamefont {V.~M.}\ \bibnamefont
  {P\'erez-Garc\'{\i}a}}, \ and\ \bibinfo {author} {\bibfnamefont
  {H.}~\bibnamefont {Ott}},\ }\href {\doibase 10.1103/PhysRevLett.102.144101}
  {\bibfield  {journal} {\bibinfo  {journal} {Phys. Rev. Lett.}\ }\textbf
  {\bibinfo {volume} {102}},\ \bibinfo {pages} {144101} (\bibinfo {year}
  {2009})}\BibitemShut {NoStop}%
\bibitem [{\citenamefont {Zezyulin}\ \emph {et~al.}(2012)\citenamefont
  {Zezyulin}, \citenamefont {Konotop}, \citenamefont {Barontini},\ and\
  \citenamefont {Ott}}]{zezyulin-2012}%
  \BibitemOpen
  \bibfield  {author} {\bibinfo {author} {\bibfnamefont {D.~A.}\ \bibnamefont
  {Zezyulin}}, \bibinfo {author} {\bibfnamefont {V.~V.}\ \bibnamefont
  {Konotop}}, \bibinfo {author} {\bibfnamefont {G.}~\bibnamefont {Barontini}},
  \ and\ \bibinfo {author} {\bibfnamefont {H.}~\bibnamefont {Ott}},\ }\href
  {\doibase 10.1103/PhysRevLett.109.020405} {\bibfield  {journal} {\bibinfo
  {journal} {Phys. Rev. Lett.}\ }\textbf {\bibinfo {volume} {109}},\ \bibinfo
  {pages} {020405} (\bibinfo {year} {2012})}\BibitemShut {NoStop}%
\bibitem [{\citenamefont {Barontini}\ \emph {et~al.}(2013)\citenamefont
  {Barontini}, \citenamefont {Labouvie}, \citenamefont {Stubenrauch},
  \citenamefont {Vogler}, \citenamefont {Guarrera},\ and\ \citenamefont
  {Ott}}]{barontini-2013}%
  \BibitemOpen
  \bibfield  {author} {\bibinfo {author} {\bibfnamefont {G.}~\bibnamefont
  {Barontini}}, \bibinfo {author} {\bibfnamefont {R.}~\bibnamefont {Labouvie}},
  \bibinfo {author} {\bibfnamefont {F.}~\bibnamefont {Stubenrauch}}, \bibinfo
  {author} {\bibfnamefont {A.}~\bibnamefont {Vogler}}, \bibinfo {author}
  {\bibfnamefont {V.}~\bibnamefont {Guarrera}}, \ and\ \bibinfo {author}
  {\bibfnamefont {H.}~\bibnamefont {Ott}},\ }\href {\doibase
  10.1103/PhysRevLett.110.035302} {\bibfield  {journal} {\bibinfo  {journal}
  {Phys. Rev. Lett.}\ }\textbf {\bibinfo {volume} {110}},\ \bibinfo {pages}
  {035302} (\bibinfo {year} {2013})}\BibitemShut {NoStop}%
\bibitem [{\citenamefont {Patil}\ \emph {et~al.}(2015)\citenamefont {Patil},
  \citenamefont {Chakram},\ and\ \citenamefont {Vengalattore}}]{patil-2015}%
  \BibitemOpen
  \bibfield  {author} {\bibinfo {author} {\bibfnamefont {Y.~S.}\ \bibnamefont
  {Patil}}, \bibinfo {author} {\bibfnamefont {S.}~\bibnamefont {Chakram}}, \
  and\ \bibinfo {author} {\bibfnamefont {M.}~\bibnamefont {Vengalattore}},\
  }\href {\doibase 10.1103/PhysRevLett.115.140402} {\bibfield  {journal}
  {\bibinfo  {journal} {Phys. Rev. Lett.}\ }\textbf {\bibinfo {volume} {115}},\
  \bibinfo {pages} {140402} (\bibinfo {year} {2015})}\BibitemShut {NoStop}%
\bibitem [{\citenamefont {Labouvie}\ \emph {et~al.}(2016)\citenamefont
  {Labouvie}, \citenamefont {Santra}, \citenamefont {Heun},\ and\ \citenamefont
  {Ott}}]{labouvie-2016}%
  \BibitemOpen
  \bibfield  {author} {\bibinfo {author} {\bibfnamefont {R.}~\bibnamefont
  {Labouvie}}, \bibinfo {author} {\bibfnamefont {B.}~\bibnamefont {Santra}},
  \bibinfo {author} {\bibfnamefont {S.}~\bibnamefont {Heun}}, \ and\ \bibinfo
  {author} {\bibfnamefont {H.}~\bibnamefont {Ott}},\ }\href {\doibase
  10.1103/PhysRevLett.116.235302} {\bibfield  {journal} {\bibinfo  {journal}
  {Phys. Rev. Lett.}\ }\textbf {\bibinfo {volume} {116}},\ \bibinfo {pages}
  {235302} (\bibinfo {year} {2016})}\BibitemShut {NoStop}%
\bibitem [{\citenamefont {Lebrat}\ \emph {et~al.}(2019)\citenamefont {Lebrat},
  \citenamefont {H\"ausler}, \citenamefont {Fabritius}, \citenamefont
  {Husmann}, \citenamefont {Corman},\ and\ \citenamefont
  {Esslinger}}]{lebrat2019quantized}%
  \BibitemOpen
  \bibfield  {author} {\bibinfo {author} {\bibfnamefont {M.}~\bibnamefont
  {Lebrat}}, \bibinfo {author} {\bibfnamefont {S.}~\bibnamefont {H\"ausler}},
  \bibinfo {author} {\bibfnamefont {P.}~\bibnamefont {Fabritius}}, \bibinfo
  {author} {\bibfnamefont {D.}~\bibnamefont {Husmann}}, \bibinfo {author}
  {\bibfnamefont {L.}~\bibnamefont {Corman}}, \ and\ \bibinfo {author}
  {\bibfnamefont {T.}~\bibnamefont {Esslinger}},\ }\href {\doibase
  10.1103/PhysRevLett.123.193605} {\bibfield  {journal} {\bibinfo  {journal}
  {Phys. Rev. Lett.}\ }\textbf {\bibinfo {volume} {123}},\ \bibinfo {pages}
  {193605} (\bibinfo {year} {2019})}\BibitemShut {NoStop}%
\bibitem [{\citenamefont {Corman}\ \emph {et~al.}(2019)\citenamefont {Corman},
  \citenamefont {Fabritius}, \citenamefont {H\"ausler}, \citenamefont {Mohan},
  \citenamefont {Dogra}, \citenamefont {Husmann}, \citenamefont {Lebrat},\ and\
  \citenamefont {Esslinger}}]{corman2019quantized}%
  \BibitemOpen
  \bibfield  {author} {\bibinfo {author} {\bibfnamefont {L.}~\bibnamefont
  {Corman}}, \bibinfo {author} {\bibfnamefont {P.}~\bibnamefont {Fabritius}},
  \bibinfo {author} {\bibfnamefont {S.}~\bibnamefont {H\"ausler}}, \bibinfo
  {author} {\bibfnamefont {J.}~\bibnamefont {Mohan}}, \bibinfo {author}
  {\bibfnamefont {L.~H.}\ \bibnamefont {Dogra}}, \bibinfo {author}
  {\bibfnamefont {D.}~\bibnamefont {Husmann}}, \bibinfo {author} {\bibfnamefont
  {M.}~\bibnamefont {Lebrat}}, \ and\ \bibinfo {author} {\bibfnamefont
  {T.}~\bibnamefont {Esslinger}},\ }\href {\doibase
  10.1103/PhysRevA.100.053605} {\bibfield  {journal} {\bibinfo  {journal}
  {Phys. Rev. A}\ }\textbf {\bibinfo {volume} {100}},\ \bibinfo {pages}
  {053605} (\bibinfo {year} {2019})}\BibitemShut {NoStop}%
\bibitem [{\citenamefont {Viti}\ \emph {et~al.}(2016)\citenamefont {Viti},
  \citenamefont {St{\'{e}}phan}, \citenamefont {Dubail},\ and\ \citenamefont
  {Haque}}]{viti-2016}%
  \BibitemOpen
  \bibfield  {author} {\bibinfo {author} {\bibfnamefont {J.}~\bibnamefont
  {Viti}}, \bibinfo {author} {\bibfnamefont {J.~M.}\ \bibnamefont
  {St{\'{e}}phan}}, \bibinfo {author} {\bibfnamefont {J.}~\bibnamefont
  {Dubail}}, \ and\ \bibinfo {author} {\bibfnamefont {M.}~\bibnamefont
  {Haque}},\ }\href {\doibase 10.1209/0295-5075/115/40011} {\bibfield
  {journal} {\bibinfo  {journal} {Epl}\ }\textbf {\bibinfo {volume} {115}},\
  \bibinfo {pages} {40011} (\bibinfo {year} {2016})},\ \Eprint
  {http://arxiv.org/abs/1507.08132} {arXiv:1507.08132} \BibitemShut {NoStop}%
\bibitem [{\citenamefont {Mossel}\ \emph {et~al.}(2010)\citenamefont {Mossel},
  \citenamefont {Palacios},\ and\ \citenamefont {Caux}}]{mossel-2010}%
  \BibitemOpen
  \bibfield  {author} {\bibinfo {author} {\bibfnamefont {J.}~\bibnamefont
  {Mossel}}, \bibinfo {author} {\bibfnamefont {G.}~\bibnamefont {Palacios}}, \
  and\ \bibinfo {author} {\bibfnamefont {J.-S.}\ \bibnamefont {Caux}},\ }\href
  {\doibase 10.1088/1742-5468/2010/09/l09001} {\bibfield  {journal} {\bibinfo
  {journal} {Journal of Statistical Mechanics: Theory and Experiment}\ }\textbf
  {\bibinfo {volume} {2010}},\ \bibinfo {pages} {L09001} (\bibinfo {year}
  {2010})}\BibitemShut {NoStop}%
\bibitem [{\citenamefont {Antal}\ \emph {et~al.}(1999)\citenamefont {Antal},
  \citenamefont {R\'acz}, \citenamefont {R\'akos},\ and\ \citenamefont
  {Sch\"utz}}]{antal-1999}%
  \BibitemOpen
  \bibfield  {author} {\bibinfo {author} {\bibfnamefont {T.}~\bibnamefont
  {Antal}}, \bibinfo {author} {\bibfnamefont {Z.}~\bibnamefont {R\'acz}},
  \bibinfo {author} {\bibfnamefont {A.}~\bibnamefont {R\'akos}}, \ and\
  \bibinfo {author} {\bibfnamefont {G.~M.}\ \bibnamefont {Sch\"utz}},\ }\href
  {\doibase 10.1103/PhysRevE.59.4912} {\bibfield  {journal} {\bibinfo
  {journal} {Phys. Rev. E}\ }\textbf {\bibinfo {volume} {59}},\ \bibinfo
  {pages} {4912} (\bibinfo {year} {1999})}\BibitemShut {NoStop}%
\bibitem [{\citenamefont {Sabetta}\ and\ \citenamefont
  {Misguich}(2013)}]{sabetta-2013}%
  \BibitemOpen
  \bibfield  {author} {\bibinfo {author} {\bibfnamefont {T.}~\bibnamefont
  {Sabetta}}\ and\ \bibinfo {author} {\bibfnamefont {G.}~\bibnamefont
  {Misguich}},\ }\href {\doibase 10.1103/PhysRevB.88.245114} {\bibfield
  {journal} {\bibinfo  {journal} {Phys. Rev. B}\ }\textbf {\bibinfo {volume}
  {88}},\ \bibinfo {pages} {245114} (\bibinfo {year} {2013})}\BibitemShut
  {NoStop}%
\bibitem [{\citenamefont {Press}\ \emph {et~al.}(2007)\citenamefont {Press},
  \citenamefont {Teukolsky}, \citenamefont {Vetterling},\ and\ \citenamefont
  {Flannery}}]{press2007numerical}%
  \BibitemOpen
  \bibfield  {author} {\bibinfo {author} {\bibfnamefont {W.}~\bibnamefont
  {Press}}, \bibinfo {author} {\bibfnamefont {S.}~\bibnamefont {Teukolsky}},
  \bibinfo {author} {\bibfnamefont {W.}~\bibnamefont {Vetterling}}, \ and\
  \bibinfo {author} {\bibfnamefont {B.}~\bibnamefont {Flannery}},\ }\href
  {http://nr.com/} {\emph {\bibinfo {title} {Numerical Recipes: The Art of
  Scientific Computing}}},\ \bibinfo {edition} {3rd}\ ed.\ (\bibinfo
  {publisher} {Cambridge University Press},\ \bibinfo {year}
  {2007})\BibitemShut {NoStop}%
\bibitem [{\citenamefont {Polkovnikov}\ \emph {et~al.}(2011)\citenamefont
  {Polkovnikov}, \citenamefont {Sengupta}, \citenamefont {Silva},\ and\
  \citenamefont {Vengalattore}}]{polkovnikov2011colloquium}%
  \BibitemOpen
  \bibfield  {author} {\bibinfo {author} {\bibfnamefont {A.}~\bibnamefont
  {Polkovnikov}}, \bibinfo {author} {\bibfnamefont {K.}~\bibnamefont
  {Sengupta}}, \bibinfo {author} {\bibfnamefont {A.}~\bibnamefont {Silva}}, \
  and\ \bibinfo {author} {\bibfnamefont {M.}~\bibnamefont {Vengalattore}},\
  }\href {\doibase 10.1103/RevModPhys.83.863} {\bibfield  {journal} {\bibinfo
  {journal} {Rev. Mod. Phys.}\ }\textbf {\bibinfo {volume} {83}},\ \bibinfo
  {pages} {863} (\bibinfo {year} {2011})}\BibitemShut {NoStop}%
\bibitem [{\citenamefont {Eisert}\ \emph {et~al.}(2015)\citenamefont {Eisert},
  \citenamefont {Friesdorf},\ and\ \citenamefont
  {Gogolin}}]{eisert2015quantum}%
  \BibitemOpen
  \bibfield  {author} {\bibinfo {author} {\bibfnamefont {J.}~\bibnamefont
  {Eisert}}, \bibinfo {author} {\bibfnamefont {M.}~\bibnamefont {Friesdorf}}, \
  and\ \bibinfo {author} {\bibfnamefont {C.}~\bibnamefont {Gogolin}},\ }\href
  {\doibase 10.1038/nphys3215} {\bibfield  {journal} {\bibinfo  {journal}
  {Nature Phys.}\ }\textbf {\bibinfo {volume} {11}},\ \bibinfo {pages} {124}
  (\bibinfo {year} {2015})}\BibitemShut {NoStop}%
\bibitem [{\citenamefont {Gogolin}\ and\ \citenamefont
  {Eisert}(2016)}]{gogolin2016equilibration}%
  \BibitemOpen
  \bibfield  {author} {\bibinfo {author} {\bibfnamefont {C.}~\bibnamefont
  {Gogolin}}\ and\ \bibinfo {author} {\bibfnamefont {J.}~\bibnamefont
  {Eisert}},\ }\href {\doibase 10.1088/0034-4885/79/5/056001} {\bibfield
  {journal} {\bibinfo  {journal} {Rep. Progr. Phys.}\ }\textbf {\bibinfo
  {volume} {79}},\ \bibinfo {pages} {056001} (\bibinfo {year}
  {2016})}\BibitemShut {NoStop}%
\bibitem [{\citenamefont {D'Alessio}\ \emph {et~al.}(2016)\citenamefont
  {D'Alessio}, \citenamefont {Kafri}, \citenamefont {Polkovnikov},\ and\
  \citenamefont {Rigol}}]{dalessio2016quantum}%
  \BibitemOpen
  \bibfield  {author} {\bibinfo {author} {\bibfnamefont {L.}~\bibnamefont
  {D'Alessio}}, \bibinfo {author} {\bibfnamefont {Y.}~\bibnamefont {Kafri}},
  \bibinfo {author} {\bibfnamefont {A.}~\bibnamefont {Polkovnikov}}, \ and\
  \bibinfo {author} {\bibfnamefont {M.}~\bibnamefont {Rigol}},\ }\href
  {\doibase 10.1088/0034-4885/79/5/056001} {\bibfield  {journal} {\bibinfo
  {journal} {Adv. Phys.}\ }\textbf {\bibinfo {volume} {65}},\ \bibinfo {pages}
  {239} (\bibinfo {year} {2016})}\BibitemShut {NoStop}%
\bibitem [{\citenamefont {Calabrese}\ \emph {et~al.}(2016)\citenamefont
  {Calabrese}, \citenamefont {Essler},\ and\ \citenamefont
  {Mussardo}}]{calabrese-2016}%
  \BibitemOpen
  \bibfield  {author} {\bibinfo {author} {\bibfnamefont {P.}~\bibnamefont
  {Calabrese}}, \bibinfo {author} {\bibfnamefont {F.~H.~L.}\ \bibnamefont
  {Essler}}, \ and\ \bibinfo {author} {\bibfnamefont {G.}~\bibnamefont
  {Mussardo}},\ }\href {\doibase 10.1088/1742-5468/2016/06/064001} {\bibfield
  {journal} {\bibinfo  {journal} {J. Stat. Mech.}\ }\textbf {\bibinfo {volume}
  {2016}},\ \bibinfo {pages} {64001} (\bibinfo {year} {2016})}\BibitemShut
  {NoStop}%
\bibitem [{\citenamefont {Gobert}\ \emph {et~al.}(2005)\citenamefont {Gobert},
  \citenamefont {Kollath}, \citenamefont {Schollw\"ock},\ and\ \citenamefont
  {Sch\"utz}}]{gobert-2005}%
  \BibitemOpen
  \bibfield  {author} {\bibinfo {author} {\bibfnamefont {D.}~\bibnamefont
  {Gobert}}, \bibinfo {author} {\bibfnamefont {C.}~\bibnamefont {Kollath}},
  \bibinfo {author} {\bibfnamefont {U.}~\bibnamefont {Schollw\"ock}}, \ and\
  \bibinfo {author} {\bibfnamefont {G.}~\bibnamefont {Sch\"utz}},\ }\href
  {\doibase 10.1103/PhysRevE.71.036102} {\bibfield  {journal} {\bibinfo
  {journal} {Phys. Rev. E}\ }\textbf {\bibinfo {volume} {71}},\ \bibinfo
  {pages} {036102} (\bibinfo {year} {2005})}\BibitemShut {NoStop}%
\bibitem [{\citenamefont {Antal}\ \emph {et~al.}(2008)\citenamefont {Antal},
  \citenamefont {Krapivsky},\ and\ \citenamefont {R\'akos}}]{antal-2008}%
  \BibitemOpen
  \bibfield  {author} {\bibinfo {author} {\bibfnamefont {T.}~\bibnamefont
  {Antal}}, \bibinfo {author} {\bibfnamefont {P.~L.}\ \bibnamefont
  {Krapivsky}}, \ and\ \bibinfo {author} {\bibfnamefont {A.}~\bibnamefont
  {R\'akos}},\ }\href {\doibase 10.1103/PhysRevE.78.061115} {\bibfield
  {journal} {\bibinfo  {journal} {Phys. Rev. E}\ }\textbf {\bibinfo {volume}
  {78}},\ \bibinfo {pages} {061115} (\bibinfo {year} {2008})}\BibitemShut
  {NoStop}%
\bibitem [{\citenamefont {Allegra}\ \emph {et~al.}(2016)\citenamefont
  {Allegra}, \citenamefont {Dubail}, \citenamefont {St{\'{e}}phan},\ and\
  \citenamefont {Viti}}]{allegra-2016}%
  \BibitemOpen
  \bibfield  {author} {\bibinfo {author} {\bibfnamefont {N.}~\bibnamefont
  {Allegra}}, \bibinfo {author} {\bibfnamefont {J.}~\bibnamefont {Dubail}},
  \bibinfo {author} {\bibfnamefont {J.-M.}\ \bibnamefont {St{\'{e}}phan}}, \
  and\ \bibinfo {author} {\bibfnamefont {J.}~\bibnamefont {Viti}},\ }\href
  {\doibase 10.1088/1742-5468/2016/05/053108} {\bibfield  {journal} {\bibinfo
  {journal} {Journal of Statistical Mechanics: Theory and Experiment}\ }\textbf
  {\bibinfo {volume} {2016}},\ \bibinfo {pages} {53108} (\bibinfo {year}
  {2016})}\BibitemShut {NoStop}%
\bibitem [{\citenamefont {Gamayun}\ \emph {et~al.}(2020)\citenamefont
  {Gamayun}, \citenamefont {Lychkovskiy},\ and\ \citenamefont
  {Caux}}]{gamayun-2020}%
  \BibitemOpen
  \bibfield  {author} {\bibinfo {author} {\bibfnamefont {O.}~\bibnamefont
  {Gamayun}}, \bibinfo {author} {\bibfnamefont {O.}~\bibnamefont
  {Lychkovskiy}}, \ and\ \bibinfo {author} {\bibfnamefont {J.-S.}\ \bibnamefont
  {Caux}},\ }\href {\doibase 10.21468/SciPostPhys.8.3.036} {\bibfield
  {journal} {\bibinfo  {journal} {SciPost Phys.}\ }\textbf {\bibinfo {volume}
  {8}},\ \bibinfo {pages} {36} (\bibinfo {year} {2020})}\BibitemShut {NoStop}%
\bibitem [{\citenamefont {Dubail}\ \emph
  {et~al.}(2017{\natexlab{a}})\citenamefont {Dubail}, \citenamefont
  {St{\'e}phan}, \citenamefont {Viti},\ and\ \citenamefont
  {Calabrese}}]{dubail-2017}%
  \BibitemOpen
  \bibfield  {author} {\bibinfo {author} {\bibfnamefont {J.}~\bibnamefont
  {Dubail}}, \bibinfo {author} {\bibfnamefont {J.-M.}\ \bibnamefont
  {St{\'e}phan}}, \bibinfo {author} {\bibfnamefont {J.}~\bibnamefont {Viti}}, \
  and\ \bibinfo {author} {\bibfnamefont {P.}~\bibnamefont {Calabrese}},\ }\href
  {\doibase 10.21468/SciPostPhys.2.1.002} {\bibfield  {journal} {\bibinfo
  {journal} {SciPost Phys.}\ }\textbf {\bibinfo {volume} {2}},\ \bibinfo
  {pages} {002} (\bibinfo {year} {2017}{\natexlab{a}})}\BibitemShut {NoStop}%
\bibitem [{\citenamefont {Brun}\ and\ \citenamefont
  {Dubail}(2017)}]{brun-2017}%
  \BibitemOpen
  \bibfield  {author} {\bibinfo {author} {\bibfnamefont {Y.}~\bibnamefont
  {Brun}}\ and\ \bibinfo {author} {\bibfnamefont {J.}~\bibnamefont {Dubail}},\
  }\href {\doibase 10.21468/SciPostPhys.2.2.012} {\bibfield  {journal}
  {\bibinfo  {journal} {SciPost Phys.}\ }\textbf {\bibinfo {volume} {2}},\
  \bibinfo {pages} {012} (\bibinfo {year} {2017})}\BibitemShut {NoStop}%
\bibitem [{\citenamefont {Dubail}\ \emph
  {et~al.}(2017{\natexlab{b}})\citenamefont {Dubail}, \citenamefont
  {St{\'e}phan},\ and\ \citenamefont {Calabrese}}]{dubail-2017a}%
  \BibitemOpen
  \bibfield  {author} {\bibinfo {author} {\bibfnamefont {J.}~\bibnamefont
  {Dubail}}, \bibinfo {author} {\bibfnamefont {J.-M.}\ \bibnamefont
  {St{\'e}phan}}, \ and\ \bibinfo {author} {\bibfnamefont {P.}~\bibnamefont
  {Calabrese}},\ }\href {\doibase 10.21468/SciPostPhys.3.3.019} {\bibfield
  {journal} {\bibinfo  {journal} {SciPost Phys.}\ }\textbf {\bibinfo {volume}
  {3}},\ \bibinfo {pages} {019} (\bibinfo {year}
  {2017}{\natexlab{b}})}\BibitemShut {NoStop}%
\bibitem [{\citenamefont {Brun}\ and\ \citenamefont
  {Dubail}(2018)}]{brun-2018}%
  \BibitemOpen
  \bibfield  {author} {\bibinfo {author} {\bibfnamefont {Y.}~\bibnamefont
  {Brun}}\ and\ \bibinfo {author} {\bibfnamefont {J.}~\bibnamefont {Dubail}},\
  }\href {\doibase 10.21468/SciPostPhys.4.6.037} {\bibfield  {journal}
  {\bibinfo  {journal} {SciPost Phys.}\ }\textbf {\bibinfo {volume} {4}},\
  \bibinfo {pages} {37} (\bibinfo {year} {2018})}\BibitemShut {NoStop}%
\bibitem [{\citenamefont {Ruggiero}\ \emph {et~al.}(2019)\citenamefont
  {Ruggiero}, \citenamefont {Brun},\ and\ \citenamefont
  {Dubail}}]{ruggiero-2019}%
  \BibitemOpen
  \bibfield  {author} {\bibinfo {author} {\bibfnamefont {P.}~\bibnamefont
  {Ruggiero}}, \bibinfo {author} {\bibfnamefont {Y.}~\bibnamefont {Brun}}, \
  and\ \bibinfo {author} {\bibfnamefont {J.}~\bibnamefont {Dubail}},\ }\href
  {\doibase 10.21468/SciPostPhys.6.4.051} {\bibfield  {journal} {\bibinfo
  {journal} {SciPost Phys.}\ }\textbf {\bibinfo {volume} {6}},\ \bibinfo
  {pages} {51} (\bibinfo {year} {2019})}\BibitemShut {NoStop}%
\bibitem [{\citenamefont {Collura}\ \emph {et~al.}(2020)\citenamefont
  {Collura}, \citenamefont {De~Luca}, \citenamefont {Calabrese},\ and\
  \citenamefont {Dubail}}]{collura-2020}%
  \BibitemOpen
  \bibfield  {author} {\bibinfo {author} {\bibfnamefont {M.}~\bibnamefont
  {Collura}}, \bibinfo {author} {\bibfnamefont {A.}~\bibnamefont {De~Luca}},
  \bibinfo {author} {\bibfnamefont {P.}~\bibnamefont {Calabrese}}, \ and\
  \bibinfo {author} {\bibfnamefont {J.}~\bibnamefont {Dubail}},\ }\href
  {\doibase 10.1103/PhysRevB.102.180409} {\bibfield  {journal} {\bibinfo
  {journal} {Phys. Rev. B}\ }\textbf {\bibinfo {volume} {102}},\ \bibinfo
  {pages} {180409} (\bibinfo {year} {2020})}\BibitemShut {NoStop}%
\bibitem [{\citenamefont {Bettelheim}\ and\ \citenamefont
  {Glazman}(2012)}]{bettelheim-2012}%
  \BibitemOpen
  \bibfield  {author} {\bibinfo {author} {\bibfnamefont {E.}~\bibnamefont
  {Bettelheim}}\ and\ \bibinfo {author} {\bibfnamefont {L.}~\bibnamefont
  {Glazman}},\ }\href {\doibase 10.1103/PhysRevLett.109.260602} {\bibfield
  {journal} {\bibinfo  {journal} {Phys. Rev. Lett.}\ }\textbf {\bibinfo
  {volume} {109}},\ \bibinfo {pages} {260602} (\bibinfo {year}
  {2012})}\BibitemShut {NoStop}%
\bibitem [{\citenamefont {Collura}\ \emph {et~al.}(2018)\citenamefont
  {Collura}, \citenamefont {De~Luca},\ and\ \citenamefont
  {Viti}}]{collura-2018}%
  \BibitemOpen
  \bibfield  {author} {\bibinfo {author} {\bibfnamefont {M.}~\bibnamefont
  {Collura}}, \bibinfo {author} {\bibfnamefont {A.}~\bibnamefont {De~Luca}}, \
  and\ \bibinfo {author} {\bibfnamefont {J.}~\bibnamefont {Viti}},\ }\href
  {\doibase 10.1103/PhysRevB.97.081111} {\bibfield  {journal} {\bibinfo
  {journal} {Phys. Rev. B}\ }\textbf {\bibinfo {volume} {97}},\ \bibinfo
  {pages} {081111} (\bibinfo {year} {2018})}\BibitemShut {NoStop}%
\bibitem [{\citenamefont {Burke}\ \emph {et~al.}(2020)\citenamefont {Burke},
  \citenamefont {Wiersig},\ and\ \citenamefont {Haque}}]{burke-2020}%
  \BibitemOpen
  \bibfield  {author} {\bibinfo {author} {\bibfnamefont {P.~C.}\ \bibnamefont
  {Burke}}, \bibinfo {author} {\bibfnamefont {J.}~\bibnamefont {Wiersig}}, \
  and\ \bibinfo {author} {\bibfnamefont {M.}~\bibnamefont {Haque}},\ }\href
  {\doibase 10.1103/PhysRevA.102.012212} {\bibfield  {journal} {\bibinfo
  {journal} {Phys. Rev. A}\ }\textbf {\bibinfo {volume} {102}},\ \bibinfo
  {pages} {012212} (\bibinfo {year} {2020})}\BibitemShut {NoStop}%
\bibitem [{\citenamefont {Wong}(2001)}]{wong}%
  \BibitemOpen
  \bibfield  {author} {\bibinfo {author} {\bibfnamefont {R.}~\bibnamefont
  {Wong}},\ }\href {\doibase 10.1137/1.9780898719260} {\emph {\bibinfo {title}
  {{Asymptotic Approximations of Integrals}}}}\ (\bibinfo  {publisher} {Society
  for Industrial and Applied Mathematics},\ \bibinfo {year} {2001})\BibitemShut
  {NoStop}%
\bibitem [{\citenamefont {Wald}\ \emph
  {et~al.}(2020{\natexlab{a}})\citenamefont {Wald}, \citenamefont {Arias},\
  and\ \citenamefont {Alba}}]{wald-2020}%
  \BibitemOpen
  \bibfield  {author} {\bibinfo {author} {\bibfnamefont {S.}~\bibnamefont
  {Wald}}, \bibinfo {author} {\bibfnamefont {R.}~\bibnamefont {Arias}}, \ and\
  \bibinfo {author} {\bibfnamefont {V.}~\bibnamefont {Alba}},\ }\href {\doibase
  10.1088/1742-5468/ab6b19} {\bibfield  {journal} {\bibinfo  {journal} {Journal
  of Statistical Mechanics: Theory and Experiment}\ }\textbf {\bibinfo {volume}
  {2020}} (\bibinfo {year} {2020}{\natexlab{a}}),\
  10.1088/1742-5468/ab6b19}\BibitemShut {NoStop}%
\bibitem [{\citenamefont {Wald}\ \emph
  {et~al.}(2020{\natexlab{b}})\citenamefont {Wald}, \citenamefont {Arias},\
  and\ \citenamefont {Alba}}]{wald-2020a}%
  \BibitemOpen
  \bibfield  {author} {\bibinfo {author} {\bibfnamefont {S.}~\bibnamefont
  {Wald}}, \bibinfo {author} {\bibfnamefont {R.}~\bibnamefont {Arias}}, \ and\
  \bibinfo {author} {\bibfnamefont {V.}~\bibnamefont {Alba}},\ }\href {\doibase
  10.1103/PhysRevResearch.2.043404} {\bibfield  {journal} {\bibinfo  {journal}
  {Physical Review Research}\ }\textbf {\bibinfo {volume} {2}},\ \bibinfo
  {pages} {043404} (\bibinfo {year} {2020}{\natexlab{b}})}\BibitemShut
  {NoStop}%
\bibitem [{\citenamefont {Alba}(2021)}]{alba2020c}%
  \BibitemOpen
  \bibfield  {author} {\bibinfo {author} {\bibfnamefont {V.}~\bibnamefont
  {Alba}},\ }\href {\doibase 10.21468/SciPostPhys.10.3.056} {\bibfield
  {journal} {\bibinfo  {journal} {SciPost Phys.}\ }\textbf {\bibinfo {volume}
  {10}},\ \bibinfo {pages} {56} (\bibinfo {year} {2021})}\BibitemShut {NoStop}%
\bibitem [{\citenamefont {Vidal}\ and\ \citenamefont
  {Werner}(2002)}]{vidal-2002}%
  \BibitemOpen
  \bibfield  {author} {\bibinfo {author} {\bibfnamefont {G.}~\bibnamefont
  {Vidal}}\ and\ \bibinfo {author} {\bibfnamefont {R.~F.}\ \bibnamefont
  {Werner}},\ }\href {\doibase 10.1103/PhysRevA.65.032314} {\bibfield
  {journal} {\bibinfo  {journal} {Phys. Rev. A}\ }\textbf {\bibinfo {volume}
  {65}},\ \bibinfo {pages} {032314} (\bibinfo {year} {2002})}\BibitemShut
  {NoStop}%
\bibitem [{\citenamefont {Plenio}(2005)}]{plenio-2005}%
  \BibitemOpen
  \bibfield  {author} {\bibinfo {author} {\bibfnamefont {M.~B.}\ \bibnamefont
  {Plenio}},\ }\href {\doibase 10.1103/PhysRevLett.95.090503} {\bibfield
  {journal} {\bibinfo  {journal} {Phys. Rev. Lett.}\ }\textbf {\bibinfo
  {volume} {95}},\ \bibinfo {pages} {090503} (\bibinfo {year}
  {2005})}\BibitemShut {NoStop}%
\bibitem [{\citenamefont {Calabrese}\ \emph {et~al.}(2012)\citenamefont
  {Calabrese}, \citenamefont {Cardy},\ and\ \citenamefont
  {Tonni}}]{calabrese-2012}%
  \BibitemOpen
  \bibfield  {author} {\bibinfo {author} {\bibfnamefont {P.}~\bibnamefont
  {Calabrese}}, \bibinfo {author} {\bibfnamefont {J.}~\bibnamefont {Cardy}}, \
  and\ \bibinfo {author} {\bibfnamefont {E.}~\bibnamefont {Tonni}},\ }\href
  {\doibase 10.1103/PhysRevLett.109.130502} {\bibfield  {journal} {\bibinfo
  {journal} {Phys. Rev. Lett.}\ }\textbf {\bibinfo {volume} {109}},\ \bibinfo
  {pages} {130502} (\bibinfo {year} {2012})}\BibitemShut {NoStop}%
\bibitem [{\citenamefont {Shapourian}\ and\ \citenamefont
  {Ryu}(2019)}]{shapourian-2019}%
  \BibitemOpen
  \bibfield  {author} {\bibinfo {author} {\bibfnamefont {H.}~\bibnamefont
  {Shapourian}}\ and\ \bibinfo {author} {\bibfnamefont {S.}~\bibnamefont
  {Ryu}},\ }\href {\doibase 10.1103/PhysRevA.99.022310} {\bibfield  {journal}
  {\bibinfo  {journal} {Phys. Rev. A}\ }\textbf {\bibinfo {volume} {99}},\
  \bibinfo {pages} {022310} (\bibinfo {year} {2019})}\BibitemShut {NoStop}%
\bibitem [{\citenamefont {Alba}\ and\ \citenamefont
  {Calabrese}(2019)}]{alba2019quantum}%
  \BibitemOpen
  \bibfield  {author} {\bibinfo {author} {\bibfnamefont {V.}~\bibnamefont
  {Alba}}\ and\ \bibinfo {author} {\bibfnamefont {P.}~\bibnamefont
  {Calabrese}},\ }\href {\doibase 10.1209/0295-5075/126/60001} {\bibfield
  {journal} {\bibinfo  {journal} {{EPL} (Europhysics Letters)}\ }\textbf
  {\bibinfo {volume} {126}},\ \bibinfo {pages} {60001} (\bibinfo {year}
  {2019})}\BibitemShut {NoStop}%
\bibitem [{\citenamefont {Alba}\ and\ \citenamefont
  {Heidrich-Meisner}(2014)}]{alba-2014}%
  \BibitemOpen
  \bibfield  {author} {\bibinfo {author} {\bibfnamefont {V.}~\bibnamefont
  {Alba}}\ and\ \bibinfo {author} {\bibfnamefont {F.}~\bibnamefont
  {Heidrich-Meisner}},\ }\href {\doibase 10.1103/PhysRevB.90.075144} {\bibfield
   {journal} {\bibinfo  {journal} {Phys. Rev. B}\ }\textbf {\bibinfo {volume}
  {90}},\ \bibinfo {pages} {075144} (\bibinfo {year} {2014})}\BibitemShut
  {NoStop}%
\bibitem [{\citenamefont {Vicari}(2012)}]{vicari-2012}%
  \BibitemOpen
  \bibfield  {author} {\bibinfo {author} {\bibfnamefont {E.}~\bibnamefont
  {Vicari}},\ }\href {\doibase 10.1103/PhysRevA.85.062324} {\bibfield
  {journal} {\bibinfo  {journal} {Phys. Rev. A}\ }\textbf {\bibinfo {volume}
  {85}},\ \bibinfo {pages} {062324} (\bibinfo {year} {2012})}\BibitemShut
  {NoStop}%
\bibitem [{\citenamefont {Nespolo}\ and\ \citenamefont
  {Vicari}(2013)}]{nespolo-2013}%
  \BibitemOpen
  \bibfield  {author} {\bibinfo {author} {\bibfnamefont {J.}~\bibnamefont
  {Nespolo}}\ and\ \bibinfo {author} {\bibfnamefont {E.}~\bibnamefont
  {Vicari}},\ }\href {\doibase 10.1103/PhysRevA.87.032316} {\bibfield
  {journal} {\bibinfo  {journal} {Phys. Rev. A}\ }\textbf {\bibinfo {volume}
  {87}},\ \bibinfo {pages} {032316} (\bibinfo {year} {2013})}\BibitemShut
  {NoStop}%
\bibitem [{\citenamefont {Ruggiero}\ \emph {et~al.}(2020)\citenamefont
  {Ruggiero}, \citenamefont {Calabrese}, \citenamefont {Doyon},\ and\
  \citenamefont {Dubail}}]{ruggiero-2020}%
  \BibitemOpen
  \bibfield  {author} {\bibinfo {author} {\bibfnamefont {P.}~\bibnamefont
  {Ruggiero}}, \bibinfo {author} {\bibfnamefont {P.}~\bibnamefont {Calabrese}},
  \bibinfo {author} {\bibfnamefont {B.}~\bibnamefont {Doyon}}, \ and\ \bibinfo
  {author} {\bibfnamefont {J.}~\bibnamefont {Dubail}},\ }\href {\doibase
  10.1103/PhysRevLett.124.140603} {\bibfield  {journal} {\bibinfo  {journal}
  {Phys. Rev. Lett.}\ }\textbf {\bibinfo {volume} {124}},\ \bibinfo {pages}
  {140603} (\bibinfo {year} {2020})}\BibitemShut {NoStop}%
\end{thebibliography}%

\end{document}